\documentclass[options,onecolumn]{mn2e}
\usepackage{epsfig}
\newif\ifAMStwofonts

\newcommand{\mt}[1]{\mbox{$\mathbfss{#1}$}}
\newcommand{\wigner}[6]
{{
\left(
\begin{array}{ccc}
#1 & #2 & #3 \\
#4 & #5 & #6 \\
\end{array}
\right)
}}

\newcommand{\VEV}[1]{\langle#1\rangle}
\newcommand{\wigo}[3]{\wigner{\ell_{#1}}{\ell'_{#2}}{\ell''_{#3}}{0}{0}{0}}

\title[Gabor transforms on the sphere]
{Gabor Transforms on the Sphere with Applications to CMB Power Spectrum Estimation}
  
\author[Frode K. Hansen ,Krzysztof M. G\'orski, and Eric Hivon]
  {{Frode K. Hansen$^1$, \thanks{E-mail: frodekh@roma2.infn.it}}, {Krzysztof M. G\'orski$^{2,3}$, \thanks{E-mail:
kgorski@eso.org}} and {Eric Hivon$^4$, \thanks{E-mail: efh@ipac.caltech.edu}}\\
$1$ Dipartimento di Fisica, Universit\`a di Roma `Tor Vergata', Via della Ricerca Scientifica 1, I-00133 Roma, Italy\\
$2$ ESO, Karl-Schwarzschild-Str.2, 85748 Garching bei M\"unchen, Germany\\
$^3$ Warsaw University Observatory, Aleje Ujazdowskie 4,00-478 Warszawa, Poland\\
$4$ IPAC/Caltech, Mail Code 100-22, 770 S. Wilson Av., Pasadena, CA 91125, USA\\
}
\pagerange{\pageref{firstpage}--\pageref{lastpage}}
\pubyear{2002}

\begin{document}

\label{firstpage}

\maketitle

\begin{abstract}
The Fourier transform of a dataset apodised with a window function is known as the Gabor transform. In this paper
we extend the Gabor transform formalism to the sphere with the intention of applying it to CMB data analysis. The
Gabor coefficients on the sphere known as the pseudo power spectrum is studied for windows of different size. By
assuming that the pseudo power spectrum coefficients are Gaussian distributed, we formulate a likelihood ansatz
using these as input parameters to estimate the full sky power spectrum from a patch on the sky. Since this
likelihood can be calculated quickly without having to invert huge matrices, this allows for fast power spectrum
estimation. By using the pseudo power spectrum from several patches on the sky together, the full sky power
spectrum can be estimated from full-sky or nearly full-sky observations.
\end{abstract}
\begin{keywords}
methods: data analysis--methods: statistical--techniques: image processing--cosmology: observations--cosmology:
cosmological parameters
\end{keywords}

\section{introduction}

The Cosmic Microwave Background (CMB) is one of our most important sources of
information about the early universe \cite{bond,jungman,review,durrer}. The pattern of the temperature fluctuations in the CMB contains information about a number of cosmological parameters. If the temperature fluctuations are Gaussian as predicted by most models of the early universe, all this information is stored in the angular power spectrum coefficients $C_\ell$. For this reason, several
experiments have been conducted to measure the CMB power spectrum. The COBE satellite discovered the fluctuations in 1992
\cite{cobe}, and since then several ground based and balloon borne
experiments \cite{boom1,maxima1,boom2,maxima2,dasi1,dasi2} have been made to study the CMB at an ever increasing
resolution. As the amount of CMB data from these experiments is rapidly
growing, the task of extracting the power spectrum from the data is
getting harder.\\

Analysing the CMB data from a given experiment consists of several steps
as the data consists of several components not belonging to the CMB \cite{comp1,comp2}. In this paper, we will
concentrate on extracting the power
spectrum from a CMB map with foregrounds removed.
The standard method of extracting the power spectrum from a sky map is the method of maximum
likelihood. This method gives the smallest error bars on the power
spectrum estimates, but has the drawback that the number of operations
needed to perform the estimation, scales as $N_\mathrm{pix}^3$, where
$N_\mathrm{pix}$ is the number of pixels in the map. For experiments with
high resolution the number of pixels can be up to several million and this method becomes infeasible using current
computers \cite{borrill}.\\

In \cite{OhSpergelHinshaw}, it is shown how the likelihood analysis can be speeded up to
scale as $N_\mathrm{pix}^2$ with assumptions about azimuthal symmetry and
uncorrelated noise. Another $N_\mathrm{pix}^{3/2}$ method for large
azimuthally symmetric parts of the sky with uncorrelated noise was presented in
\cite{pseudo}. The likelihood problem can also be solved exact in $~N_\mathrm{pix}^2$
operations with correlated noise for special scanning strategies as
demonstrated in \cite{ringtorus1,ringtorus2}. In
\cite{bond,BJK,bartlett} it is shown how one can
approximate the likelihood to speed up the calculations, but still an
$~N_\mathrm{pix}^3$ operation is needed. This has led people to find other
estimators than the maximum likelihood estimator to extract the power
spectrum. In \cite{tegmark} an optimal estimator was found but the
calculation scales as $N_\mathrm{pix}^2$ times a huge prefactor. Recently some near optimal estimators have been found
which can be
calculated in $~N_\mathrm{pix}^2$ operations \cite{dore,szapudi,master}
The data from the BOOMERANG \cite{boom1,boom2} experiment was analysed
using the MASTER method \cite{master}. In this method, the power spectrum was
extracted by a quadratic estimator based on the pseudo power spectrum (the power spectrum on the
cut sky). A similar method was suggested by \cite{amad} for the Planck surveyor. Here we propose to use the pseudo
power spectrum ($\tilde C_\ell$) for
likelihood estimation. This principle was also used in \cite{pseudo} but then
for large sky coverage so that the correlations between the $\tilde
C_\ell$ coefficients could be neglected.\\

In this paper, we study the effect of {\it Gabor transforms} on the
sphere. Gabor transforms, or windowed Fourier transforms are just
Fourier transforms where the function $f(x)$ to be Fourier transformed is
multiplied with a {\it Gabor window} $W(x)$ \cite{gabor}. In the
discrete case $f(x_i)$ can be a data stream. If parts of the data
stream are of poor quality or is missing, this can be represented as
$W(x_i)f(x_i)$ where the window $W$ is zero where there are missing
parts. The window can also be formed so that it smoothes the edges
close to the missing parts and in this way avoid ringing in the
Fourier spectrum.\\

We will study the effect of Gabor transforms on the sphere and use it for fast CMB power spectrum estimation.
The Gabor transform in this context is just the multiplication
of the CMB sky with
a window function before using the spherical harmonic transform to get
the Gabor transform coefficients in this case called the {\it pseudo power spectrum}. The window can be a top-hat
to take out
certain parts of the sky in the case of limited sky coverage. Another window can
be a Gaussian Gabor window for smoothing the transition between the
observed and unobserved area of the sky. The Gabor window can also be
designed in such a way as to increase signal-to-noise by giving pixels
with high signal-to-noise higher significance in the analysis. The use
of the windowed Fourier transform was already studied in
\cite{hobson} in the flat-sky approximation. We show that some of
their results are also valid on the sphere.\\

In the standard likelihood approach of power spectrum estimation, the pixels on the CMB sky or the spherical
harmonic coefficients $a_{\ell m}$ are used as elements in the data vector in which case the correlation matrix
will have dimensions of the order $N_\mathrm{pix}\times N_\mathrm{pix}$. A matrix of this size can not be inverted in a
reasonable amount of time with current computers. We propose to use the pseudo power spectrum coefficients $\tilde
C_\ell$ as elements of the data vector in the likelihood. In this case the size of the correlation matrix will at
most be $l_\mathrm{max}\times l_\mathrm{max}$ which can be inverted in a few seconds. The most time consuming part is the
calculation of the elements of the correlation matrix of pseudo-$C_\ell$.\\  

In Section (\ref{sect:gabor}) we will first describe the one dimensional Gabor transform and then define the Gabor
transform on the sphere. We will define the pseudo power spectrum which is just the Gabor coefficients on the CMB
sky. The kernel relating the full sky power spectrum and the pseudo power spectrum for a Gaussian and top-hat Gabor
window will be discussed. Then in Section (\ref{sect:lik}) we will use the pseudo power spectrum as input values to
a maximum likelihood estimation of the full sky power spectrum. The probability distribution of the pseudo power
spectrum coefficients will be assumed Gaussian and we will show that this is a good approximation at high
multipoles ($\ell>100$). Some examples of likelihood estimations of the power spectrum with different noise
patterns will be shown. In Section (\ref{sect:ext}) two extensions of the method will be discussed. First the use
of the pseudo power spectrum from different Gabor windows centred at different points on the sphere simultaneously
is demonstrated. In this way full-sky or nearly full-sky observations can be analysed. The second extension of the
method is the use of Monte Carlo simulations to obtain noise properties in the case where this is faster than using
the analytic expression or where the noise is correlated. Finally in Section (\ref{sect:disc}) the results and
further extensions are discussed.

\section{The Gabor Transformation and the Temperature Power Spectrum}
\label{sect:gabor}

In this section we will first describe the Gabor transform for
functions on a one dimensional line. Then we extend the formalism to functions on
the sphere and the properties of the Gabor transform coefficients on
the CMB sky, the pseudo-$C_\ell$, are discussed. As most CMB experiment will not be able to observe the full sky,
it is important to study the properties of the power spectrum on the sky apodised with a window function. As we will show later, the best
way to construct the window is not always to set it to $1$ in the observed area and to $0$ in the
non-observed area of the sky. For this reason we will study the Gabor transform for windows with different profiles. On the cut sky the pseudo power spectrum coefficients will get coupled \cite{pseudo,master}. We will
study how strong this coupling is for different window sizes and for different windows. We will in particular study
the top-hat and the Gaussian windows. The top-hat window is important, as it is the window
which preserves most of the information in the observed data set. The Gaussian window smoothes of the edges between
the observed and unobserved areas of the sky and in this way cuts off long range correlations between pseudo
$C_\ell$.

\subsection{The one dimensional Gabor transform}
For a data set $d_j$ with N elements, the normal Fourier transform is defined as,
\begin{equation}
\tilde d_k=\sum_jd_j\mathrm{e}^{\mathrm{i}2\pi jk/N}.
\end{equation}
A tilde on $\tilde d$ shows that these are the Fourier coefficients. The inverse transform is then,
\begin{equation}
d_j=\frac{1}{N}\sum_k\tilde d_k\mathrm{e}^{-\mathrm{i}2\pi jk/N}.
\end{equation}
Sometimes it is useful to study the spectrum of just a part
of the data set. This could be if some parts are of poor
quality or the spectrum is changing along the data set. In this case,
one can multiply the data set with a function, removing the unwanted
parts and taking out a segment to be studied. The function can be a step function cutting out the segment
to study with sharp edges or a function which smoothes
the edges of the segment to avoid ringing (typically a Gaussian).\\

The Fourier transform with such a multiplication was studied by Gabor
\cite{gabor} and is called the {\it
Gabor Transform}. It is defined for a segment centred at $j=M$ and with
wavenumber $k$ as,
\begin{equation}
\tilde d_{kM}=\sum_jd_jG_{j-M}\mathrm{e}^{\mathrm{i}2\pi jk/N}.
\end{equation}
Here $G_{j-M}$ is the {\it Gabor window}, the function to multiply the
data set with. The transform is similar to the Wavelet transform. The
difference is that the window function in the Wavelet transform is
frequency dependent so that the size of the segment is changing with
frequency.\\

Analogously to the Fourier transform, there is also an inverse Gabor
transform. To recover the whole data set from a Gabor transform, one
needs the Fourier coefficients taken with several windows $G_{j-M}$
being centred at different points $M$. This means that the
data set has to be split into several segments. The centre of each
segment is set to
$M=mK$ where $K$ determines the density of segments and $m$ is an
integer specifying the
segment number. One then has for the inverse transform
\begin{equation}
d_j=\sum_m\sum_k\tilde d_{km}g_{km}.
\end{equation}
Due to the non-orthogonality of the Gabor transform, the {\it dual
Gabor window} $g_{km}$ is not trivial to find, but several techniques
have been developed for calculating this dual window (e.g. \cite{dual}
and references therein).\\

In this paper we will study the Gabor transform on the sphere and
apply it to CMB analysis. We will take out a disc on the CMB sky,
using either top-hat or Gaussian apodisation and then derive the pseudo
power spectrum $\tilde C_\ell$ on the apodised sky. The $\tilde
C_\ell$ will be used for likelihood estimation of the underlying full
sky power
spectrum. We also show
how several discs (segments) centred at different points can be
combined to yield the full sky power spectrum.

\subsection{Gabor transform on the sphere}
We start by defining the $\tilde C_\ell$ for a Gabor window
$G({\mathbf{\hat n}})$ as,
\begin{equation}
\tilde C_\ell=\sum_m \frac{\tilde a_{\ell m}^*\tilde a_{\ell m}}{2\ell+1},
\end{equation}
where
\begin{equation}
\label{eq:psalm}
\tilde a_{\ell m}=\int d{\mathbf{\hat n}} T({\mathbf{\hat n}})G({\mathbf{\hat n}})Y^*_{\ell m}({\mathbf{\hat n}}).
\end{equation}
Here $T({\mathbf{\hat n}})$ is the observed temperature in the direction of the
unit vector ${\mathbf{\hat n}}$, $Y_{\ell m}({\mathbf{\hat n}})$ is the
spherical Harmonic function and $G({\mathbf{\hat n}})$ is the Gabor window. We now find an expression for the expectation value of $\tilde C_\ell$.

We will here use a Gabor window which is azimuthally symmetric about a point
${\mathbf{\hat n}}_0$ on the sphere, so that the window is only a function of
the angular
distance from this point on the sphere
$\cos\theta={\mathbf{\hat n}}\cdot{\mathbf{\hat n}}_0$. Then one can write the Legendre expansion
of the window as,
\begin{equation}
G(\theta)=\sum_{\ell}\frac{2\ell+1}{4\pi}g_\ell P_\ell(\cos\theta)=\sum_{\ell
m}g_\ell Y_{\ell m}({\mathbf{\hat n}})Y^*_{\ell m}({\mathbf{\hat n}}_0).
\end{equation}
One can also write,
\begin{equation}
T({\mathbf{\hat n}})=\sum_{\ell m}a_{\ell m}Y_{\ell m}({\mathbf{\hat n}}).
\end{equation}
Inserting these two expressions in equation (\ref{eq:psalm}) one gets
\begin{eqnarray}
\tilde a_{\ell
m}&=&\sum_{\ell'm'}a_{\ell'm'}\sum_{\ell''m''}g_{\ell''}Y^*_{\ell''m''}({\mathbf{n}}_0)\int Y^*_{\ell
m}({\mathbf{\hat n}})Y_{\ell'm'}({\mathbf{\hat n}})Y_{\ell''m''}({\mathbf{\hat n}})d{\mathbf{\hat n}}\\
&=&\sum_{\ell'm'}a_{\ell'm'}\sum_{\ell''m''}g_{\ell''}Y^*_{\ell''m''}({\mathbf{\hat
n}}_0)\sqrt{\frac{(2\ell+1)(2\ell'+1)(2\ell''+1)}{4\pi}}\\
&&\times\wigner{\ell}{\ell'}{\ell''}{-m}{m'}{m''}\wigo{}{}{}(-1)^m,
\end{eqnarray}
where relation (\ref{eq:wigy}) for Wigner 3j Symbols were used.
Using this expression, the relation $\VEV{a_{\ell
m}^*a_{\ell'm'}}=C_\ell\delta_{\ell\ell'}\delta_{mm'}$ and the
orthogonality of Wigner symbols (equation \ref{eq:wigort}),
one can write $\VEV{\tilde C_\ell}$ as,
\begin{equation}
\label{eq:kernrel}
\VEV{\tilde C_\ell}=\sum_{\ell'}C_{\ell'}K(\ell,\ell').
\end{equation}
With $C_\ell$ we will always mean $\VEV{C_\ell}$ when we are referring to the
full sky power spectrum. In this expression, $K(\ell,\ell')$ is the Gabor kernel, 
\begin{equation}
\label{eq:kernexp}
K(\ell,\ell')=
(2\ell'+1)\sum_{\ell''}g_{\ell''}^2\frac{(2\ell''+1)}{(4\pi)^2}\wigo{}{}{}^2.
\end{equation}
The Legendre coefficients  $g_\ell$, are found by the inverse Legendre
transformation,
\begin{equation}
g_\ell=2\pi\int_{\theta=0}^{\theta=\theta_\mathrm{C}} G(\theta)P_\ell(\cos\theta)d\cos\theta,
\end{equation}
where $\theta_\mathrm{C}$ is the cut-off angle where the window goes to zero.
One sees from the expression for the kernel, that there is no dependency
on ${\mathbf{\hat n}}_0$. This means that $\VEV{\tilde C_\ell}$ is the same,
independent on where the Gabor window is centred. In the rest of this section we will study the shape of this kernel which couples the $\tilde C_\ell$ on the apodised sphere.

\begin{figure}
\begin{center}
\leavevmode
\epsfig {file=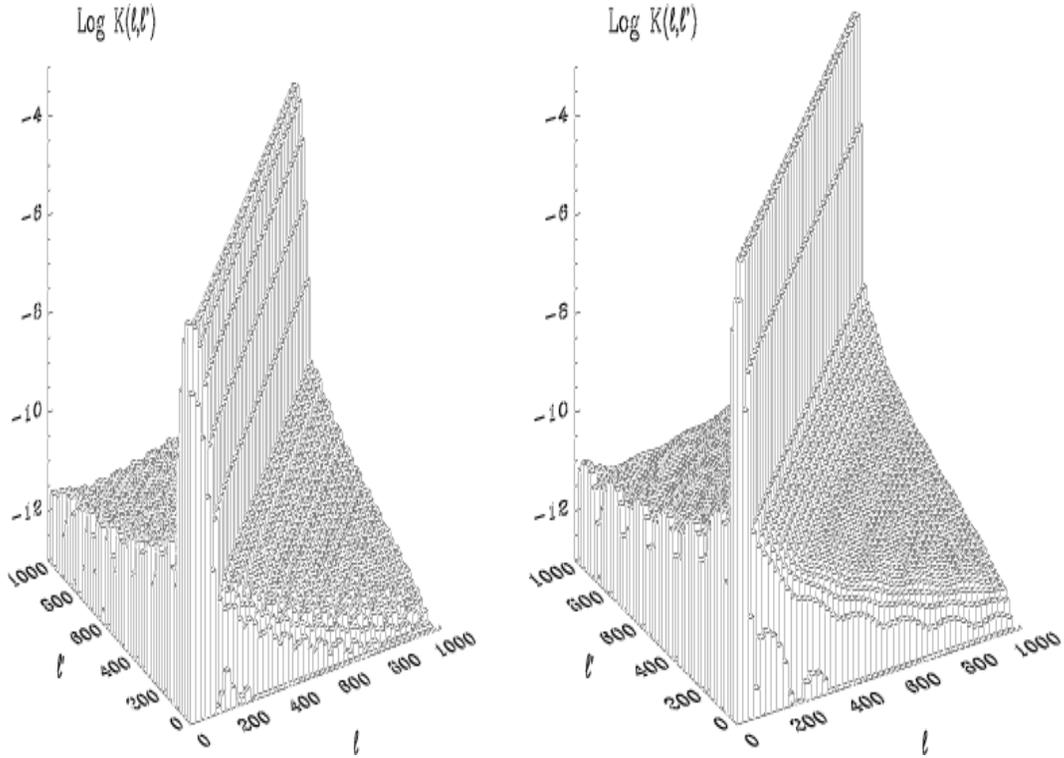,height=10cm,width=14cm}
\caption{The logarithm of the kernel \protect{$K(\ell,\ell')$} describing the
connection between the spherical harmonic coefficients \protect{$C_\ell$} on the
full sky and the corresponding coefficients \protect{$\tilde C_\ell$} on the
apodised sky via the relation \protect{$\tilde
C_\ell=\sum_{\ell'}K(\ell,\ell')C_{\ell'}$}. The figure shows the
kernel for a \protect{$5^\circ$} and
\protect{$15^\circ$} FWHM Gaussian Gabor window with \protect{$\theta_\mathrm{C}=3\sigma$}
(left and right respectively).}
\label{fig:kernelg}
\end{center}
\end{figure}

\begin{figure}
\begin{center}
\leavevmode
\epsfig {file=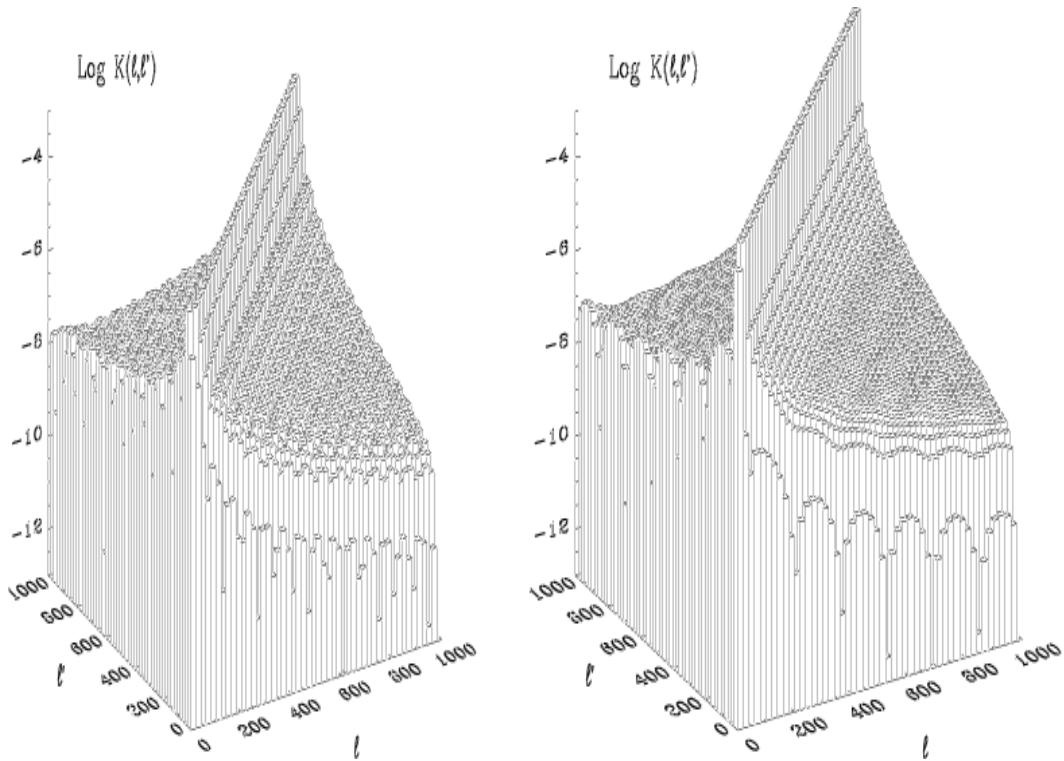,height=10cm,width=14cm}
\caption{Same as Fig. \protect{\ref{fig:kernelg}} for top-hat
Gabor windows covering the same area on the sky as the Gaussian windows.}
\label{fig:kernelth}
\end{center}
\end{figure}

\begin{figure}
\begin{center}
\leavevmode
\epsfig {file=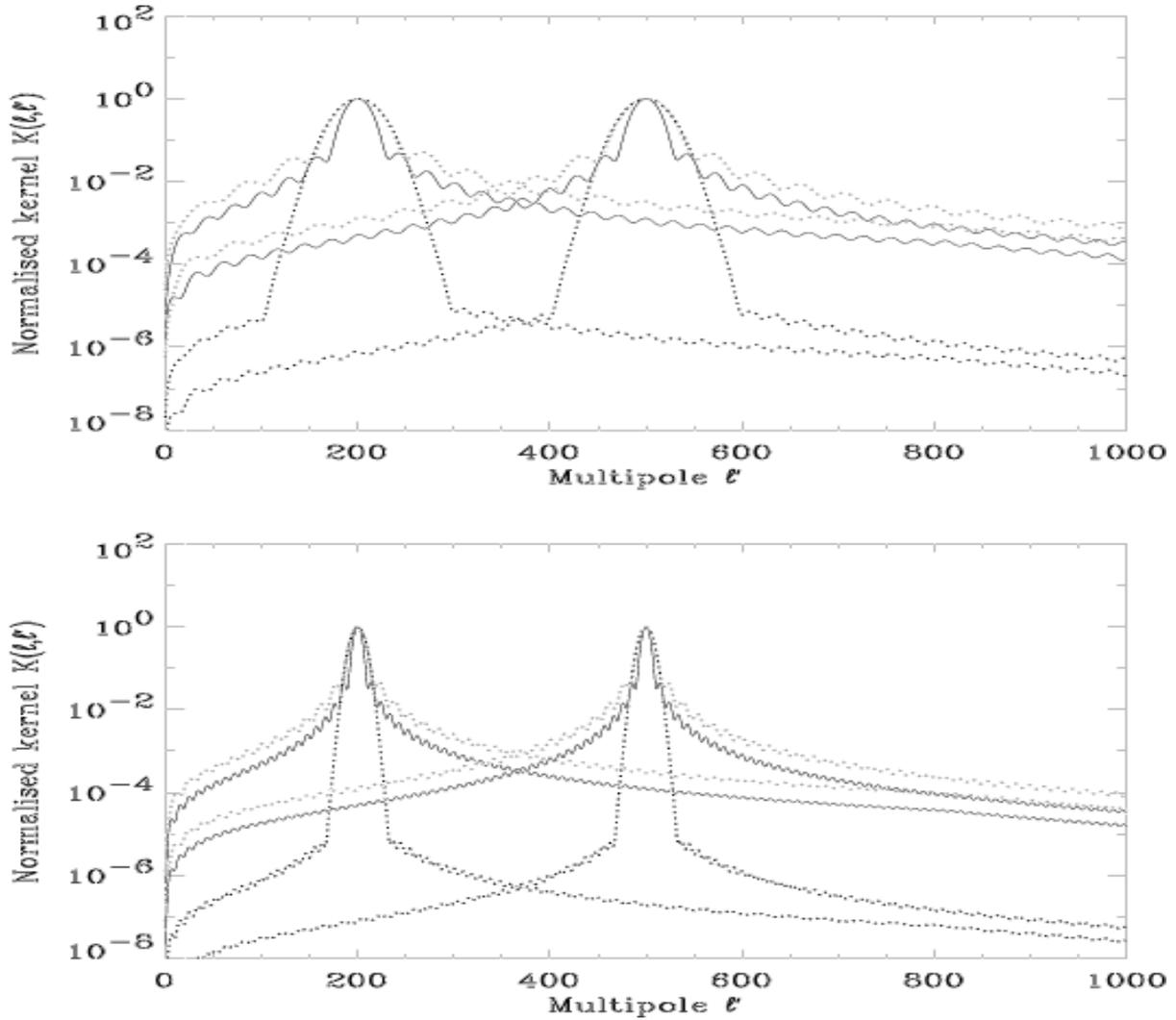,height=14cm,width=16cm}
\caption{The figures show slices of the kernel $K(\ell,\ell')$
connecting the full sky and cut sky spherical harmonic
coefficients. The full kernels are shown in Figs.
\ref{fig:kernelg} and \ref{fig:kernelth}. The slices are taken at \protect{$\ell=200$ and $\ell=500$} for the 5
(upper plot)
and 15 (lower plot) degree Gaussian Gabor window (dotted black line). The solid line is
for the corresponding (same area on the sky) top-hat window $W_\mathrm{A}$. The dotted coloured line is for the top-hat window $W_\mathrm{I}$ having the
same integrated area as the Gaussian window. The
kernels are here normalised so that the peak in the given slice has its
maximum at $1$. In this way one can easier compare the shape of the kernels.}
\label{fig:cutkern}
\end{center}
\end{figure}

\begin{figure}
\begin{center}
\leavevmode
\epsfig {file=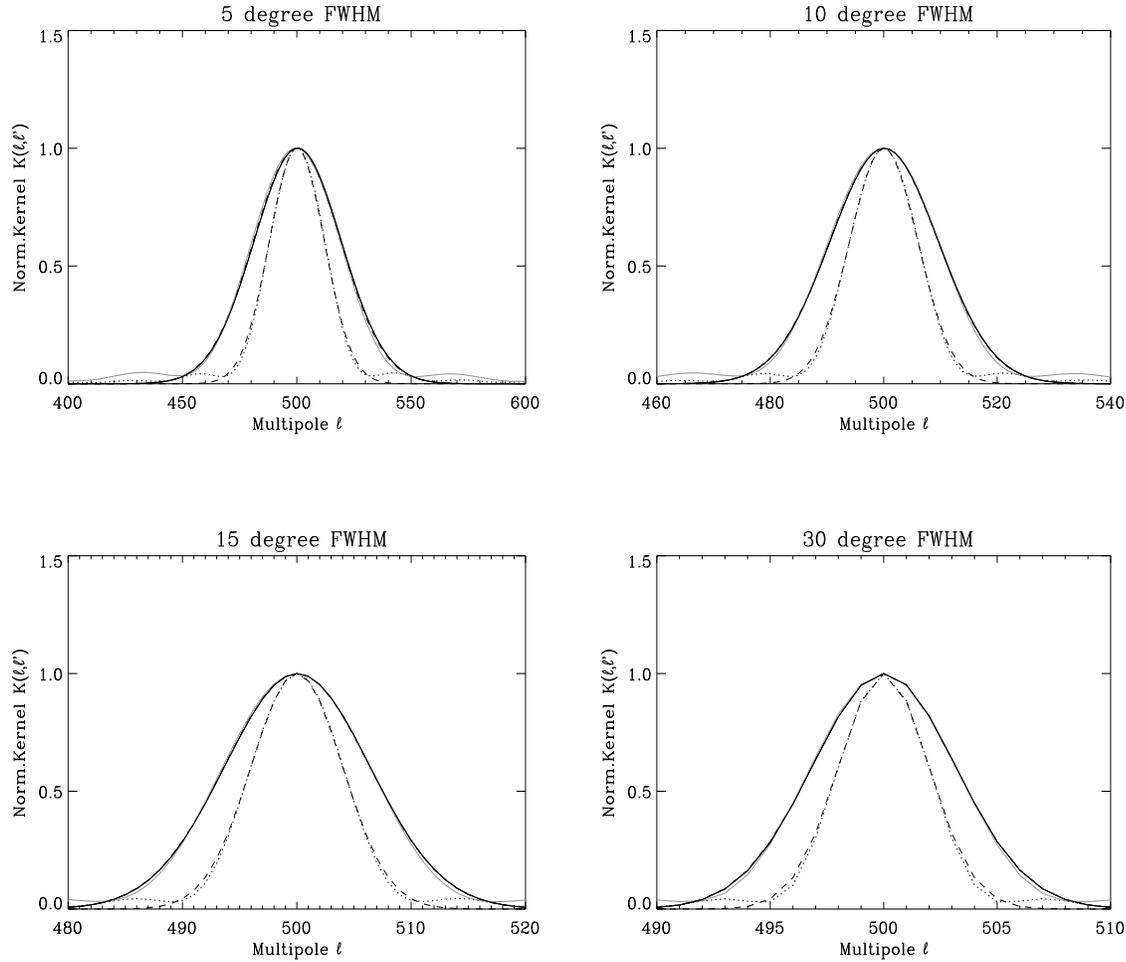,bbllx=30pt,bblly=0pt,bburx=596pt,bbury=800pt,height=14cm,width=16cm}

\caption{The figures show a slice of the kernel $K(\ell,\ell')$
connecting the full sky and cut sky spherical harmonic
coefficients. The slices are taken at \protect{$\ell=500$} for a
\protect{$5$}, \protect{$10$}, \protect{$15$} and \protect{$30$} degree FWHM Gaussian Gabor window (solid
line) with a \protect{$\theta_\mathrm{C}=3\sigma$} cut-off. The dotted
line shows the kernel for a top-hat window $W_\mathrm{A}$ covering the same area on
the sky. The coloured lines which are almost on top of the lines for the Gaussian Gabor windows show the kernels
for a top-hat window $W_\mathrm{I}$ which has the same integrated area as the Gaussian windows $W_\mathrm{G}$. The dashed lines which are almost on top of the dotted and solid line (and for this reason not so easily seen in the plot) are Gaussian fits to the curves.}
\label{fig:widths}
\end{center}
\end{figure}

\begin{figure}
\begin{center}
\leavevmode
\epsfig {file=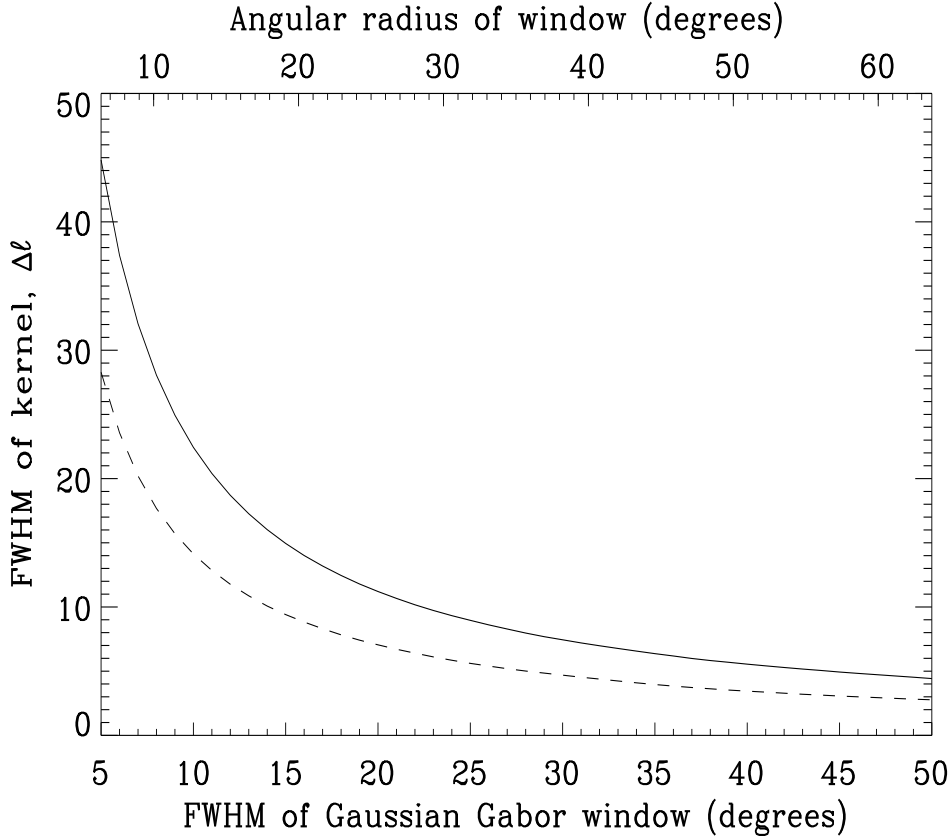,height=12cm,width=14cm}
\caption{ The figure shows the uncertainty relation
\protect{$\Delta\ell\Delta\theta=\mathrm{constant}$} for a Gabor
transform on the sphere. The solid line shows the width \protect{$\Delta\ell$}
of the Gabor kernel \protect{$K(\ell,\ell')$} connecting the full sky
and the cut sky power spectra when applying a Gaussian Gabor
window with a cut \protect{$\theta_\mathrm{C}=3\sigma$}. The FWHM is shown on the lower abscissa. The dashed
line shows the width of the kernel for a top-hat window. The full radius of the top-hat window is shown on
the upper abscissa. The curves are well described by
\protect{$\Delta\ell=220/\theta_\mathrm{FWHM}$} and
\protect{$\Delta\ell=175/\theta_\mathrm{radius}$} for the Gaussian and top-hat
windows respectively. As will be discussed in Section \ref{sect:results}, this relation also describes the width of
the correlation matrix of $\tilde C_\ell$. The width of the correlation matrix using a Gaussian window follows the
relation in this plot times a factor $1.42$. For the top-hat window, the width of the correlation matrix is the
same as for the kernel shown in this figure.}
\label{fig:fwhmrel}
\end{center}
\end{figure}

\begin{figure}
\begin{center}
\leavevmode
\epsfig {file=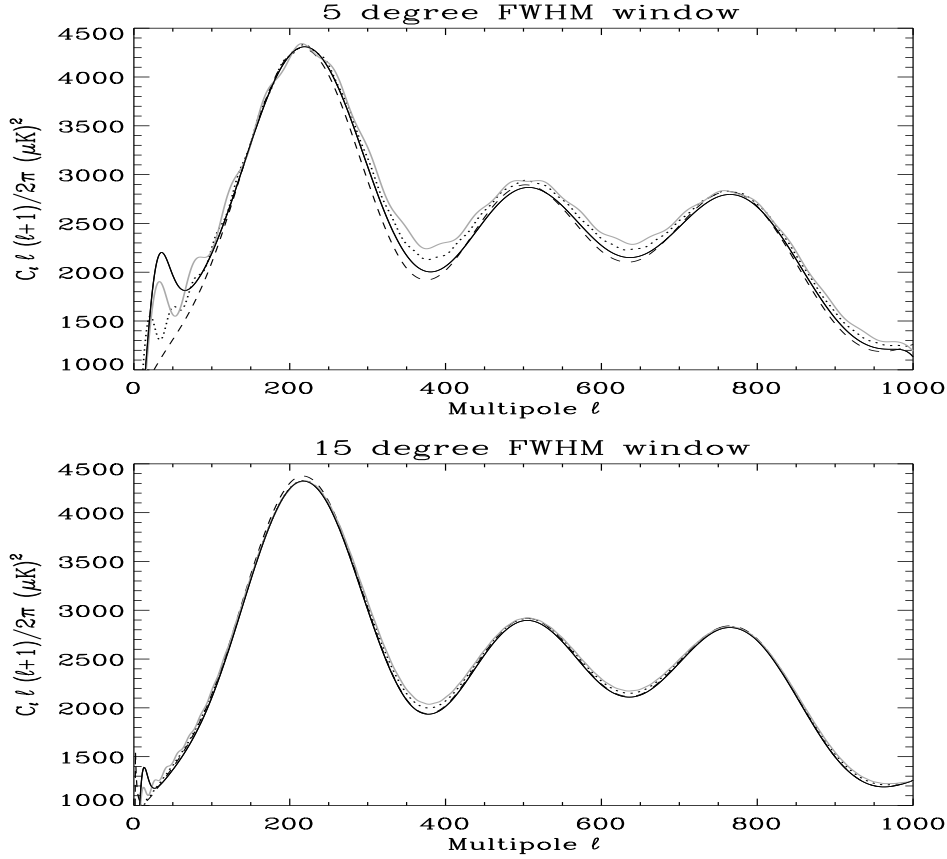,bbllx=0pt,bblly=150pt,bburx=696pt,bbury=800pt,height=12cm,width=16cm}
\caption{The windowed power spectra \protect{$\tilde C_\ell$} for a $5$ and $15$ degree FWHM Gaussian
Gabor Gabor window $W_\mathrm{G}$ cut at $\theta_\mathrm{C}=3\sigma$ (solid line) and for a top-hat window $W_\mathrm{A}$ covering the same area
on the sky (dotted line). The spectrum for the top-hat window $W_\mathrm{I}$ for which the integrated area of the window
corresponds to the Gaussian is shown as a coloured line. All spectra are normalised in such a way that they can be
compared
directly with the full sky spectrum which is shown on each plot as a
dashed line. Only in the first plot are all four lines visible. In
the last plot, the full sky spectrum and the Gaussian pseudo
spectrum (dashed and solid line) are only distinguishable in the first
few multipoles.}
\label{fig:pcl}
\end{center}
\end{figure}

\begin{figure}
\begin{center}
\leavevmode
\epsfig {file=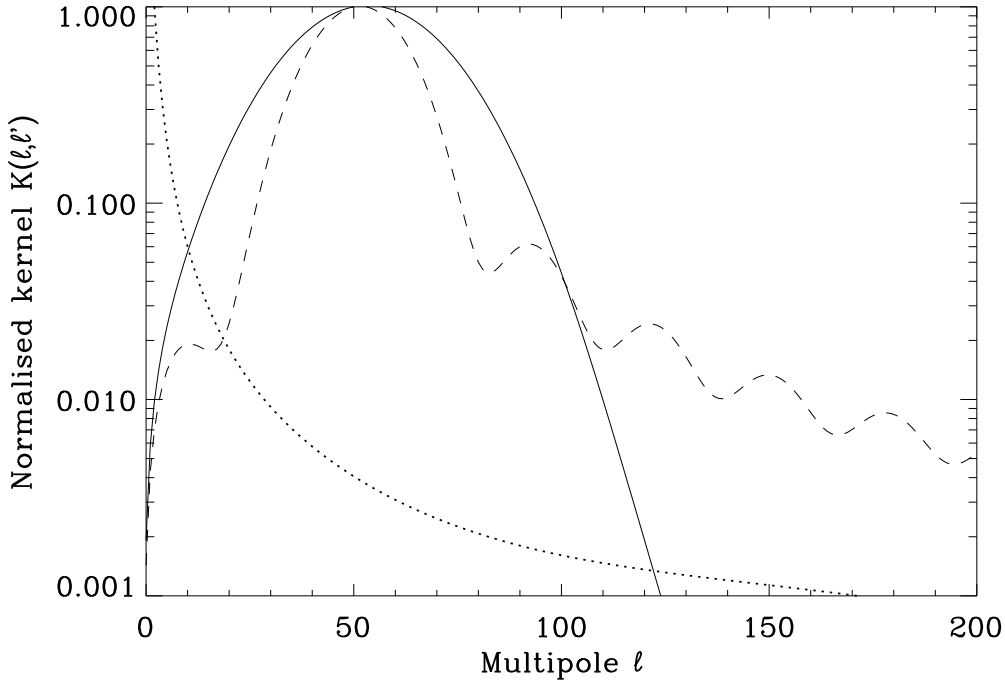,height=10cm,width=14cm}
\caption{The figure shows a slice of the kernel \protect{$K(\ell,\ell')$}
connecting the full sky and cut sky spherical harmonic
coefficients. The slice is taken at \protect{$\ell=50$} for a
\protect{$5$} degree FWHM Gaussian Gabor window (solid
line) and a corresponding top-hat window $W_\mathrm{A}$ (dashed line).  The kernels are normalised to one at
the peak. A typical power spectrum normalised to one at the quadrupole
is plotted as a dotted line. The figure aims at explaining the extra
peak in the pseudo power spectrum at low multipoles for the Gaussian
Gabor window shown in Fig. \ref{fig:pcl}.}
\label{fig:kerncutlowl}
\end{center}
\end{figure}

\begin{figure}
\begin{center}
\leavevmode
\epsfig {file=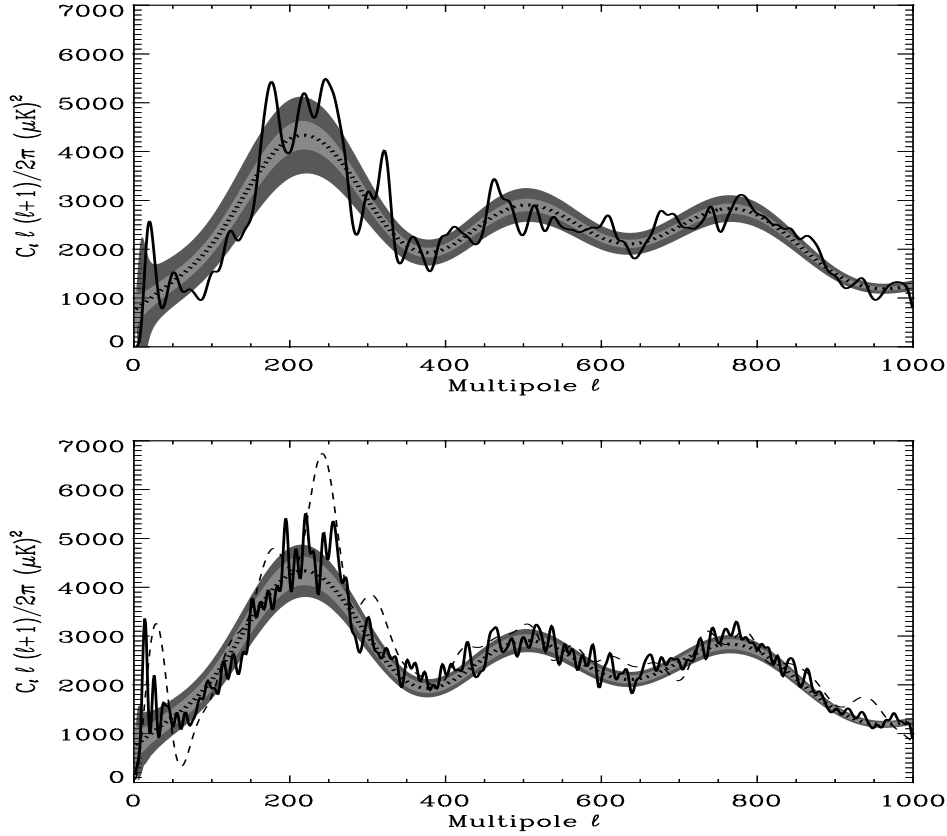,bbllx=0pt,bblly=150pt,bburx=696pt,bbury=800pt,height=12cm,width=16cm}
\caption{One realisation of the windowed power spectra. The upper plot
shows a realisation of a pseudo power spectrum using a $15$ degree
FWHM Gaussian Gabor window. The pseudo spectrum is normalised in such
a way that it can be compared directly to the full sky spectrum which
average is shown as a dotted line. The lower plot shows the same
realisation using a corresponding top-hat window. The light shaded area
shows $1\sigma$ cosmic variance around the full sky average spectrum. The darker
area shows $1\sigma$ cosmic and sampling variance taken from the theoretical formula. On the lower plot, the pseudo
spectrum with a top-hat window $W_\mathrm{A}$ having the same integrated area as in the upper plot is shown as a dashed line.}
\label{fig:pclex}
\end{center}
\end{figure}

In Fig. \ref{fig:kernelg} we have plotted the kernel for a Gaussian
Gabor window,
\begin{eqnarray}
G(\theta)=&\mathrm{e}^{-\theta^2/(2 \sigma^2)}&\theta\leq\theta_\mathrm{C},\\
G(\theta)=&0&\theta>\theta_\mathrm{C},
\end{eqnarray}
with $5$ and $15$ degrees FWHM (corresponding to
$\sigma=2.12^\circ$ and $\sigma=6.38^\circ$) and $\theta_\mathrm{C}=3\sigma$. One sees that the kernel is
centred about $\ell=\ell'$, and falls off rapidly. Fig.
(\ref{fig:kernelth}) shows the same for the corresponding top-hat Gabor
windows,
\begin{eqnarray}
G(\theta)=&1&\theta\leq\theta_\mathrm{C},\\
G(\theta)=&0&\theta>\theta_\mathrm{C}.
\end{eqnarray}
The top-hat windows are covering the same area on the sky as
the corresponding Gaussian windows in Fig. \ref{fig:kernelg}
($\theta_\mathrm{C}$ is the same). Ones
sees that the diagonal is broader for the smaller windows
indicating stronger couplings. Another thing to notice is that whereas
the kernel for the top-hat Gabor window only falls by about 4 orders of magnitude from the
diagonal to the far off-diagonal elements, the Gaussian Gabor kernel falls
by about 8 orders of magnitude (the vertical axis on the four plots
are the same). The smooth cut-off of the Gaussian
Gabor window cuts off long range correlations in spherical harmonic space. One of the aims of the first part of
this paper is to see how the pseudo power spectrum of a given disc on
the sky (top-hat window) is affected by the multiplication with a
Gaussian Gabor window. For this reason the pseudo spectrum will be
studied for a top-hat and a Gaussian covering the same area on the sky. We will also study a top-hat window which has  the same integrated area as the Gaussian window. The cut-off angle $\theta_\mathrm{C}=\theta_\mathrm{int}$ for these windows is given by
\begin{equation}
\int_{\theta=0}^{\theta=3\sigma}G(\theta)d\cos\theta=\int_{\theta=0}^{\theta=\theta_\mathrm{int}}d\cos\theta.
\end{equation}
In this section we will be comparing a Gaussian window (called $W_\mathrm{G}$) having $\theta_\mathrm{C}=3\sigma$ with a
top-hat window (called $W_\mathrm{A}$) having the same area on the sky ($\theta_\mathrm{C}=3\sigma$) and with a top-hat window (called $W_\mathrm{I}$) having the same
integrated area ($\theta_\mathrm{C}=\theta_\mathrm{int}$).\\

 In Fig.
\ref{fig:cutkern}, we have plotted slices of the kernel at $\ell=200$ and
$\ell=500$ for the 5 and 15 degree
FWHM Gaussian Gabor windows (dashed line). The
solid line is the corresponding kernel (same area on the sky) when
using the top-hat Gabor window ($W_\mathrm{A}$). One sees that the Gaussian window
effectively cuts off long range correlations whereas the top-hat window
is narrower close to the diagonal. The Gaussian window has larger
short range correlations. The coloured lines show the slice of the kernel for a top-hat window having the same
integrated area as the Gaussian window ($W_\mathrm{I}$). These kernels have the same widths as the kernels for the Gaussian windows, but the long range correlations are
significantly larger. In \cite{hobson} it was shown that in the flat-sky approximation, the long range correlations are significant when the window has a sharp cut-off. On the sphere we see that even for a sharp top-hat window the long range correlations are damped.\\

Fig. \ref{fig:widths} shows how the width of the kernel gets
narrower and the correlations smaller as the Gabor window opens
up. The four kernels are shown for $\ell=500$ and the Gaussian
windows have 5, 10, 15 and 30 degree FWHM with $\theta_\mathrm{C}=3\sigma$. The same kernels for the
top-hat windows $W_\mathrm{A}$ (dotted lines) and $W_\mathrm{I}$ (coloured line) are plotted on top.
Gaussian fits are plotted on top of the kernels and
show that the kernels are very close to Gaussian functions near the
diagonal.\\

In Fig. \ref{fig:fwhmrel} we have plotted the relation between the
FWHM width $\Delta\ell$ of the kernel and the size $\Delta\theta$ of the window for
Gaussian and top-hat windows. The two curves are very
well described by $\Delta\ell=220/\theta_\mathrm{FWHM}$ for the Gaussian
window ($\theta_\mathrm{FWHM}$ in degrees) and $\Delta\ell=175/\theta_\mathrm{radius}$ ($\theta_\mathrm{radius}$ being the radius of the
top-hat window in degrees) for
the top-hat window. Clearly for a given observed area of the sky,
multiplying with a Gaussian will increase the FWHM of the kernel. This
is also what was seen in Figs \ref{fig:cutkern} and
\ref{fig:widths}. We will see that this results in a lower spectral resolution for the Gaussian window compared to the top-hat window. But the lower long range correlations of the Gaussian window makes the shape of the pseudo power spectrum closer to that of the full sky power spectrum.

In Fig. \ref{fig:pcl}, we show the shapes of the $\tilde C_\ell$
for Gaussian and top-hat windows compared to the full sky spectrum. The plots which were
made using the analytical formula (\ref{eq:kernrel}) show
$\tilde C_\ell$ for a $5$ and $15$ degree FWHM Gaussian
Gabor window (solid line) cut at $3\sigma$. The corresponding spectrum for the top-hat
 Gabor window $W_\mathrm{A}$ is shown as dotted lines and for the top-hat window $W_\mathrm{I}$ as coloured lines. The spectra are normalised in such a way that they can be
compared to the full sky power spectrum (dashed line). For the
$5^\circ$ FWHM window one can still distinguish the four lines. At
this window size the pseudo spectra are very similar to the full sky
spectra but with small deviations depending on the shapes of the
kernel and the shape of the power spectrum. In this case the spectrum
for the Gaussian window seems to be smaller at the peaks and larger at the troughs whereas
the spectrum for the top-hat windows is always larger.

For the $15^\circ$
FWHM windows the pseudo spectrum using the Gaussian Gabor window are on top
of the full sky power spectrum. For the top-hat windows it is still possible to
distinguish the pseudo spectrum from the full sky power spectrum
although the lines are still very close. The plot implies that the $\tilde C_\ell$ could be
good estimators of the underlying full sky $C_\ell$ provided that the
window is big enough. Note that for small windows, the Gaussian Gabor
window makes the pseudo spectrum a better estimator than the pseudo-spectrum for a top-hat window at higher
multipoles.  In \cite{hobson} it was shown in the flat-sky approximation that the pseudo power spectrum for small fields get significantly distorted, but that the shape of the pseudo spectrum gradually approaches the shape of the full sky power spectrum when the window gets larger. We see here that the same results applies to the treatment on the sphere. In the flat-sky approximation however, the error in estimating the average power spectrum from the pseudo power spectrum from one single realisation is bigger due to the long range correlations of the pseudo power spectrum coefficients in the flat-sky approximation.
\\

One feature which is very prominent is the additional peak at low
$\ell$ for the Gaussian window. The reason for this peak comes from the fact
that the diagonal in the Gaussian kernel is broader than in the
top-hat kernel for a top-hat window with corresponding area. For the low multipoles the power spectrum is dropping
rapidly
because of the Sachs-Wolfe effect and the lowest multipole $C_\ell$ are much bigger than the $C_\ell$ for higher
multipoles. Since the Gaussian kernel is broad,
the $\tilde C_\ell$ at low multipoles will pick up more from the
$C_\ell$ at lower multipoles than the narrower top-hat kernel (see
Fig. \ref{fig:cutkern}). These low multipole $C_\ell$ have very
high values compared to the higher multipole $C_\ell$ and for that
reason the $\tilde C_\ell$ for the Gaussian window will get a higher
value. This is illustrated in Fig. \ref{fig:kerncutlowl} where a
slice of the kernel at $\ell=50$ is shown for the $5^\circ$ FWHM
Gaussian Gabor window (solid line) and the corresponding top-hat $W_\mathrm{A}$
(dashed line) normalised to one at the peak. The dotted line shows a
typical power spectrum. Clearly the Gaussian kernel will pick up more
of the high value $C_\ell$ at low multipoles. Note that for the pseudo spectrum for the top-hat window $W_\mathrm{I}$ where the
integral of the top-hat window corresponds to the integrated Gaussian window (coloured line), there is also a peak
at low multipole. The reason is that the width of the kernel is the same as for the Gaussian.

In Fig. \ref{fig:pclex} we show the pseudo power spectra for a
particular realisation using a $15$ degree FWHM Gaussian window (upper
plot) and a top-hat window $W_\mathrm{A}$(lower plot). The pseudo spectra are
compared to the average full sky spectra shown as a dashed line. The dark shaded area shows
the expected $1\sigma$ cosmic and sample variance on the pseudo
spectra taken from the formulae to be developed in the next
sections. The lighter shaded area shows only cosmic variance. Note
that the pseudo spectrum for the Gaussian window is smoother than the
pseudo spectrum for the top-hat window. This shows the lower spectral resolution of the Gaussian window due to the broader kernel.

\section{Likelihood Analysis}
\label{sect:lik}

In this section we will show how the pseudo power spectrum can be used
as input to a likelihood analysis for estimating the full sky power
spectrum from an observed disc on the sky multiplied with a Gabor
window. We will in this section concentrate on a Gaussian Gabor $W_\mathrm{G}$
window, but the formalism is valid for any azimuthally symmetric Gabor
window. We start by showing that the pseudo-$C_\ell$ are close to Gaussian distributed which allows for a Gaussian
form of the likelihood function. Then we show how the correlation matrices can be calculated quickly for an
axisymmetric patch on the sky with uncorrelated noise. The extension to the more realistic situation with
correlated noise and non-axisymmetric sky patches will be made in the next section. We will show the results of
power spectrum estimations with different noise profiles and window sizes. We will also show that the use of a
window different from the top-hat window can be advantageous for some noise profiles, even if the window has a lower spectral resolution than the top-hat window.\\

\subsection{The form of the likelihood function}

\begin{figure}
\begin{center}
\leavevmode
\epsfig {file=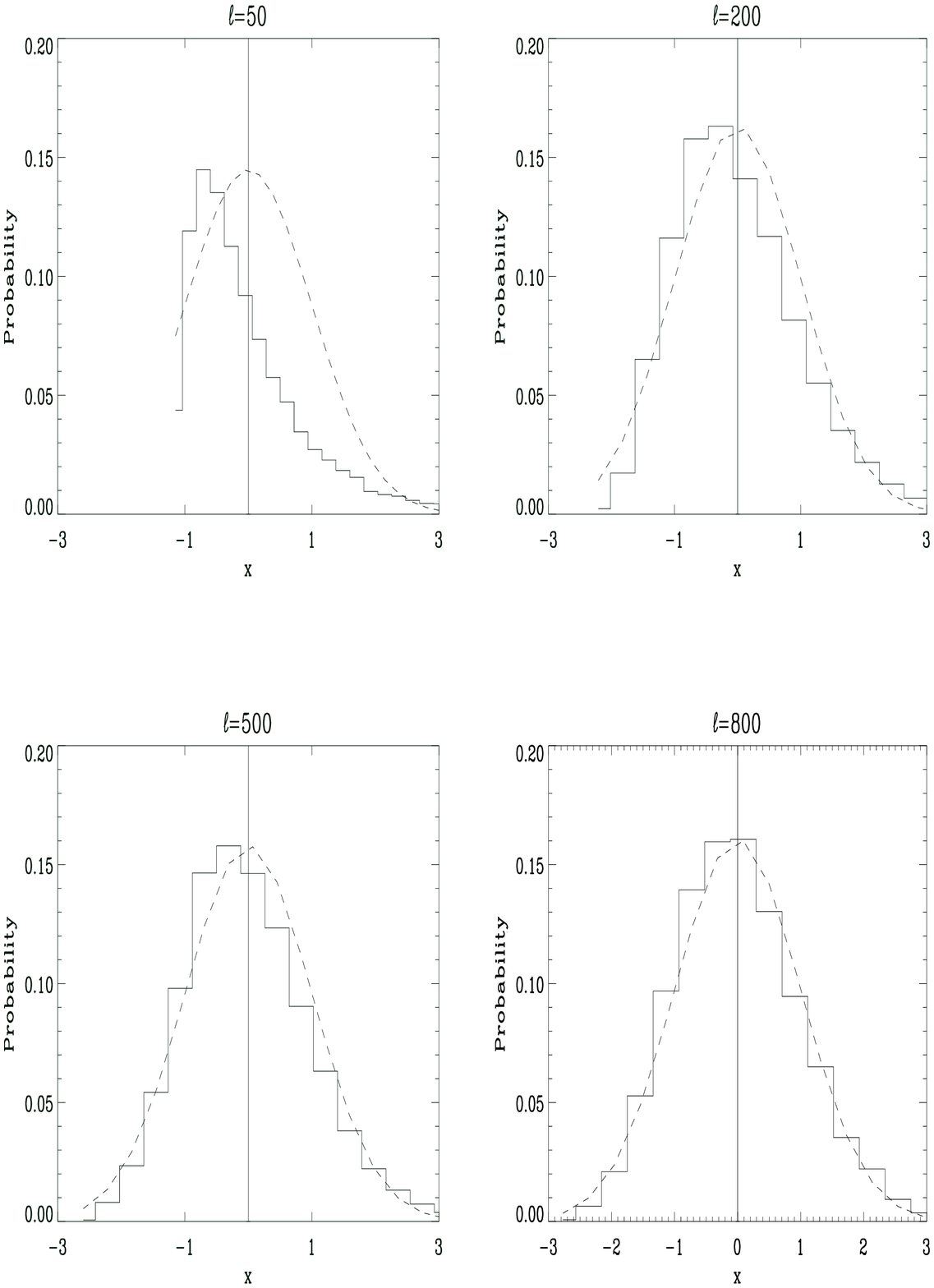,bbllx=0pt,bblly=0pt,bburx=650pt,bbury=800pt,height=12cm,width=16cm}
\caption{The probability distribution of \protect{$\tilde C_\ell$}
taken from 10000 simulations with a \protect{$5^\circ$} FWHM Gaussian
Gabor window truncated at $\theta_\mathrm{C}=3\sigma$. The variable $x$ is given as
\protect{$x=(\tilde C_\ell-\VEV{\tilde C_\ell})/\sqrt{\VEV{(\tilde C_\ell-\VEV{\tilde
C_\ell})^2}}$}. The dashed line is a Gaussian
with the theoretical mean and standard deviation of the
\protect{$\tilde C_\ell$}. The plot shows the \protect{$\tilde C_\ell$}
distribution for \protect{$\ell=50$, $\ell=200$, $\ell=500$},
and \protect{$\ell=800$}. The probabilities are normalised such that
the integral over $x$ is $1$.}
\label{fig:prob5}
\end{center}
\end{figure}

\begin{figure}
\begin{center}
\leavevmode
\epsfig {file=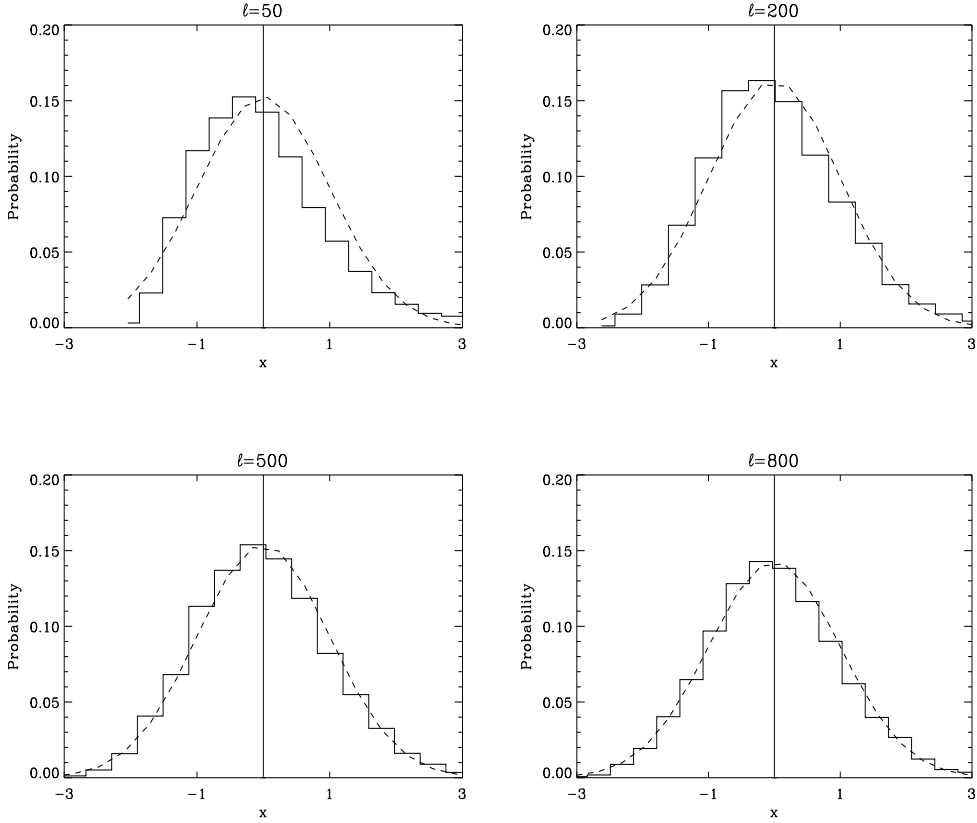,bbllx=0pt,bblly=0pt,bburx=650pt,bbury=800pt,height=12cm,width=16cm}
\caption{Same as Fig. \ref{fig:prob5} but for a $15^\circ$ FWHM
Gaussian Gabor window.}
\label{fig:prob15}
\end{center}
\end{figure}

\begin{figure}
\begin{center}
\leavevmode
\epsfig {file=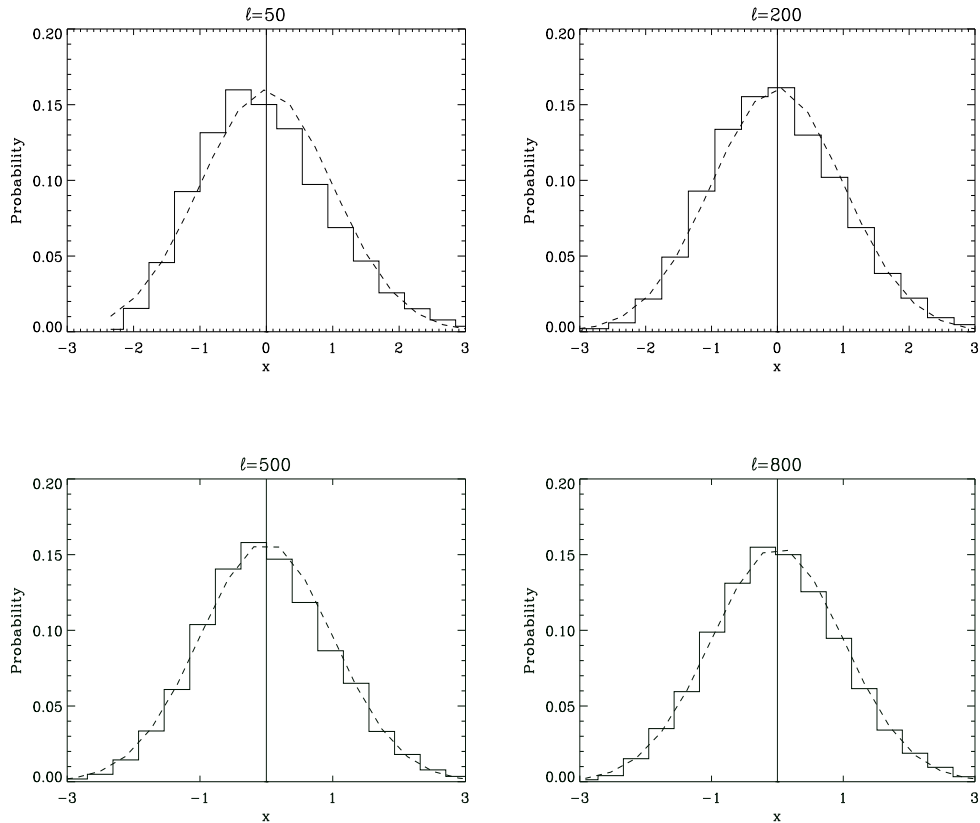,bbllx=0pt,bblly=0pt,bburx=650pt,bbury=800pt,height=12cm,width=16cm}
\caption{Same as Fig. \ref{fig:prob15} but for a top-hat Gabor window $W_\mathrm{A}$
covering the same area on the sky.}
\label{fig:probth}
\end{center}
\end{figure}

To know the form of the
likelihood function, one needs to know the probability distribution of
$\tilde C_\ell$. In Figs \ref{fig:prob5} and \ref{fig:prob15} we show the probability
distribution from 10000 simulations with a $5^\circ$ and $15^\circ$ FWHM Gaussian
Gabor window respectively. The dashed line shows a
Gaussian with mean value and standard deviation found from the
formulae given in the previous and next section. One can see that the
probability distribution is slightly skewed for low $\ell$, but for high $\ell$ it seems
to be very well approximated by a Gaussian. Also the small window
shows more deviations from a Gaussian than the bigger window. In
Fig. \ref{fig:probth} we show that this result is not limited to the Gaussian window. The plot shows the probability distribution from a
simulation with a top-hat Gabor window $W_\mathrm{A}$ covering the same area on the sky as
the $15^\circ$ FWHM Gaussian window. Also for this window the
probability distribution is close to Gaussian.\\

\begin{figure}
\begin{center}
\leavevmode
\epsfig {file=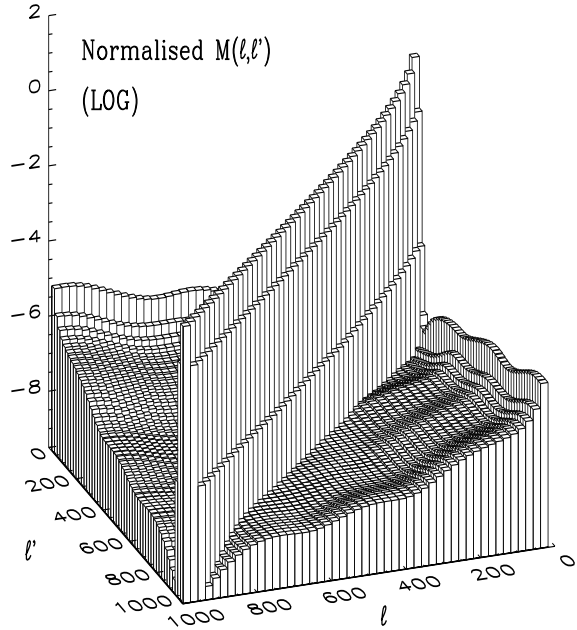,height=10cm,width=10cm}
\caption{The figure shows the correlation matrix
\protect{$M_{\ell\ell'}$} between pseudo spectrum coefficients
normalised with the pseudo spectrum \protect{$(\VEV{\tilde C^\mathrm{T}_\ell\tilde C^\mathrm{T}_{\ell'}}-\VEV{\tilde C^\mathrm{T}_\ell}\VEV{\tilde
C^\mathrm{T}_{\ell'}})/(\VEV{\tilde C^\mathrm{T}_\ell}\VEV{\tilde C^\mathrm{T}_{\ell'}})$} for
a 15 degree FWHM Gaussian Gabor window. A standard CDM power spectrum
was used to produce this matrix.}
\label{fig:cormat}
\end{center}
\end{figure}

\begin{figure}
\begin{center}
\leavevmode
\epsfig {file=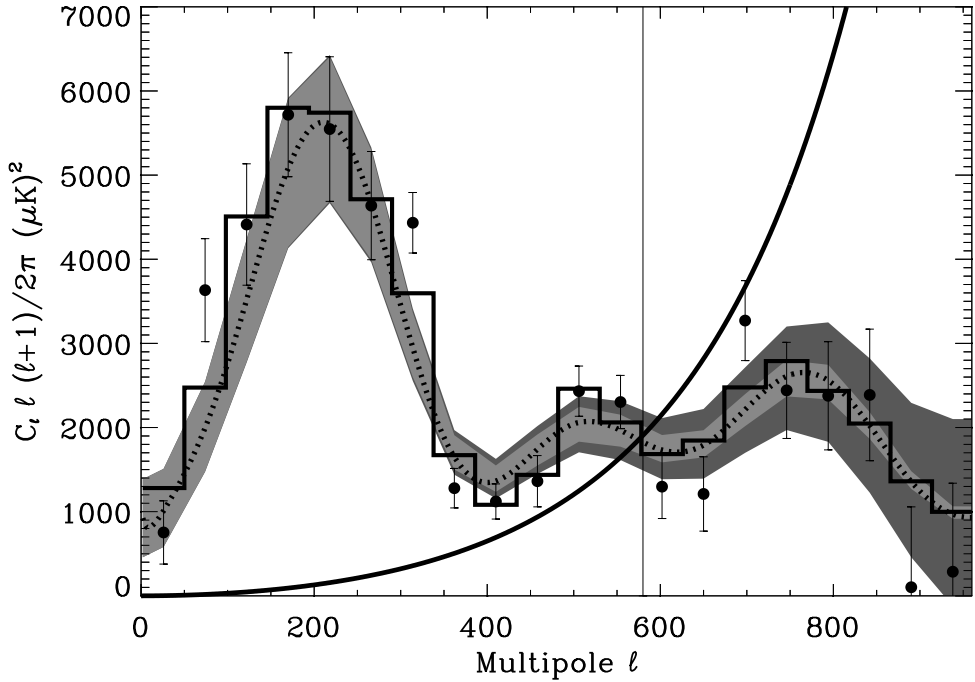,height=10cm,width=14cm}
\caption{The analysis of an input model with
\protect{$\Omega_\mathrm{total}=1$}, \protect{$\Omega_\Lambda=0.7$},
\protect{$\Omega_\mathrm{b}h^2=0.03$} and \protect{$n_\mathrm{s}=0.975$}. We used a
non-uniform white noise model with \protect{$S/N=1$} at
\protect{$\ell=575$}. The dotted line is the input average full sky power spectrum and
the histogram shows the binned pseudo power spectrum for this
realisation (without noise). We used $N^\mathrm{bin}=20$ bins and $N^\mathrm{in}=100$ input sample
points to the likelihood. The shaded areas around the binned average
full sky spectrum (which is not plotted) are the theoretical variance
with and without noise. The bright shaded area shows cosmic and sample
variance and the dark shaded area also has variance due to noise
included. The variance due to noise was calculated using formula (\ref{eq:uninoiseerror}) for uniform noise. The
\protect{$1\sigma$} error bars on the estimates are taken from the
inverse Fisher matrix. The solid line increasing from the left to the
right is the noise power spectrum.}
\label{fig:master1}
\end{center}
\end{figure}

\begin{figure}
\begin{center}
\leavevmode
\epsfig {file=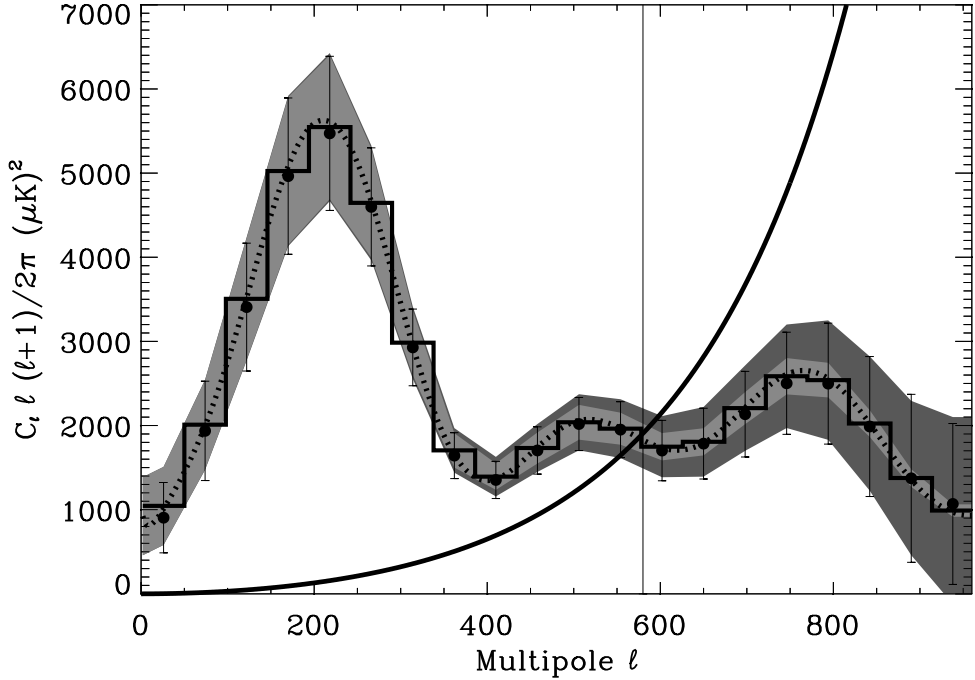,height=10cm,width=14cm}
\caption{Same as in Fig. \ref{fig:master1} but this is an average
over 1000 simulations and estimations. The histogram is now the binned
average full sky power spectrum. The error bars on this plot are
the \protect{$1\sigma$} variances taken from Monte Carlo.}
\label{fig:master2}
\end{center}
\end{figure}

\begin{figure}
\begin{center}
\leavevmode
\epsfig {file=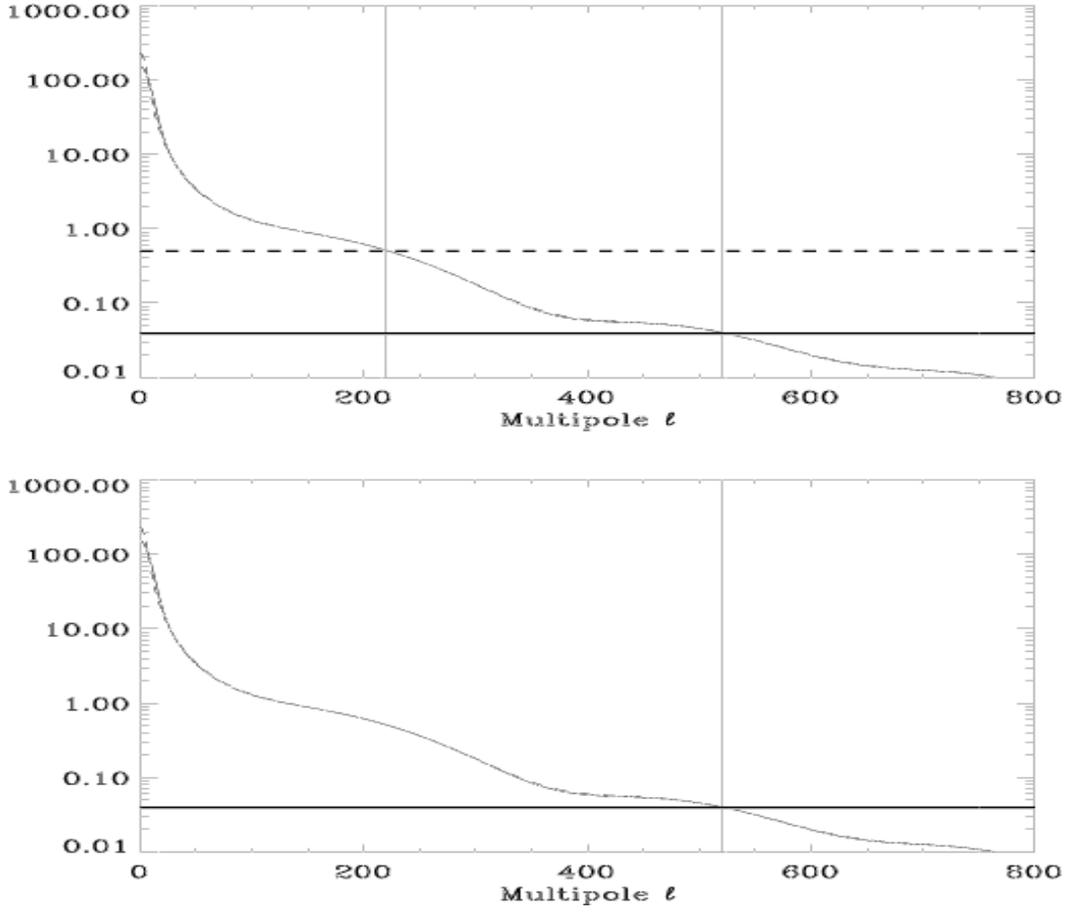,height=12cm,width=14cm}
\caption{The plots show average signal and noise pseudo power spectra plotted
separately. The spectra are normalised so that they can be compared to
the full sky power spectrum. The solid and dashed curves which almost fall together are
the signal pseudo power spectra for a $15$ degree FWHM Gaussian Gabor
window $W_\mathrm{G}$ and a corresponding top-hat window $W_\mathrm{A}$ respectively. In the upper
plot the noise model shown in Fig. \ref{fig:noisemap} was used. This
noise model is increasing from the north pole and down to the edges of
the patch. This is opposite of the Gaussian window and for this reason the
Gaussian window gets higher signal to noise ratio. The solid horizontal
line in the upper plot
shows the noise pseudo power spectrum for the Gaussian window and the
dashed horizontal line shows the noise pseudo power spectrum for the
top-hat window. In the lower plot a uniform noise model was used so
that the noise pseudo power spectra fall together and are shown as a
solid vertical line. The figure shows how a Gabor window different from a top-hat can be
used to increase signal to noise.}
\label{fig:sngaussth}
\end{center}
\end{figure}

\begin{figure}
\begin{center}
\leavevmode
\epsfig {file=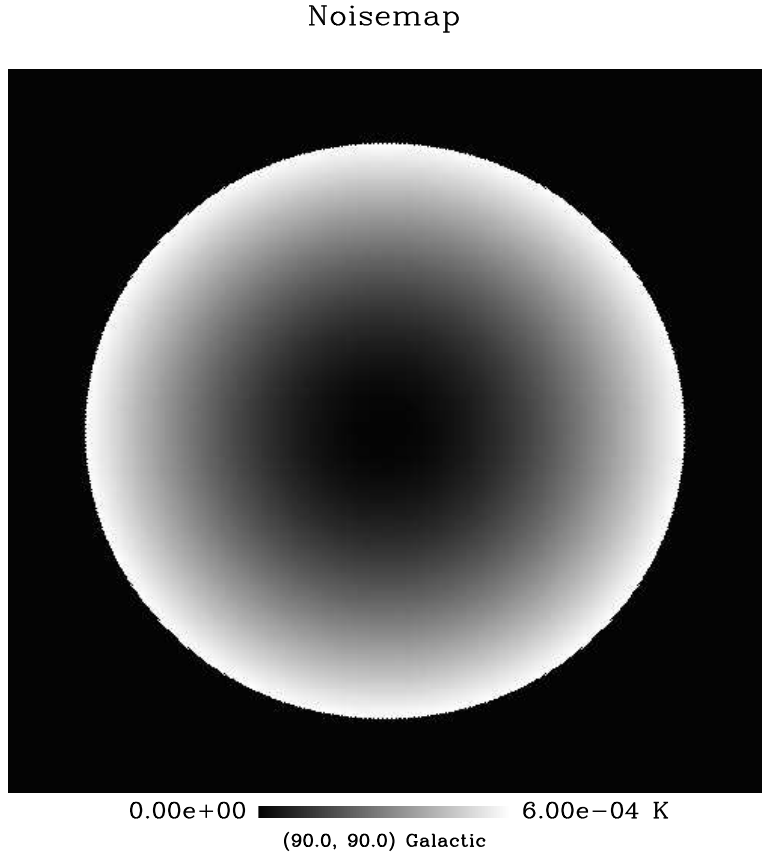,height=12cm,width=10cm}
\caption{The noise map with noise increasing from the north pole and
downwards. The figure shows a gnomic projection with the north pole in the centre.}
\label{fig:noisemap}
\end{center}
\end{figure}

\begin{figure}
\begin{center}
\leavevmode
\epsfig {file=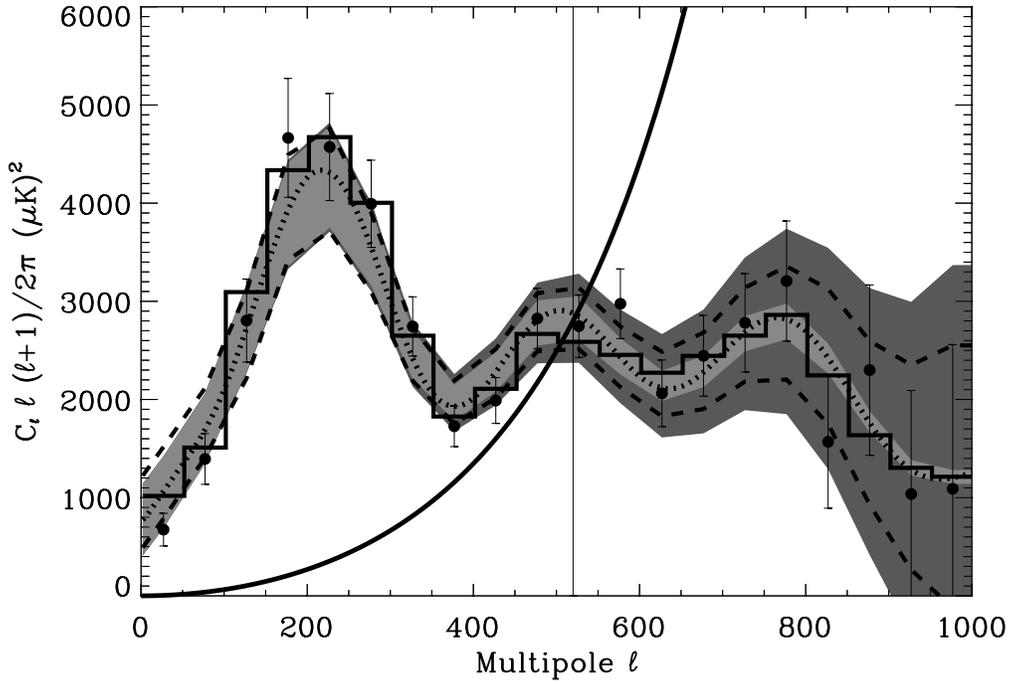,height=10cm,width=14cm}
\caption{Same as Fig. \ref{fig:master1} but for a standard CDM
model. The noise is increasing from the centre and out to the edges
while the Gaussian Gabor window has the opposite effect, giving an
increased significance to pixels with less noise. As in Fig.
\ref{fig:master1} the shaded areas
show the analytically calculated variance using the `naive' formula for the uniform noise case. The
dashed lines show the expected variance using the inverse of the
Fisher matrix.}
\label{fig:cdm1}
\end{center}
\end{figure}

\begin{figure}
\begin{center}
\leavevmode
\epsfig {file=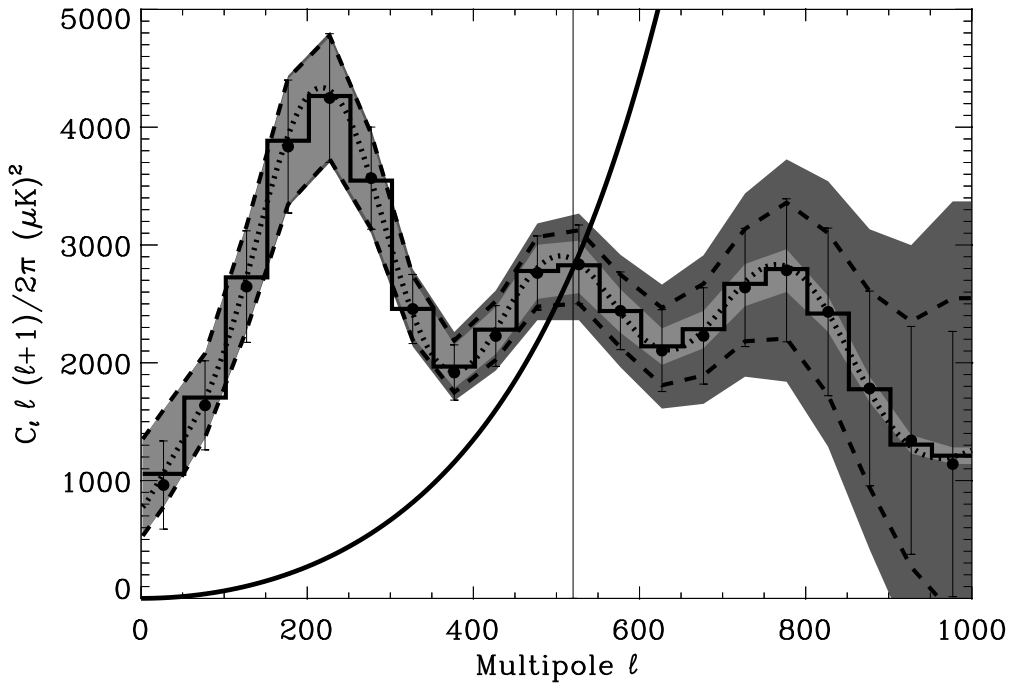,height=10cm,width=14cm}
\caption{Same as Fig. \ref{fig:cdm1} but with 5000 simulations and
estimations averaged. The histogram now shows the binned average full
sky power spectrum. The error bars on the estimates are here the \protect{$1\sigma$}
averages from Monte Carlo.}
\label{fig:cdm2}
\end{center}
\end{figure}

\begin{figure}
\begin{center}
\leavevmode
\epsfig {file=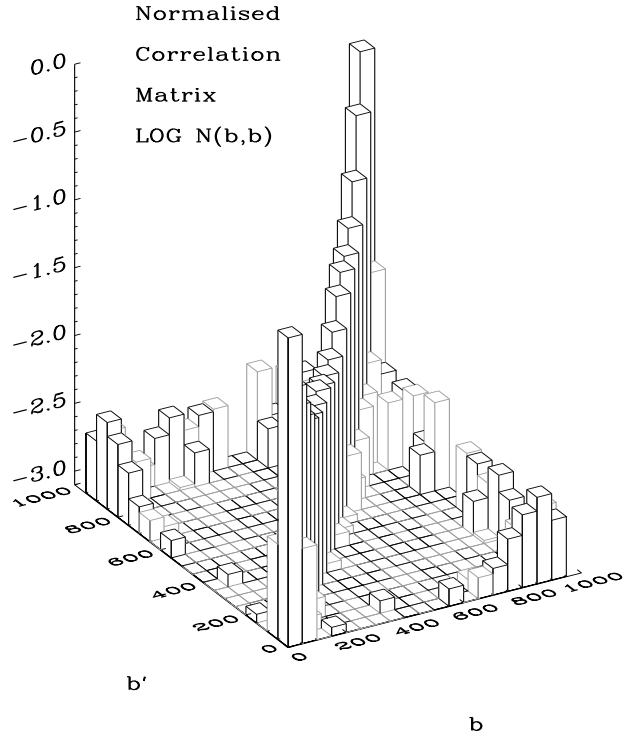,height=10cm,width=10cm}
\caption{The average correlation matrix
 \protect{$N(b,b')\equiv\VEV{D_bD_{b'}}/(\VEV{D_b}\VEV{D_{b'}})-1$} of the estimates in Fig.
\protect{\ref{fig:cdm2}}. The negative elements are coloured.}
\label{fig:estcor}
\end{center}
\end{figure}

\begin{figure}
\begin{center}
\leavevmode
\epsfig {file=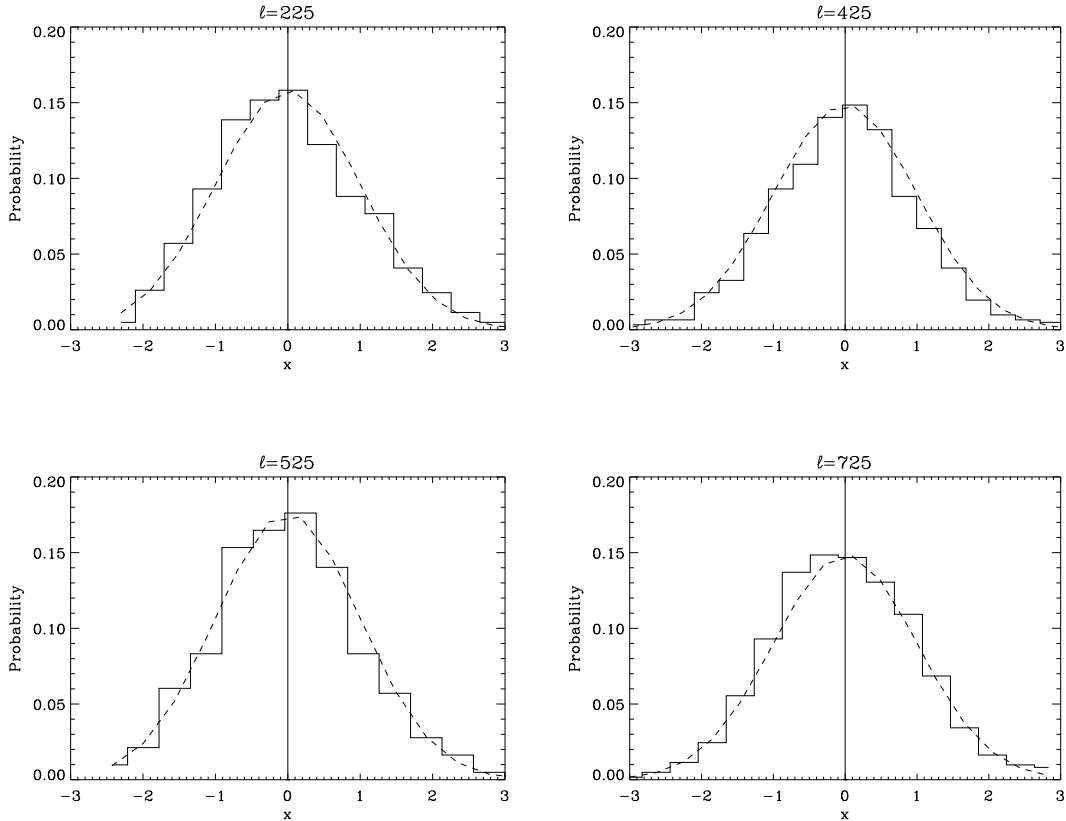,bbllx=0pt,bblly=0pt,bburx=596pt,bbury=800pt,height=12cm,width=16cm}
\caption{The probability distribution of the estimates $D_b$
in Fig. \protect{\ref{fig:cdm2}}.
 The variable $x$ is given as
\protect{$x=(D_b-\VEV{D_b})/\sqrt{\VEV{(D_b-\VEV{D_b})^2}}$}. The dashed line is a Gaussian
with mean and standard deviation taken from Monte Carlo. The plot
shows bin estimates centred at \protect{$\ell=225$, $\ell=425$,
$\ell=525$} and \protect{$\ell=725$}.}
\label{fig:estprob}
\end{center}
\end{figure}

\begin{figure}
\begin{center}
\leavevmode
\epsfig {file=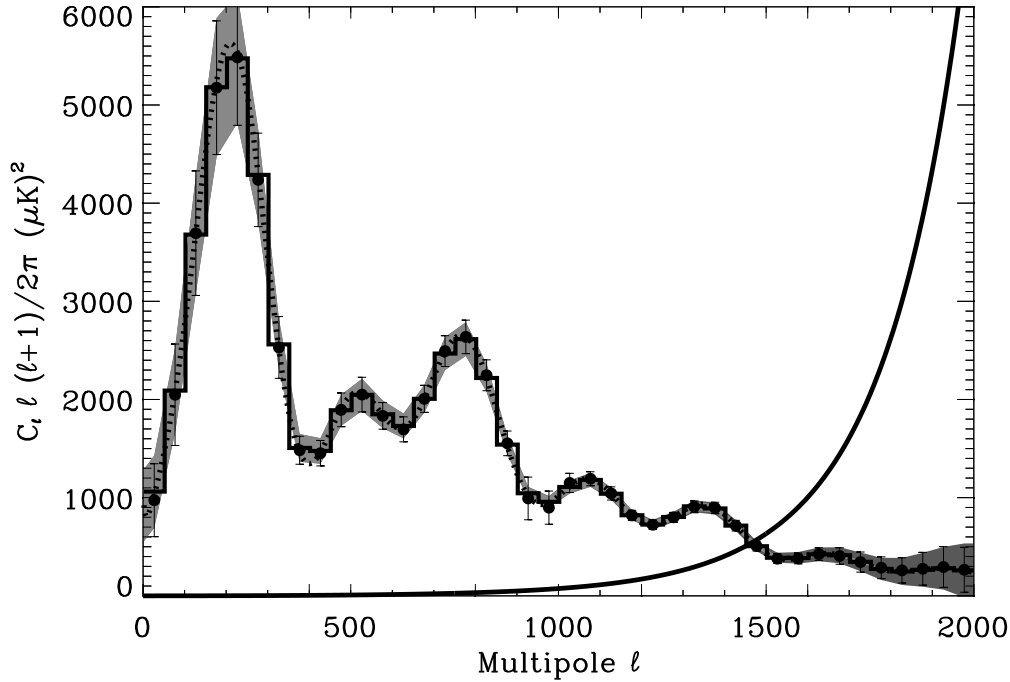,height=10cm,width=14cm}
\caption{Same as Fig. \protect{\ref{fig:master2}} for \protect{$100$}
simulations where the beam and noise level was set according to the
specifications of the Planck \protect{$143 \mathrm{GHz}$}
channel. Again a \protect{$15^\circ$} FWHM  Gaussian Gabor window was
used. The power spectrum was estimated in $40$ bins between \protect{$\ell=2$}
and \protect{$\ell=2048$}}
\label{fig:l2048}
\end{center}
\end{figure}

\begin{figure}
\begin{center}
\leavevmode
\epsfig {file=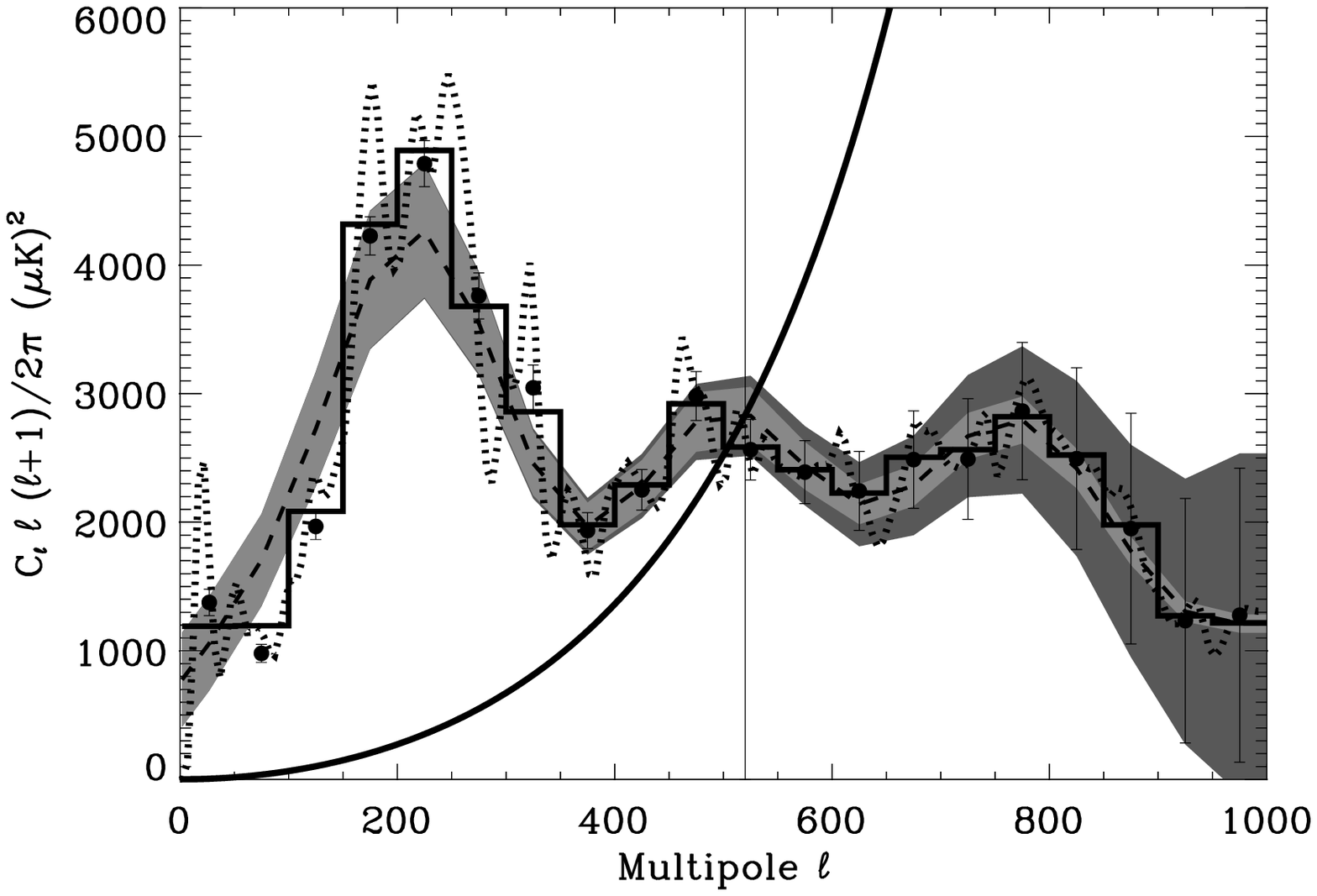,height=10cm,width=14cm}
\caption{The average of 300 estimations where the input
\protect{$\tilde C_\ell$} were taken from simulations with a fixed CMB realisation but
varying noise realisations. The dotted line is the \protect{$N^\mathrm{in}$}
input $\tilde C_\ell$ from the CMB realisation without
noise. The histogram is the same spectrum binned in $N^\mathrm{bin}$ bins. The dashed line is the binned
average full sky power spectrum from which this realisation was
made. The shaded areas around the binned full sky spectrum show the
variance with (dark) and without (bright) noise. The solid line rising
from the left to the right is the noise power spectrum.}
\label{fig:onerea}
\end{center}
\end{figure}

From the above plots it seems to be reasonable to approximate the
likelihood function with a Gaussian provided the window is big enough
and multipoles at high enough $\ell$ values are used,
\begin{equation}
\mathcal{L}=\frac{\mathrm{e}^{-\frac{1}{2}{\mathbf{ d}}^\mathrm{T} {\mt{M}}^{-1}{\mathbf{
d}}}}{\sqrt{2\pi \det{\mt{M}}}}.
\end{equation}

Omitting all constant terms and factors, the log-likelihood can then
be written:

\begin{equation}
L={\mathbf{ d}}^\mathrm{T} {\mt{M}}^{-1}{\mathbf{ d}}+\ln{\det{{\mt{M}}}}.
\end{equation}

Here ${\mathbf{ d}}$ is the data vector vector
which contains the observed $\tilde{C}_\ell$ for the set of sample
$\ell$-values $\ell_i$. The data is taken from the observed windowed sky in the
following way:
\begin{equation}
d_i=\tilde{C}_{\ell_i}-\VEV{\tilde{C}_{\ell_i}}.
\end{equation}

The matrix ${\mt{M}}$ is the covariance between pseudo-$C_l$ which elements are given
by:
\begin{equation}
M_{ij}=\VEV{\tilde{C}_{\ell_i}\tilde{C}_{\ell_j}}-\VEV{\tilde{C}_{\ell_i}}\VEV{\tilde{C}_{\ell_j}}.
\end{equation}

\begin{figure}
\begin{center}
\leavevmode
\epsfig {file=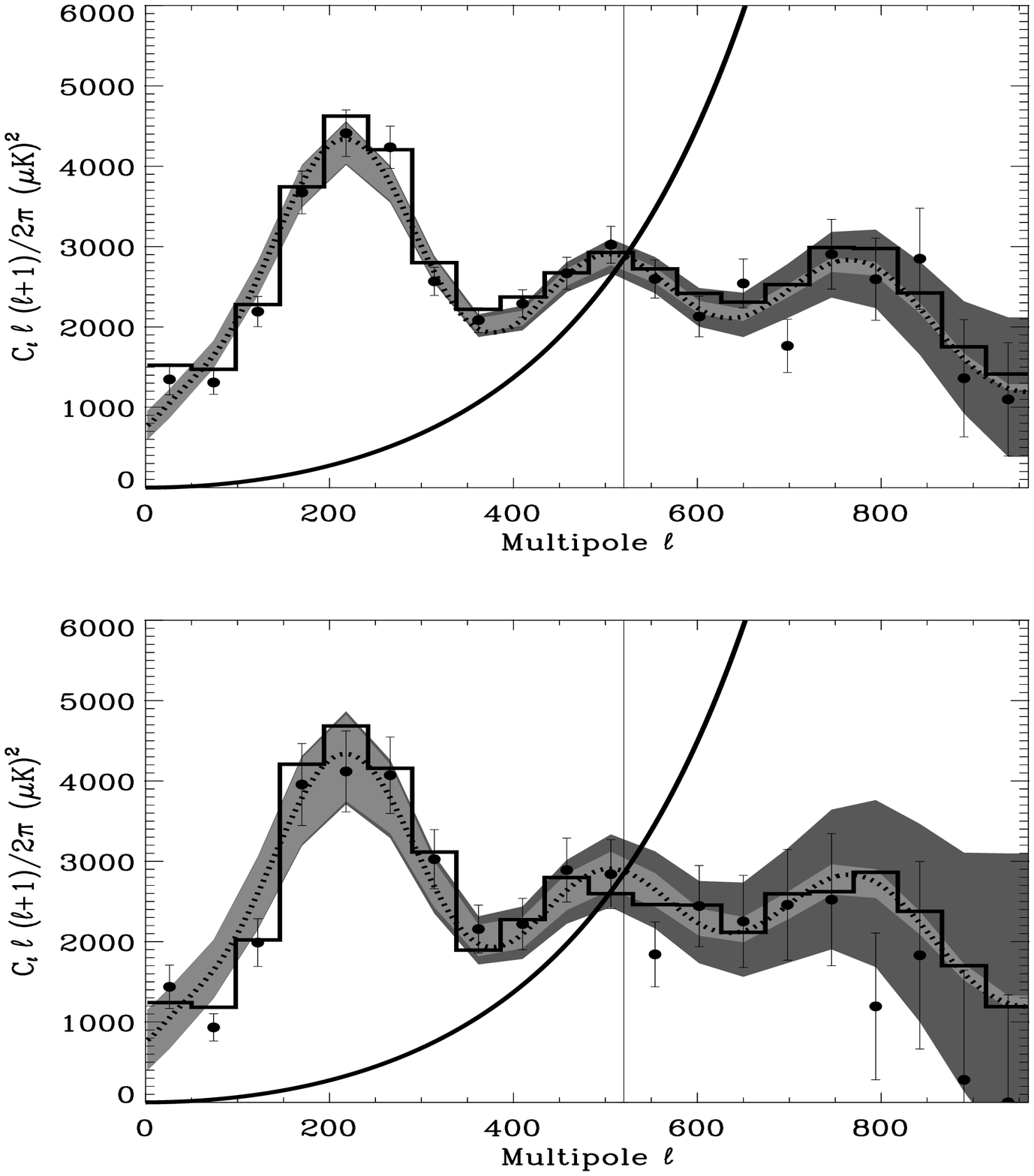,bbllx=40pt,bblly=100pt,bburx=496pt,bbury=800pt,height=12cm,width=10cm}
\caption{Estimates of \protect$\tilde C_\ell$ using a Gaussian Gabor
window (lower plot) and a top-hat window (upper plot). Here we used a
uniform noise model with \protect{$S/N=1$ at $\ell=520$}. The dotted
line shows the average full sky power spectrum and the histogram shows the
input pseudo power spectrum without noise for this realisation, binned in the same way as the estimates. The
bright shaded area shows the cosmic and sample variance around the
binned average spectrum (not plotted). The dark shaded area has the variance due to
noise included. The $1\sigma$ error bars on the estimates are taken from the
inverse Fisher matrix. The solid line increasing from left to right is the
noise power spectrum.}
\label{fig:thvsg}
\end{center}
\end{figure}

In Appendix (\ref{app:cor}) and (\ref{app:noise}) we have found expressions which enable fast evaluations of $d_i$
and $M_{ij}$ for signal and noise. These major results are are given in equations (\ref{eq:cormat}), (\ref{eq:noisecormat}) and (\ref{eq:crosscormat}) and the recursion which enables fast calculation of these expressions is given in equation (\ref{eq:rec}).
In the derivations of the expressions for $M_{ij}$, the rotational invariance of the (not-averaged) $\tilde C_\ell$
shown in Appendix (\ref{app:rotin}) was used. Because of this rotational invariance, all derivations can be done
with the Gabor window centred at the north pole. In Fig. \ref{fig:cormat} we used the full formula (equation
\ref{eq:cormat}) from Appendix (\ref{app:cor}) to calculate the signal correlation matrix for a
typical power spectrum with a $15^\circ$ FWHM Gaussian Gabor window
(note that in the figure, the correlation matrix is normalised with
the pseudo power spectrum). The
correlation between $\tilde C_\ell$ of different multipoles is falling of rapidly with the distance from the
diagonal. Only exception being the small 'wall' at low
multipoles which again comes from the coupling to the smallest
multipoles which have very high values.\\\\

\subsection{Likelihood estimation and results}
\label{sect:results}

Because of the limited information content in one
patch of the sky one can not estimate the full sky $C_\ell$ for all
multipoles $\ell$. For
this reason the full sky power spectrum has to be estimated in
$N^\mathrm{bin}$ bins. Also the algorithm to minimise the
log-likelihood needs the different numbers to be estimated to be of roughly the
same order of magnitude. For this reason we estimate for some
parameters $D_b$ which for bin $b$ is defined as 

\begin{equation}
\label{eq:binning}
C_\ell=\frac{D_b}{\ell(\ell+1)},\ \
\ell_b\leq\ell<\ell_{b+1},\\
\end{equation}
where $\ell_b$ is the first multipole in bin $b$.

Since the $\tilde C_\ell$ are coupled, one can not use all multipoles
in the data vector,
the covariance matrix would in this case become singular. One has to
choose a number $N^\mathrm{in}$ of multipoles $\ell_i$ for which one finds
$d_i$. How many multipoles to use depends on how tight the $\tilde
C_\ell$ are
coupled which depends on the width $\Delta\ell_\mathrm{kern}$ of the kernel (Fig.
\ref{fig:fwhmrel}) or the width $\Delta\ell_\mathrm{cor}$ of the correlation matrix. The width of the correlation
matrix(normalised with the psuedo power spectrum) varies with window size in the same way as the width of the
kernel varies with window size. In fact for the top-hat window these two widths are the same and for the Gaussian
window we found that $\Delta\ell_\mathrm{cor}\approx1.42\Delta\ell_\mathrm{kern}$.  The optimal number of $N^\mathrm{in}$ to use seems
to be $N^\mathrm{in}\approx3/2\ \ell_\mathrm{max}/\Delta\ell_\mathrm{kern}$. To
use a lower $N^\mathrm{in}$ increases the error bars on the estimates and a
higher $N^\mathrm{in}$ does not improve the estimates. One can at most fit for as many $C_\ell$s as the
number of $\tilde C_\ell$ ($N^\mathrm{in}$) one has used in the analysis. So one needs to find a number
$N^\mathrm{bin}\le N^\mathrm{in}$ of bin
values $D_b$ from which one can construct the full sky power spectrum $C_\ell$.\\

In Appendix (\ref{app:cor}) and (\ref{app:noise}) we found that the full correlation matrix can be written as
\begin{equation}
M_{ij}=M_{ij}^\mathrm{S}+M_{ij}^\mathrm{N}+M_{ij}^\mathrm{X},
\end{equation}
where $M_{ij}^\mathrm{N}$ is the noise correlation matrix which has to be precomputed for a specific noise model
(analytically or by Monte Carlo as will be shown in Section (\ref{sect:ext})). The signal and signal-noise cross
correlation matrices are on the form
\begin{eqnarray}
M^\mathrm{S}_{ij}&=&\sum_b\sum_{b'}D_bD_{b'}\chi(b,b',i,j),\\
M^\mathrm{X}_{ij}&=&\sum_kD_b\chi'(b,i,j),
\end{eqnarray}
where the $\chi$-functions can be precomputed using formulae (\ref{eq:chis}) and (\ref{eq:chix}).

We will now describe some test simulations to show how the method
works. As a first test, we used the same model as was used in
\cite{master}, with $\Omega_\mathrm{total}=1$, $\Omega_\Lambda=0.7$,
$\Omega_\mathrm{b}h^2=0.03$ and $n_\mathrm{s}=0.975$. These are the parameters from
the combined Maxima-Boomerang analysis \cite{jaffe}. We used a
circular patch with $15.5^\circ$ radius covering the same fraction of the
sky as in \cite{master}. Using HEALPix we simulated a CMB sky using a
standard CDM power spectrum with $l_\mathrm{max}=1024$ and a $7'$ pixel size
($N_\mathrm{side}=512$ in HEALPix language). We smoothed the map with a
$10'$ beam and added non-correlated non-uniform noise to it. Here a Gaussian Gabor window with $FWHM=12^\circ$ was
used with a cut-off $\theta_\mathrm{C}=3\sigma$. For the likelihood estimation, we had $N^\mathrm{bin}=20$ full sky
$C_\ell$ bins and
$N^\mathrm{in}=100$ $\tilde C_\ell$ values between $\ell=2$ and
$\ell=960$. In Fig. \ref{fig:master1} one can see the result. The shaded
areas are the expected $1\sigma$ variance with and without noise. These were
found from the theoretical formula

\begin{equation}
\label{eq:uninoiseerror}
\Delta C_b=\sqrt{\frac{2}{\nu_b}}(C_b+N_b),
\end{equation}
where $N_b$ is the noise `on the sky', $\nu_b$ is the effective number of degrees of freedom given as
\begin{equation}
(2\ell_b+1)\Delta\ell f_\mathrm{sky}\frac{w_2^2}{w_4},
\end{equation}
and the $w_i$ factors are dependent on the window according to
\begin{equation}
f_\mathrm{sky}w_i=\frac{1}{4\pi}\int_{4\pi}d{\hat\mathbf{n}}G^i({\hat\mathbf{n}}).
\end{equation}

This formula is exact for a uniform noise model \cite{master} and is similar to the
one used in most publications. It is in this case a very good
approximation even with non-uniform noise. In the next example however
we will show that the formula has to be used with care. In the figure, the error bars
on the estimates are taken from the
Fisher matrix and the signal-to-noise ratio $S/N=1$ at $\ell=575$.\\

In Fig. \ref{fig:master2}, we have plotted the average of 1000
such simulations, with different noise and sky realisations. From the plot,
the method seems to give an unbiased estimate of the power spectrum bins
$D_b$. For the lowest multipoles the estimates are slightly lower
than the binned input spectrum. This is a result of the slightly
skewed probability distribution of $\tilde C_\ell$ for small windows at these low
multipoles (see Figs \ref{fig:prob5} and \ref{fig:prob15}). The
probability that the $\tilde C_\ell$ at lower multipoles have a value
lower than the average $\VEV{\tilde C_\ell}$ is high and the assumption
about a Gaussian distribution about this average leads the estimates
to be lower. When a bigger area of the sky is available such that
several patches can be analysed jointly to give the full sky power
spectrum, this bias seems to disappear. This will be shown in Section
({\ref{sect:multiple}).\\

In this example one can see that the $1\sigma$
error
bars from Monte Carlo coincide very well with the theoretical error
shown as shaded areas from the formula in \cite{master}. Note that
the error bars on the
higher $\ell$ are smaller than in \cite{master} because the noise model
used in that paper was not white. Also they took into account errors
due to map making which is not considered here.\\

As a next test, we used a simulation with the same resolution and beam
size. The power spectrum was this time a standard CDM power spectrum. We used
an axisymmetric noise model with noise increasing from the centre and
outwards to the edges (see Fig. \ref{fig:noisemap}). This is the kind of noise model which could be
expected from an experiment scanning on rings, with the rings crossing
in the centre. We now use a circular patch with $18.5^\circ$ radius and
a $FWHM=15^\circ$ Gaussian Gabor window cut at $\theta_\mathrm{C}=3\sigma$. An interesting point now is
that the Gabor window is decreasing from the centre and outwards,
which is opposite of the noise pattern.  This gives the pixels with low
noise high significant in the analysis and the pixels with high noise
low significance. One sees from the expressions for the signal and
noise pseudo power spectra that the Gabor window will work differently
on both. This means that $S/N$ is different depending on the Gabor
window. For this case, we have plotted the average pseudo power spectrum for
signal and noise separately in Fig.
\ref{fig:sngaussth}. This shows the described effect. The
\protect{$S/N$} ratio is much higher for the Gaussian Gabor window in
this case, favouring the use of this window for the analysis.\\

For this example we used again $N^\mathrm{bin}=20$ and $N^\mathrm{in}=100$. The result is
shown in Fig. \ref{fig:cdm1}. In Fig. \ref{fig:cdm2} the
average over 5000 simulations and estimations is shown. One can see that the estimate also
does well beyond $\ell=520$ which is where the effective $S/N=1$. The method is still unbiased. The error bars in
the
part where noise dominates are
here lower than the theoretical approximation (\ref{eq:uninoiseerror}) shown
as the dark shaded area. The
dashed lines show the theoretical $1\sigma$ variance taken from the
inverse Fisher matrix which here gives a very good agreement with
Monte Carlo.\\

In Fig. \ref{fig:estcor} we show the average (over 5000
estimations) of the correlation between the estimates $D_b$ between different bins. The figure shows that the
correlations between
estimates are low and in fact in each line all off-diagonal elements
are more than an order of magnitude lower than the diagonal element of
that line. In Fig. \ref{fig:estprob} we show that the probability
distribution of the estimates in Fig. \ref{fig:cdm2} is almost
Gaussian.\\

To test the method at higher multipoles we also did one estimation up
to multipole $\ell=2048$. We used HEALPix resolution $N_\mathrm{side}=1024$ and
simulated a sky with a $8'$ Gaussian beam and added noise from a strongly varying non-uniform
noise model. Both the beam and noise level were adjusted according to the
specifications for the Planck HFI $143 \mathrm{GHz}$ detector
\cite{plancka}. We used again a $15^\circ$ FWHM Gaussian Gabor window
cut $3\sigma$ away from the centre. In the estimation we used
$N^\mathrm{bin}=40$ bins and $N^\mathrm{in}=200$ input $\tilde C_\ell$ between
$\ell=7$ and $\ell=2048$. The average of 100 such simulations is
shown in Fig. \ref{fig:l2048}. Each complete likelihood estimation
(which includes a
total of about $25$ likelihood evaluations) took about $8$ minutes on
a single processor on a $500 \mathrm{MHz}$ DEC Alpha work station.

In Fig. \ref{fig:onerea}, we have plotted the average of 300
estimations where the input data was the $\tilde
C_\ell$ from simulations with a fixed CMB realisation and varying noise
realisation. The dotted line shows the $N^\mathrm{in}$ $\tilde C_\ell$s (without noise) used
as input to the likelihood. The histogram is as before the input
pseudo spectrum without noise binned in $N^\mathrm{bin}$ bins. This means
that each histogram line shows the average of the dotted line over the
bin. One can see that the estimated power spectrum is partly
following the $N^\mathrm{in}$ input $\tilde C_\ell$ and partly the binned
power spectrum. 

Finally, we made a comparison between a top-hat window and a Gaussian Gabor
window. In this case we used uniform noise, so that the Gaussian and
top-hat Gabor windows have the same $S/N$ ratio which we set to $1$ at
$\ell=520$. We used a disc with $18^\circ$ radius, $N^\mathrm{in}=200$
and $N^\mathrm{bin}=20$. In Fig. \ref{fig:thvsg} one can see the result. The
lower plot shows the estimates with the Gaussian Gabor window
($15^\circ$ FWHM) and the upper with the
top-hat window. The Gaussian window is suppressing parts of the data
and for this reason gets a higher sample variance than the top-hat. This
effect is seen in the plot. Clearly when no noise weighting is
required the top-hat window seems to be the preferred window (which was
also discussed in \cite{master}). This chapter has been concentrating
on the Gaussian window to study power spectrum estimation in the
presence of a window different from a top-hat. It has been
shown that a different window can be advantageous when the noise is not
uniformly distributed as one can then give data with different quality
different significance.

\section{Extensions of the Method}
\label{sect:ext}

A real CMB experiment usually does not observe an axisymmetric patch of the sky. Usually the noise between pixels
is also correlated. In order to take these two issues into account we will discuss two extensions of the
method. The formalism for the extensions are worked out and some
simple examples are shown. Further investigations of these extensions
are left for a future paper, where the analysis of MAP and Planck data will be discussed. To be able to analyse
non-axisymmetric parts of the sky, we propose to split the area up into several axisymmetric pieces and use the
pseudo-$C_\ell$ from all these patches in the data vector of the likelihood and in this way analyse all patches
jointly. We show that if the patches are not overlapping, the correlation between pseudo-$C_\ell$ from different
patches is so weak that it can be neglected. To deal with correlated noise, we propose to use Monte Carlo
simulations to find the noise correlation matrix. We demonstrate that for uncorrelated noise, one needs a few
thousand simulations in order for the error bars on the $C_\ell$ estimates not to get larger than when using the
analytic expression.

\subsection{Multiple patches}
\label{sect:multiple}

It has been shown how one can do power spectrum estimation on one axisymmetric patch on the sky. The next question
that arises is what to do
when the observed area on the sky is not axisymmetric. In this case
one can split the area into several axisymmetric pieces and
use the $\tilde C_\ell$ from each piece. Then the $\tilde C_\ell$
from all the patches are used together in the likelihood
maximisation. The first thing to check before embarking on this idea
is the correlation between $\tilde C_\ell$s in different patches. In Appendix (\ref{app:multiple}) the analytical
formula for the correlation matrix describing the correlations between $\tilde C_\ell$ for different patches was
derived (equation \ref{eq:spot2spot} and \ref{eq:spot2spotalm}). With these expressions we can check how the correlations decrease as the distance between the two patches
increase.\\

After the expression (\ref{eq:spot2spot}) was tested with Monte
Carlo simulations, we computed the correlations between $\tilde
C_\ell$ for two patches $A$ and $B$ where we varied the distance
$\theta$ between the centres of $A$ and $B$. We used a
standard CDM power spectrum and both patches $A$ and $B$ had a
radius of $18^\circ$ apodised with a $15^\circ$ FWHM Gaussian Gabor
window. In Fig. \ref{fig:spot2spot-diag} we have plotted the
diagonal of the normalised correlation matrix $(\VEV{\tilde C_\ell^\mathrm{A}\tilde
C_\ell^\mathrm{B}}-\VEV{\tilde C_\ell}^2)/\VEV{\tilde C_\ell}^2$. The angels $\theta$ we used were
$6^\circ$, $12^\circ$, $24^\circ$, $30^\circ$,
$36^\circ$ and $180^\circ$. One sees clearly how the correlations drop
with the distance. In the two last cases there were no common pixels
in the patches. As one could expect, the correlations for the largest
angels (the first few multipoles) do not drop that fast.\\

\begin{figure}
\begin{center}
\leavevmode
\epsfig {file=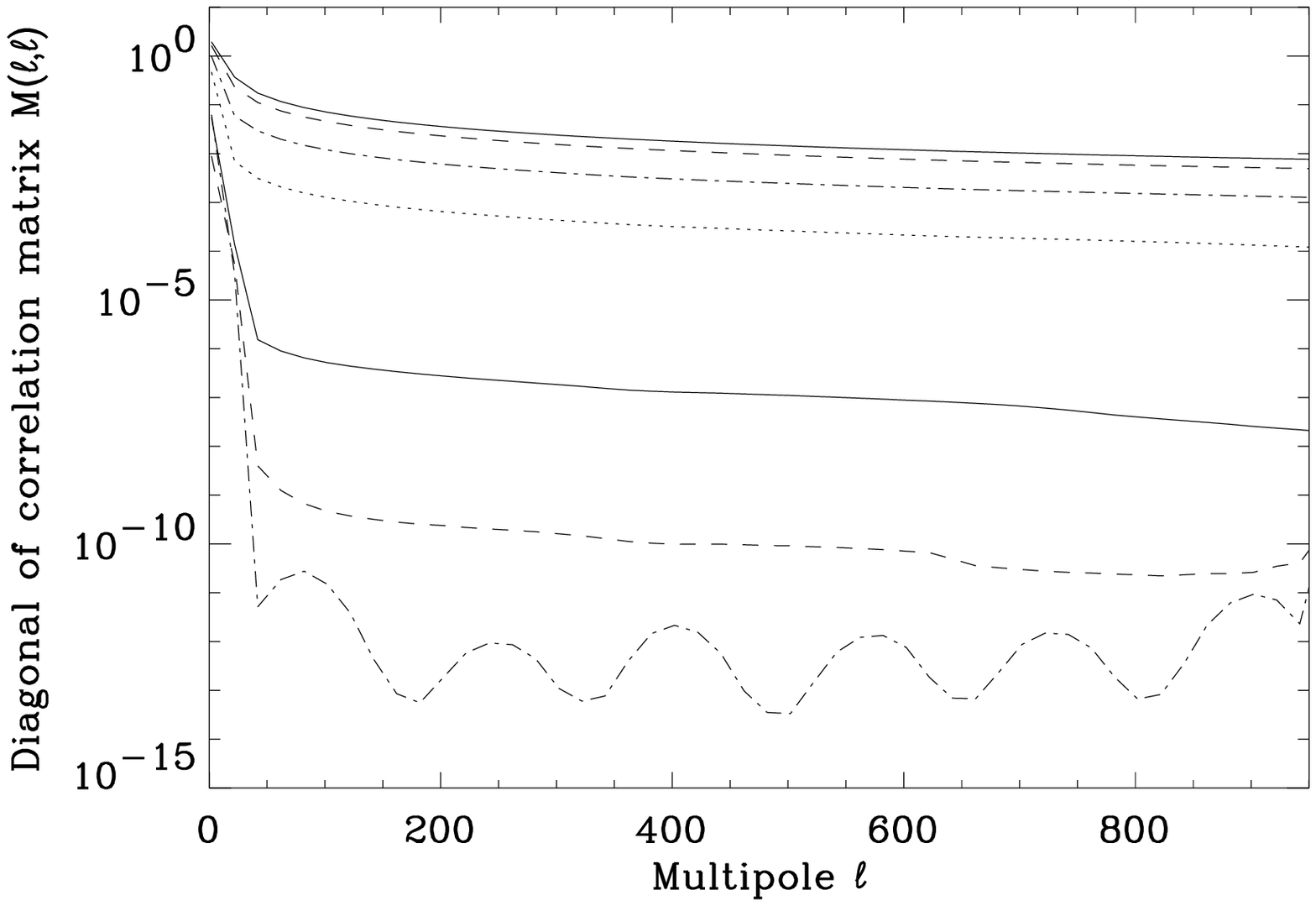,height=10cm,width=14cm}
\caption{The correlation between \protect{$\tilde C_\ell$} between two
patches $A$ and $B$ with an angular distance \protect{$\theta$} between
the centres. A normal CDM power spectrum was used and the patches had an
$18^\circ$ radius apodised with a $15^\circ$ FWHM Gaussian Gabor window. The
figure shows the diagonal of the normalised correlation matrix
\protect{$(\VEV{\tilde C_\ell^\mathrm{A}\tilde C_\ell^\mathrm{B}}-\VEV{\tilde C_\ell}^2)/\VEV{\tilde
C_\ell}^2$} where of course \protect{$\VEV{C_\ell}=\VEV{C_\ell^\mathrm{A}}=\VEV{C_\ell^\mathrm{B}}$}
The angles used are (from top to bottom on the figure)
\protect{$0^\circ$}, \protect{$6^\circ$}, \protect{$12^\circ$},
\protect{$18^\circ$}, \protect{$30^\circ$}, \protect{$36^\circ$} and \protect{$180^\circ$}.}
\label{fig:spot2spot-diag}
\end{center}
\end{figure}

In Fig. \ref{fig:spot2spot-cut} we have plotted two slices of
the correlation matrix of $\tilde C_\ell$ for a single patch at
$\ell=400$ and $\ell=800$. On the top we plotted the diagonals of the
correlation matrices for separation angle $\theta=30^\circ$,
$\theta=36^\circ$ and $\theta=180^\circ$. One sees that for the case
where the patches do not have overlapping pixels, the whole diagonals have
the same level as the far-off-diagonal elements in the
$\theta=0^\circ$ matrix. When doing power spectrum estimation on one
patch, the result did not change significantly when these far-off-diagonal elements
were set to zero. For this reason one expects that when analysing
several patches which do not overlap, simultaneously, the correlations
between non-overlapping patches do not need to be taken into account. Note however
that for the $\theta=30^\circ$ which means that there are only a few
overlapping pixels, the approximation will not be that good as the
level is orders of magnitude above the far-off-diagonals of the
$\theta=0^\circ$ matrix. Another thing to note is that for the lowest
multipoles, the correlation between patches is still high but we will
also assume this part to be zero and attempt a joint analysis of
non-overlapping patches.

\begin{figure}
\begin{center}
\leavevmode
\epsfig {file=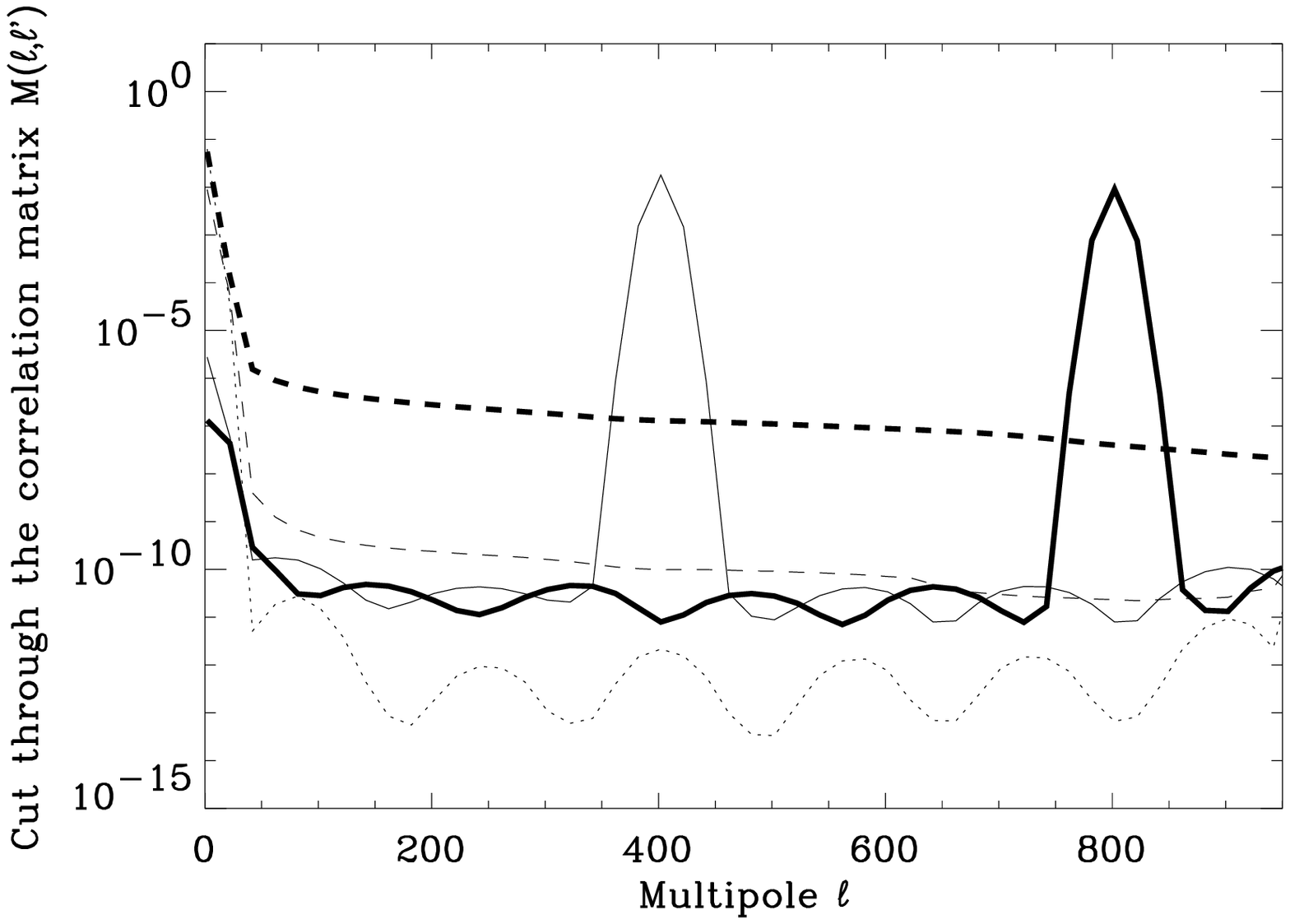,height=10cm,width=14cm}
\caption{slices of the correlation matrices which diagonals are
shown in Fig.
\protect{\ref{fig:spot2spot-diag}}. The solid lines (thin and thick)
show a slice of the \protect{$\theta=0^\circ$} correlation matrix at
\protect{$\ell=400$} and \protect{$\ell=800$} respectively. The dashed lines (thin and
thick) show the
diagonal of the correlation matrices for \protect{$\theta=36^\circ$} and
\protect{$\theta=30^\circ$} respectively. The dotted line is the diagonal of the
\protect{$\theta=180^\circ$} matrix.}
\label{fig:spot2spot-cut}
\end{center}
\end{figure}

The full correlation matrices for $0$ and $30$ degree separation are
shown in Fig. \ref{fig:cor030}. The figures show how the diagonal
is dropping relative to the far off-diagonal elements.

\begin{figure}
\begin{center}
\leavevmode
\epsfig {file=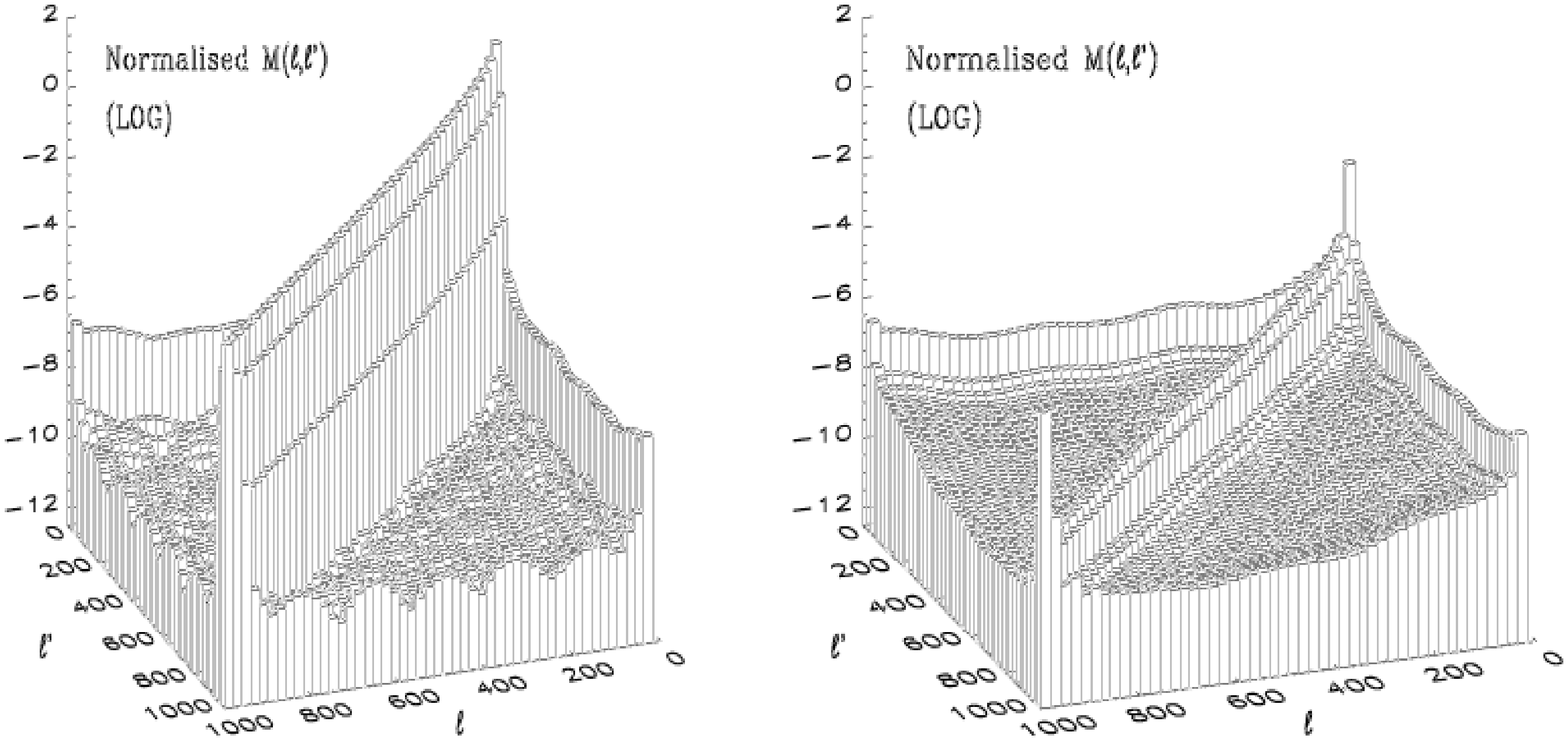,height=10cm,width=14cm}
\caption{The figure shows the normalised correlation matrices \protect{$M(\ell,\ell')=(\VEV{C^\mathrm{A}_\ell C^\mathrm{B}_\ell}-\VEV{C^\mathrm{A}_\ell}\VEV{C^\mathrm{B}_\ell})/(\VEV{C^\mathrm{A}_\ell}\VEV{C^\mathrm{N}_\ell})$}
between pseudo spectrum coefficients 
for two patches $A$ and $B$ of $18^\circ$ radius and with $0$ degree (left plot) and $30$ degree
separation. A Gaussian Gabor window with $15$ degree FWHM was
used. The aim of the plot is to show how correlations between $\tilde
 C_\ell$ from different patches drop when the distance between the two
patches is about the FWHM of the Gaussian kernel.}
\label{fig:cor030}
\end{center}
\end{figure}

In Fig. \ref{fig:cor36180} the full correlation matrices for $36$
and $180$ degree separation is shown. For $36$ degree separation one
can see that the diagonal has almost disappeared with respect to the
rest of the matrix whereas for $180$ degree the diagonal has vanished
completely. But the 'wall' at low multipoles remains.

\begin{figure}
\begin{center}
\leavevmode
\epsfig
{file=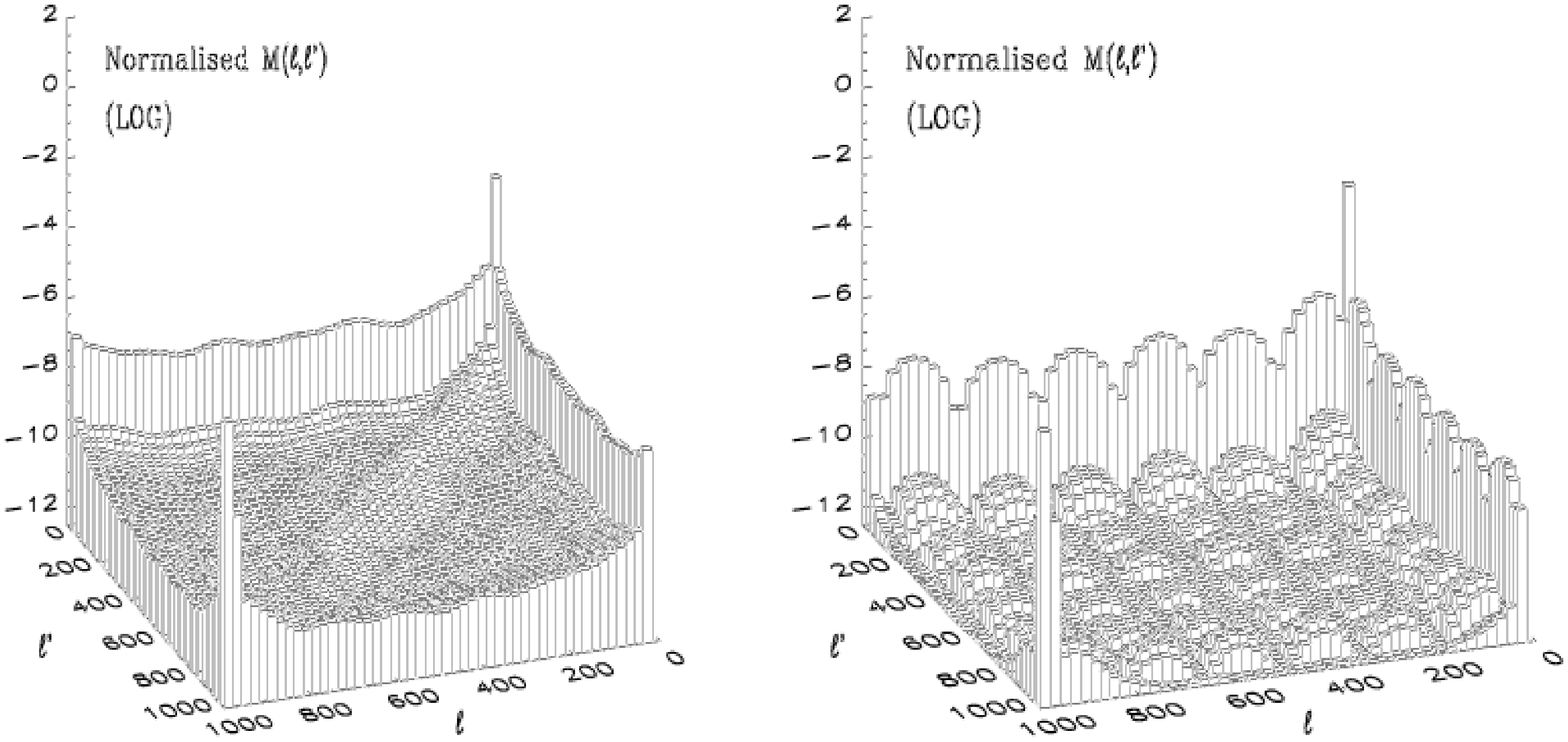,height=10cm,width=14cm}
\caption{This figure shows the same as Fig. \ref{fig:cor030} but
for $36$ and $180$ degree separation of the patches.} 
\label{fig:cor36180}
\end{center}
\end{figure}

In Fig. \ref{fig:onsphere}, we did a separate $C_\ell$ estimation on 146
non-overlapping patches with radius $18^\circ$ apodised with a $15^\circ$ FWHM
Gaussian Gabor window. The patches where uniformly distributed over the
sphere and uniform noise was added to the whole map. The figure
shows the average of the $146$ $C_\ell$ estimates. One can see that the
estimate seems to be approaching the full sky power spectrum even at
small multipoles.

\begin{figure}
\begin{center}
\leavevmode
\epsfig {file=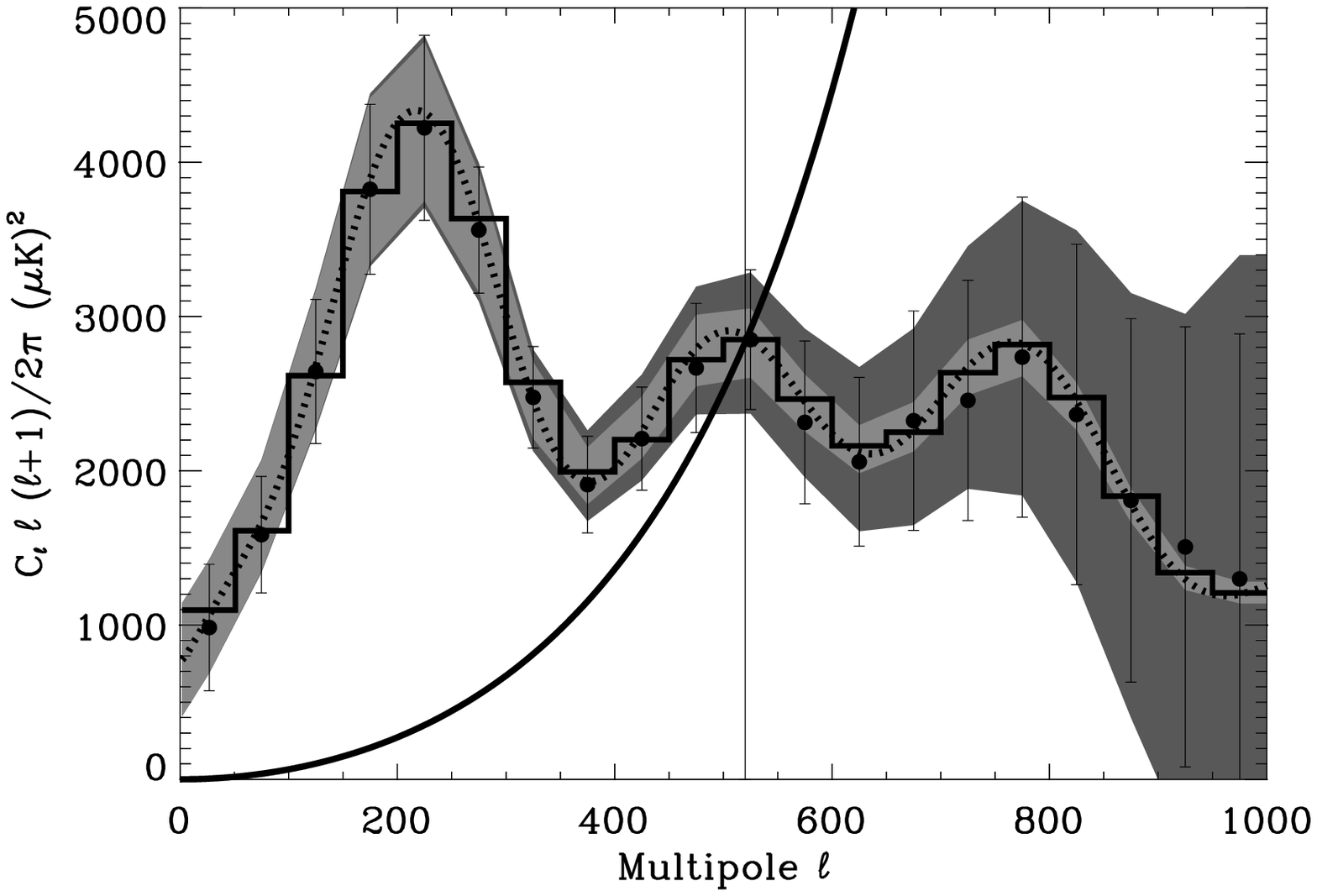,height=10cm,width=14cm}
\caption{The average of 146 individual power spectrum estimations of 146
non-overlapping patches on the same CMB map with uniform noise added
to it. The
patches all had radius \protect{$18^\circ$} degrees apodised with a \protect{$15^\circ$} FWHM Gaussian
Gabor window. The histogram shows the binned average of the
\protect{$146$} \protect{$\tilde C_\ell$} from the different
patches without noise. The dotted line is the average full sky power spectrum and the
shaded areas around the binned full sky power spectrum (not plotted) show the theoretical \protect{$1\sigma$}
variance with
(dark) and without (bright) noise. The solid line rising from the left
to the right is the noise power spectrum.}
\label{fig:onsphere}
\end{center}
\end{figure}

\begin{figure}
\begin{center}
\leavevmode
\epsfig {file=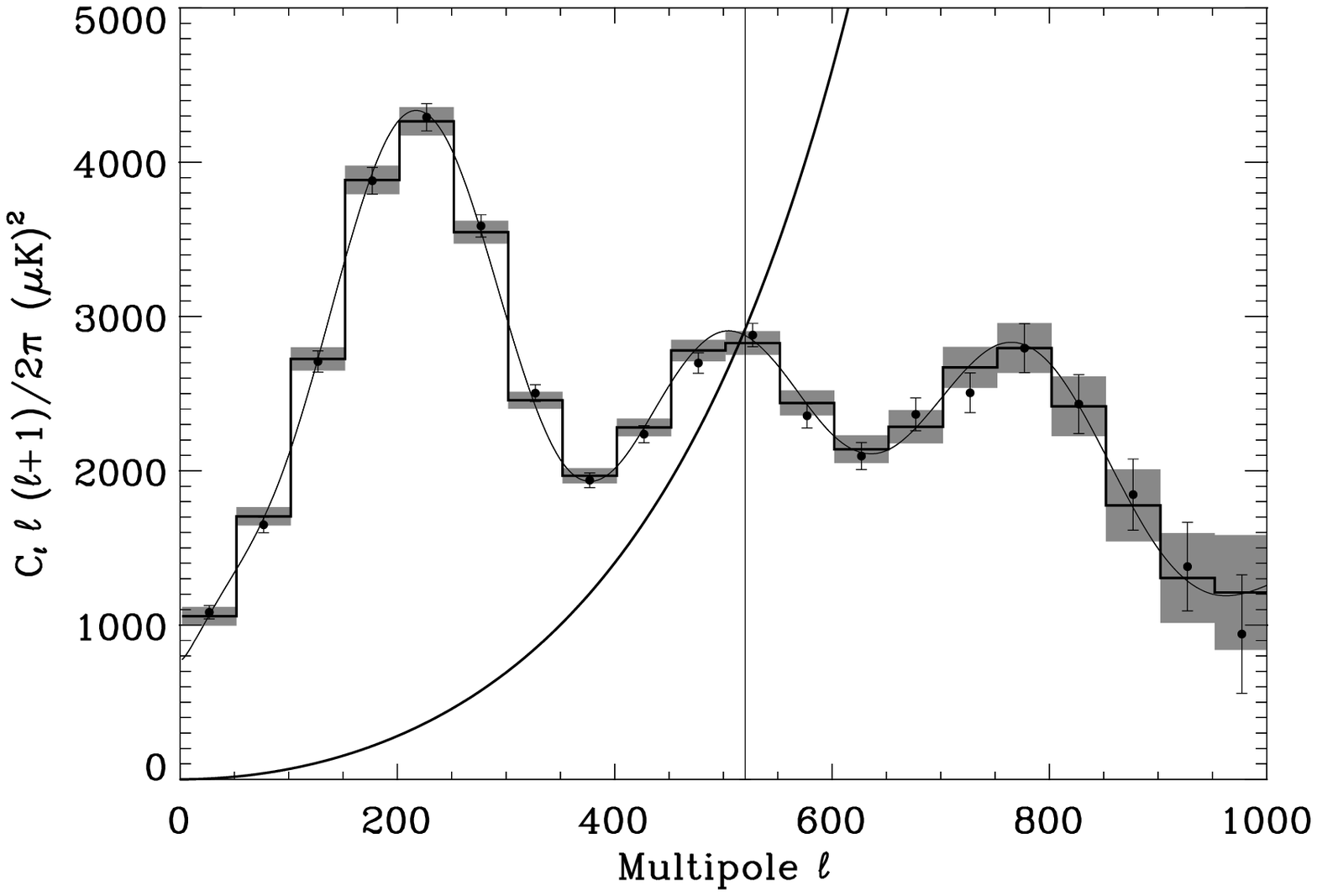,height=10cm,width=14cm}
\caption{The result of a joint analysis of 146 patches on the same
CMB map. The solid line shows the average full sky power spectrum, the
histogram shows the binned full sky power spectrum and the shaded boxes
show the expected \protect{$2\sigma$} deviations due to noise, cosmic and sample
variance. The sizes of the shaded boxes were calculated from the
formula for uniform noise (equation \ref{eq:uninoiseerror}). The dots show the estimates with
\protect{$2\sigma$} error bars taken from the Fisher matrix. As before the rising
solid line is the noise power spectrum and the vertical line shows
where \protect{$S/N=1$}.}
\label{fig:multiclest}
\end{center}
\end{figure}

Finally we made a joint analysis of all the $146$ patches. The idea was to
extend the data vector in the likelihood so that it contained the $N^\mathrm{in}$
$\tilde C_\ell$ from all the $146$ patches. The data vector can then be
written as ${\mathbf{ d}}=\{{\mathbf{ d}}_1,{\mathbf{ d}}_2,...,{\mathbf{ d}}_{146}\}$ where
${\mathbf{ d}}_i$ now denotes the
whole data vector for patch number $i$. From the results above it seems
to be a good approximation to assume that the correlation between
$\tilde C_\ell$ from different patches is zero so that the correlation
matrix will be block diagonal. Each block is then the correlation
matrix for each individual patch. The log-likelihood can then simply be
written as
\begin{equation}
L=\sum_{i=1}^{146}{\mathbf{ d}}_i^\mathrm{T}{\mt{M}}_i^{-1}{\mathbf{
d}}_i+\sum_{i=1}^{146}\ln\det {\mt{M}}_i,
\end{equation}
where ${\mt{M}}_i$ is the correlation matrix for patch number $i$. In Fig.
\ref{fig:multiclest} the result of this joint analysis is shown. One can
see that the full sky power spectrum is well within the two sigma
error bars of the estimates.\\

The method of combining patches on the sky for power spectrum analysis
will be developed further in a forthcoming paper.

\subsection{Monte Carlo simulations of the noise correlations and
extension to correlated noise}

The computation of the noise correlation matrix in the general case without any approximations 
takes $\sqrt{N_\mathrm{pix}}l_\mathrm{max}^2(N^\mathrm{in})^2$ which is approximately
$N_\mathrm{pix}^{3/2}(N^\mathrm{in})^2$ when $N_\mathrm{pix}$ is large ($N_\mathrm{pix}\approx\ell_\mathrm{max}^2$). When $N^\mathrm{in}$ is getting large this can be
calculated quicker using Monte Carlo simulations (as was done in
\cite{master}). Finding the $\tilde C_\ell^\mathrm{N}$ from one noise map takes
$\mathcal{O}(N_\mathrm{pix}^{3/2})$ operations so
using Monte Carlo simulations to find the whole noise matrix takes
$\mathcal{O}(N_\mathrm{pix}^{3/2}N_\mathrm{sim})$ operation where $N_\mathrm{sim}$ is the number of Monte Carlo
simulations needed. So when $N_\mathrm{sim}<<(N_\mathrm{in}^2)$ it will be
advantageous using MC if this gives the same result.\\

Also when the noise gets correlated, the analytic calculation of
$\VEV{\tilde a_{\ell m}^\mathrm{N}\tilde a_{\ell'm}^\mathrm{N}}$ will be very expensive. In this case
another method for computing $\VEV{\tilde a_{\ell m}^\mathrm{N}\tilde a_{\ell'm}^\mathrm{N}}$ will be
necessary and Monte Carlo simulations could also prove useful. For
a given noise model several noise realisations can be made and
averaged to yield the noise correlation matrix and the $\VEV{\tilde a_{\ell
m}^\mathrm{N}\tilde a_{\ell'm}^\mathrm{N}}$ term needed in the estimation process. This is of course dependent on a method for fast
evaluation of maps with different realisations.

\begin{figure}
\begin{center}
\leavevmode
\epsfig {file=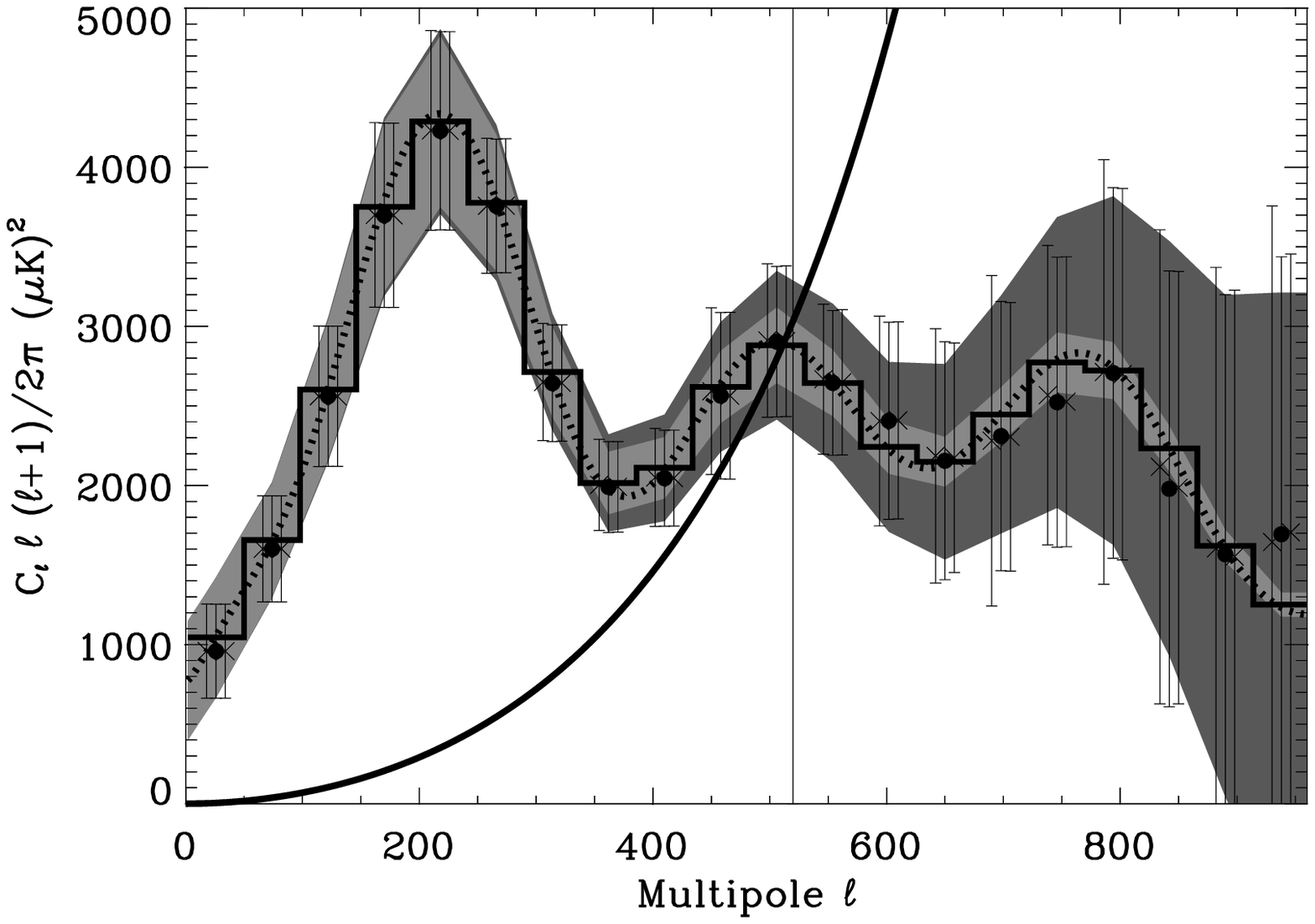,height=10cm,width=14cm}
\caption{Same as Fig. \ref{fig:cdm2} but with a different noise
model. Here the average of 100 estimations is shown. The big dots in
the middle of the bins show
the result using analytical expressions for the noise matrices. The
crosses on the left side of each big dot show the results of using
noise matrices from \protect{$N_\mathrm{sim}=1000$} MC simulations. The
crosses on the right side are for \protect{$N_\mathrm{sim}=20000$} noise simulations.}
\label{fig:mcnoise}
\end{center}
\end{figure}

In figure \ref{fig:mcnoise},  the result of $C_\ell$
estimation with noise matrix and $\VEV{a_{\ell m}^\mathrm{N}a_{\ell'm}^\mathrm{N}}$ computed
with Monte Carlo is shown. Again a standard CDM power spectrum was used with a
non-uniform white noise model and a $15^\circ$ FWHM Gaussian Gabor
window. In the $C_\ell$ estimation $N^\mathrm{in}=200$ $\tilde C_\ell$ were
used and $N^\mathrm{bin}=20$ power spectrum bins were estimated. The noise
matrices were calculated using (1) the analytical expression, (2) MC
with $N_\mathrm{sim}=20000$ and (3) MC with $N_\mathrm{sim}=1000$. In Fig.
\ref{fig:mcnoisecor}, a slice of the correlation matrices for
the different cases is shown for $\ell=500$. The dashed line (case
(3)) follows the solid line (case (1)) to a level of about $10^{-2}$
of the diagonal. The dotted line (case (2)) is roughly correct to about
$10^{-1}$ times the value at the diagonal.\\ 

\begin{figure}
\begin{center}
\leavevmode
\epsfig {file=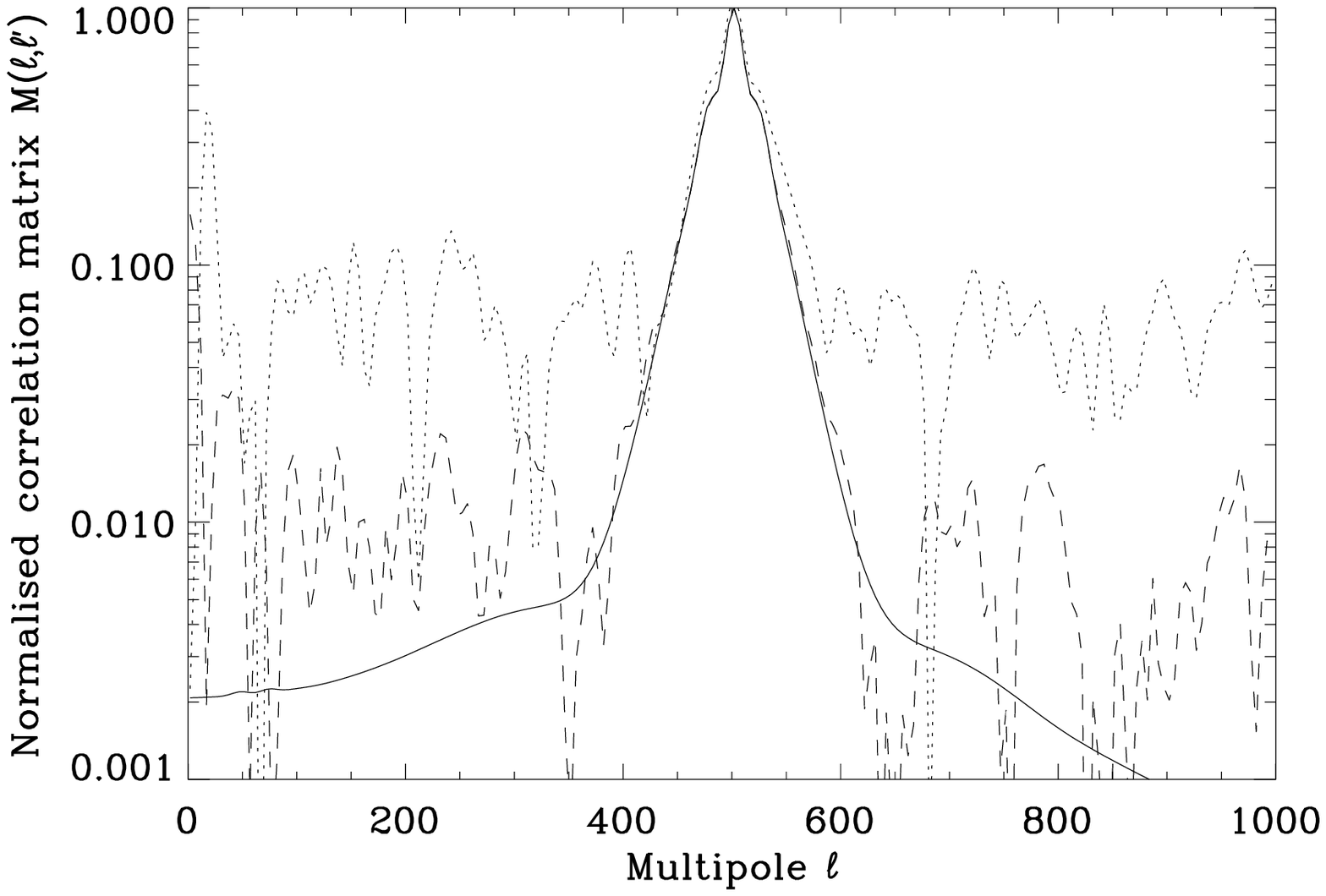,height=10cm,width=14cm}
\caption{A slice of the noise correlation matrix at multipole
\protect{$\ell=500$}. The correlation matrix was evaluated using the
analytical formulae (solid line), 20000 MC simulations (dashed line)
and 1000 MC simulations (dotted line). The matrix is here normalised
to be $1$ at the diagonal.}
\label{fig:mcnoisecor}
\end{center}
\end{figure}

We did 100
estimations for each case and the average result is plotted in Fig.
\ref{fig:mcnoise}. The big dots are the results from case (1), the
crosses on the right hand side are the results from (2) and the crosses
on the left hand side the results from (3). The average estimates seem
to be consistent, only in the highly noise dominated regime they
start to deviate. For case (2), the error bars are for some multipoles
higher and for some lower than the analytic case. The differences
are at most $3\%$. We conclude that using this many simulations, the
error bars do not increase significantly over the analytic case. For
case (3) the error bars are up to $17\%$ higher (and only higher) than
the analytic case. It seems that $1000$ simulations was not sufficient
to keep the same accuracy of the estimates as when using analytic
noise matrices. To
keep the error bars close to the analytic case, it seems that a few thousand simulations are necessary.

\section{Conclusion}
\label{sect:disc}

In this paper, we propose to use the spherical harmonic transform of the sky
apodised by a window function, or Gabor transform, as a fast and robust tool
to
estimate the CMB fluctuations power spectrum. It is known that the coupling
between modes resulting from the analysis on a cut sky affects the shape of
the
measured pseudo power spectrum and the statistics of the $C_\ell$ coefficients
(\cite{pseudo,master} and reference therein). In the case of axisymmetric windows  we can
compute analytically (in about $\ell_\mathrm{max}^3$ operations) the kernel relating the cut sky power
spectrum to the full sky one for a Gaussian and top-hat profile we give an analytical relation between the
spectral resolution attainable and the size of the sky window. Studying
windows
of different sizes, we show that for windows as small as $36$ degrees in
radius,
the measured power spectrum is undiscernable from the true one for $\ell$ larger
than about 50.

Noting that for large multipoles ($\ell\geq100$ for windows with radius larger than $36$ degree) the
statistics of the pseudo-$C_\ell$ coefficients measured in Monte Carlo
simulations
is close to Gaussian, we suggest the use of the pseudo power spectrum as
input
data vector in a likelihood estimation. 
For the first time, we show how the correlation matrix between the pseudo
power
spectrum coefficients obtained on an axisymmetric window of arbitrary profile
can be
computed rapidly for any input power spectrum, based on a recurrence relation.
The computation of the correlation matrix needs a precomputation (independent of the power spectrum) of
$\ell_\mathrm{max}N_m(N^\mathrm{in})^2$ operations and each calculation of the correlation matrix with a given power spectrum
takes $(N^\mathrm{in})^2(N^\mathrm{bin})^2$ operations, where $N^\mathrm{in}$ is the number of input pseudo-$C_\ell$ coefficients used,
$N^\mathrm{bin}$ is the number of estimated $C_\ell$ bins and $N_m$ is a window dependent factor ($N_m\approx200$ for a
Gaussian window and $N_m\approx400$ for a top-hat window). The noise correlation matrix can also be computed by
recurrence. For
axisymmetric noise this is very quick ($N_\mathrm{pix}\ell_\mathrm{max}$ operations). For general non-uniform noise this
takes
some more time (between $(N^\mathrm{in})^2\sqrt{N_\mathrm{pix}}\ell_\mathrm{max}$ and $(N_\mathrm{in})^2\sqrt{N_\mathrm{pix}}\ell_\mathrm{max}^2$ operations
dependent on the window profile and number of approximations). For a Gaussian window with a sharply varying noise
profile and a patch of sky similar in size to the one observed by
BOOMERANG (about $2\%$ of the sky) 
it takes about a day on
one single $500 \mathrm{MHz}$ processor. This is the computationally heaviest part of the method but this has to be
done
only once. The inversion of the correlation matrix, which is the leading
problem
when doing likelihood analysis, is now overcome, as the size of
the correlation matrix is so small that inversion is feasible. In the standard likelihood approach, the correlation
matrix has dimensions $N_\mathrm{pix}\times N_\mathrm{pix}$ which needs $N_\mathrm{pix}^3$ operations to be inverted. In our approach,
the size of the correlation matrix is $N^\mathrm{in}\times N^\mathrm{in}$ which in our example $N^\mathrm{in}=200$ is inverted in a few
seconds.

By doing Monte Carlo simulations of different experimental settings, we
shown
that the likelihood estimator is unbiased. The error bars were found using
the
inverse Fisher matrix and compared to the error bars obtained from Monte
Carlo. There was an excellent agreement between the two sets of error bars.
In \cite{master} it was shown that using a Gaussian apodisation suppresses the signal such that the error bars on
the estimated power spectrum becomes larger. In this paper we have shown that using a different window than the
top-hat window can be important for increasing the signal-to-noise ratio in data with non-uniform noise. We applied
a Gaussian window to an observed disc which had the 
noise level increasing from the
centre of the disc and outwards similar to what one can expect around the
ecliptic poles in scanning strategies like the ones of MAP and Planck.
In this case the Gaussian window has a high value in the centre where signal-to-noise per pixel is high and a low
value close to the more noise dominated edges. We shown that for this noise profile
using a Gaussian window increased the signal-to-noise ratio significantly
over
the top-hat window, showing that adapting the window to downweight noisy
pixels
gives better performance than a simple uniform weighting.

Finally two extensions of the power spectrum estimation method were
discussed. First it was shown that for observed areas on the sky which are
not
axisymmetric, one can cover the area by several axisymmetric patches and
make a joint analysis of the pseudo power spectrum coefficients from all the
patches. Each of the patches can have a different window in order to optimise
signal-to-noise in each patch. This method will be extended in a forthcoming
paper where we will discuss the use of the method for analysing MAP and
Planck
data sets. We also shown that the calculation of the noise correlation
matrix
can be quicker by Monte Carlo simulations if the number of pixels in the
observed area is huge (about $10^6$ pixels but dependent on the window shape). This may also be used in the case of
correlated noise. We
shown that a few thousand simulations are necessary to get the same accuracy
in
the power spectrum estimates as when using the analytic formula for the noise
correlation matrix.

In \cite{polarisation} we show that the power spectrum estimation method presented in this paper can easily be
extended to polarisation. By extending the data vector in the likelihood to have also the pseudo-$C_\ell$ from
polarisation, one can in a similar way estimate for the temperature and polarisation power spectra jointly.

\section*{Acknowledgements}

We would like to thank A. J. Banday and B. D. Wandelt for helpful discussions. We acknowledge the use of HEALPix \cite{healpix}
and CMBFAST \cite{cmbfast}. FKH was supported by a grant from the Norwegian Research Council.

\begin{appendix}

\section{Rotation Matrices}
\label{app:rotmat}
A spherical function $T({\mathbf{\hat n}})$ is rotated by the operator $\hat
D(\alpha\beta\gamma)$ where $\alpha\beta\gamma$ are the three Euler
angles for rotations (See T.Risbo 1996) and the inverse rotation is
$\hat D(-\gamma-\beta-\alpha)$. For the spherical harmonic functions,
this operator takes the form,
\begin{equation}
Y_{\ell m}({\mathbf{\hat n}}')=\sum_{m'=-\ell}^\ell
D_{m'm}^\ell(\alpha\beta\gamma)Y_{\ell m'}({\mathbf{\hat n}}),
\end{equation}
where $D_{m'm}^\ell$ has the form
\begin{equation}
D_{m'm}^\ell(\alpha\beta\gamma)=\mathrm{e}^{\mathrm{i}m'\alpha}d_{m'm}^\ell(\beta)\mathrm{e}^{\mathrm{i}m\gamma}.
\end{equation}
Here $d_{m'm}^\ell(\beta)$ is a real coefficient with the following
property:
\begin{equation}
d_{m'm}^\ell(\beta)=d_{mm'}^\ell(-\beta).
\end{equation}
The D-functions also have the following property:
\begin{equation}
D_{m'm}^\ell(\alpha\beta\gamma)=\sum_{m''}D_{m'm''}^\ell(\alpha_2\beta_2\gamma_2)D_{m''m}^\ell(\alpha_1\beta_1\gamma_1),
\end{equation}
where $(\alpha\beta\gamma)$ is the result of the two consecutive
rotations $(\alpha_1\beta_1\gamma_1)$ and
$(\alpha_2\beta_2\gamma_2)$.

The complex conjugate of the rotation matrices can be written as
\begin{equation}
D^{\ell*}_{mm'}=(-1)^{m+m'}D^\ell_{(-m)(-m')}.
\end{equation}

\section{Some Wigner Symbol Relations}
\label{app:wig}

Throughout the paper, the Wigner 3j Symbols will be used
frequently. Here are some relations for these symbols, which are
used.
The orthogonality relation is,
\begin{equation}
\label{eq:wigort}
\sum_{mm'}\wigner{\ell}{\ell'}{\ell''}{m}{m'}{m''}\wigner{\ell}{\ell'}{L''}{m}{m'}{M''}=(2\ell''+1)^{-1}\delta_{\ell''L''}\delta_{m''M''}.
\end{equation}
The Wigner 3j Symbols can be represented as an integral of rotation
matrices (see Appendix(\ref{app:rotmat})),
\begin{equation}
\label{eq:wigd}
\frac{1}{8\pi^2}\int d\cos\theta d\phi d\gamma D_{m_1m'_1}^\ell
D_{m_2m'_2}^{\ell'}D_{m_3m'_3}^{\ell''}=\wigner{\ell}{\ell'}{\ell''}{m_1}{m_2}{m_3}\wigner{\ell}{\ell'}{\ell''}{m'_1}{m'_2}{m'_3}.
\end{equation}
This expression can be reduced to,
\begin{equation}
\label{eq:wigy}
\int d{\mathbf{\hat n}} Y_{\ell m}({\mathbf{\hat n}})
Y_{\ell'm'}({\mathbf{\hat n}})Y_{\ell''m''}({\mathbf{\hat
n}})=\sqrt{\frac{(2\ell+1)(2\ell'+1)(2\ell''+1)}{4\pi}}\wigner{\ell}{\ell'}{\ell''}{m}{m'}{m''}\wigo{}{}{}.
\end{equation}

\section{Recurrence Relation}
\label{app:rec}
It is important for the precalculations to the likelihood analysis that the calculation of
$h(\ell,\ell',m)$ is fast. For this reason a
recurrence relation for $h(\ell,\ell',m)$ would be helpful. To speed up the calculation of the noise
correlation matrix for non-axisymmetric noise, it would also help if
one had a more general recurrence relation for $h(\ell,\ell',m,m')$. We
will now show how to find such a recurrence for these objects which we
now call $A_{\ell'\ell}^{m'm}$ to simplify notation (and for the
notation to comply with \cite{pseudo}). The definition
is again,
\begin{equation}
A_{\ell'\ell}^{m'm}=\int_a^bG({\mathbf{\hat n}})Y^*_{\ell m}({\mathbf{\hat n}})Y_{\ell'm'}({\mathbf{\hat n}})d{\mathbf{\hat n}},
\end{equation}
where $G({\mathbf{\hat n}})=G(\theta,\phi)$ is a general function and $Y_{\ell
m}$ are the spherical harmonics which can be factorised into one part
dependent on $\theta$ and one dependent on $\phi$ in the following way,
\begin{equation}
Y_{\ell m}(\theta,\phi)=\lambda_{\ell m}(\cos{\theta})\mathrm{e}^{-\mathrm{i}\phi m}.
\end{equation}
Now writing,
\begin{eqnarray}
A_{\ell '\ell }^{m'm}&=&\int d\cos\theta\lambda_{\ell
m}(\cos\theta)\lambda_{\ell'm'}(\cos\theta)\int d\phi
\mathrm{e}^{-\mathrm{i}\phi(m-m')}G(\theta,\phi)\\
&\equiv&\int d\cos\theta\lambda_{\ell
m}(\cos\theta)\lambda_{\ell'm'}(\cos\theta)F_{m'm}(\theta),
\end{eqnarray}
where $F_{m'm}(\theta)$ is simply the Fourier transform of the window
at each $\theta$. The quantities $A_{\ell '\ell }^{m'm}$ and
$F_{m'm}(\theta)$ are in general complex quantities obeying,
\begin{equation}
A_{\ell '\ell }^{m'm}=(A_{\ell \ell'}^{mm'})^*\ \ \
F_{m'm}(\theta)=(F_{mm'}(\theta))^*
\end{equation}

The $A_{\ell '\ell }^{m'm}$ can be expressed as
\begin{equation}
\label{eq:atoi}
A_{\ell '\ell }^{m'm}=\frac{\sqrt{(2\ell '+1)(2\ell +1)}}{2}\sqrt{\frac{(\ell '-m')!(\ell -m)!}{(\ell '+m')!(\ell
+m)!}}I_{\ell '\ell }^{m'm},
\end{equation}
where $I_{\ell '\ell }^{m'm}$ is defined as:
\begin{equation}
I_{\ell '\ell }^{m'm}=\int_a^bF_{m'm}(x)P_\ell ^m(x)P_{\ell '}^{m'}(x)dx.
\end{equation}
The following relation for the Legendre Polynomials will
be used:
\begin{equation}
\label{eq:plmrel}
xP_\ell ^m=\frac{\ell -m+1}{2\ell +1}P_{\ell +1}^m+\frac{\ell +m}{2\ell +1}P_{\ell -1}^m
\end{equation}
We now define the object $X_{\ell '\ell }^{m'm}$ as
\begin{equation}
X_{\ell '\ell }^{m'm}=\int_a^bF_{m'm}(x)xP_\ell ^mP_{\ell '}^{m'}dx
\end{equation}
Using relation (\ref{eq:plmrel}) in this definition,  one gets,
\begin{equation}
X_{\ell '\ell }^{m'm}=\frac{\ell -m+1}{2\ell +1}I_{\ell '(\ell +1)}^{m'm}+\frac{\ell +m}{2\ell +1}I_{\ell '(\ell
-1)}^{m'm}
\end{equation}
One can also exchange $(\ell,\ell')$ and $(m,m')$ to get
\begin{equation}
X_{\ell \ell'}^{mm'}=\frac{\ell '-m'+1}{2\ell '+1}I_{\ell (\ell '+1)}^{mm'}+\frac{\ell '+m'}{2\ell '+1}I_{\ell
(\ell '-1)}^{mm'}
\end{equation}
Taking the complex conjugate of the first expression and subtracting
the last, one has
\begin{eqnarray}
(X_{\ell '\ell }^{m'm})^*-X_{\ell \ell'}^{mm'}=0&=&\frac{\ell
-m+1}{2\ell +1}(I_{\ell '(\ell +1)}^{m'm})^*+\frac{\ell +m}{2\ell
+1}(I_{\ell '(\ell -1)}^{m'm})^*\\
&&-\frac{\ell '-m+1}{2\ell '+1}I_{\ell (\ell '+1)}^{mm'}-\frac{\ell '+m}{2\ell '+1}I_{\ell (\ell '-1)}^{mm'}
\end{eqnarray}
Then setting $\ell '=\ell '-1$ one gets:
\begin{equation}
I_{\ell' \ell}^{m'm}=\frac{2\ell '-1}{\ell '-m'}\left(\frac{\ell -m+1}{2\ell
+1}I_{(\ell'-1)(\ell+1)}^{m'm}+\frac{\ell +m}{2\ell +1}I_{(\ell'-1)(\ell-1)}^{m'm}-\frac{\ell '+m-1}{2\ell
'-1}I_{(\ell'-2)\ell}^{m'm}\right)
\end{equation}
Using equation (\ref{eq:atoi}), one can express this as
\begin{eqnarray}
\label{eq:rec}
A_{\ell '\ell }^{m'm}&=&\frac{1}{\sqrt{{\ell '}^2-m'^2}}\biggl(\sqrt{\frac{(4{\ell '}^2-1)((\ell +1)^2-m^2)}{(2\ell
+1)(2\ell +3)}}
A_{(\ell '-1)(\ell +1)}^{m'm}\\
\nonumber
&& +\sqrt{\frac{(4{\ell '}^2-1)(\ell ^2-m^2)}{4\ell ^2-1}}A_{(\ell '-1)(\ell -1)}^{m'm}
-\sqrt{\frac{(2\ell '+1)((\ell '-1)^2-m'^2)}{2\ell '-3}}A_{(\ell '-2)\ell }^{m'm}\biggl),
\end{eqnarray}
which is the final recurrence relation. The $A_{m'\ell }^{m'm}$ elements must
be provided before the recurrence is started. Then for each $(m,m')$, set
$\ell '=m'+1$ and let $\ell $ go from $\ell '$ and upwards, then set $\ell '=m'+2$ and again
let $\ell $ go from $\ell '$ and upwards. Continue to the desired size of $\ell '$.
Note that, in order to get all objects up to $A_{\ell _\mathrm{max}\ell _\mathrm{max}}^{m'm}$
one always needs to go up to $\ell =2\ell _\mathrm{max}$ during recursion. This
is because of the $A_{(\ell '-1)(\ell +1)}^{m'm}$ term which demands
an object indexed $(\ell +1)$ in the previous $\ell '$ row.\\
To start the recurrence, one can precomputed the $A^{m'm}_{m'\ell}$ factors
fast and easily using FFT and a sum over rings on the grid. F.ex. for
the HEALPix grid, we did it the following way,
\begin{equation}
A_{m'\ell}^{m'm}=\sum_r\lambda^r_{m'm'}\lambda^r_{\ell m}\sum_{j=0}^{N_r-1}
\mathrm{e}^{-2\pi ij/N_r(m-m')}G_{rj},
\end{equation}
where the last part is the Fourier transform of the Gabor window,
calculated by FFT, $r$ is ring number on the grid and $j$ is azimuthal
position on each ring. Ring $r$ has $N_r$ pixels.

It turns out that the recurrence can be numerically unstable dependent
on the window and multipole, and in order to
avoid problems we (using double precision numbers) restart the
recurrence with a new set of precomputed $A^{m'm}_{\ell'\ell}$ for every
50th $\ell'$ row. However for most windows and multipoles that we tested the
recurrence can run for hundreds of $\ell$-rows without problems.\\\\

\section{Rotational Invariance}
\label{app:rotin}
It was shown that the average $\VEV{\tilde C_\ell}$ is invariant under
rotations of the Gabor window.
We will now show that the non-averaged $\tilde C_\ell$ are rotationally invariant under
any rotation of the sky AND Gabor window by the same angle. This fact
justifies that we can always put the window on the north pole
since this simplifies the calculations. In the following we will use the rotation matrices $D^\ell_{mm'}$
described in Appendix (\ref{app:rotmat}). Consider a rotation of the sky and window by the angles
$(-\gamma-\beta-\alpha)$. Then the $\tilde a_{\ell m}$ becomes,
\begin{equation}
\tilde a_{\ell m}^{\mathrm{rot}}=\int d{\mathbf{\hat n}}[\hat
D(-\gamma-\beta-\alpha) T({\mathbf{\hat n}}) G({\mathbf{\hat n}})] Y^*_{\ell m}({\mathbf{\hat n}}).
\end{equation}
If one makes the inverse rotation of the integration angle ${\mathbf{\hat n}}$, one
can write this as
\begin{equation}
\tilde a_{\ell m}^{\mathrm{rot}}=\int d{\mathbf{\hat n}} T({\mathbf{\hat n}})G({\mathbf{\hat n}})[\hat
D^*(\alpha\beta\gamma)Y^*_{\ell m}({\mathbf{\hat n}})],
\end{equation}
which is just
\begin{equation}
\tilde a_{\ell
m}^{\mathrm{rot}}=\sum_{m'}D_{m'm}^{\ell*}(\alpha\beta\gamma)\int
d{\mathbf{\hat n}}T({\mathbf{\hat n}})G({\mathbf{\hat n}})Y^*_{\ell m'}({\mathbf{\hat n}}).
\end{equation}
One can identify the last integral as the normal $\tilde a_{\ell m}$.
\begin{equation}
\tilde a_{\ell
m}^{\mathrm{rot}}=\sum_{m'}D_{m'm}^{\ell*}(\alpha\beta\gamma)\tilde
a_{\ell m}.
\end{equation}
So the $\tilde a_{\ell m}$ are NOT rotationally invariant. Rotation
mixes $m$-modes for a given $\ell$-value.\\

Now to the $\tilde C_l$. One has that
\begin{eqnarray}
\tilde C_{\ell}^\mathrm{rot}&=&\frac{1}{2\ell+1}\sum_m a^\mathrm{rot}_{\ell
m}a^\mathrm{rot*}_{\ell m}\\
&=&\frac{1}{2\ell+1}\sum_m\sum_{m'}\sum_{m''}D_{m'm}^\ell(\alpha\beta\gamma)D_{m''m}^{\ell*}(\alpha\beta\gamma)\tilde
a_{\ell m'} \tilde a_{\ell m''}^*\\
&=&\frac{1}{2\ell+1}\sum_{m'm''}\tilde
a_{\ell m'} \tilde a_{\ell m''}^*\sum_m D_{m'm}^\ell(\alpha\beta\gamma)D_{m''m}^{\ell*}(\alpha\beta\gamma).
\end{eqnarray}
Using the properties given in Appendix (\ref{app:rotmat}), one can write the last D-function on
the last line as,
\begin{eqnarray}
\hat
D_{m''m}^{\ell*}(\alpha\beta\gamma)&=&D_{m''m}^\ell(-\alpha\beta-\gamma)\\
&=&\mathrm{e}^{-\mathrm{i}m''\alpha}d_{m''m}^\ell(\beta)\mathrm{e}^{-\mathrm{i}m\gamma}\\
&=&\mathrm{e}^{-\mathrm{i}m''\alpha}d_{mm''}^\ell(-\beta)\mathrm{e}^{-\mathrm{i}m\gamma}\\
&=&D_{mm''}^\ell(-\gamma-\beta-\alpha).
\end{eqnarray}
Knowing that $(-\gamma-\beta-\alpha)$ is the inverse rotation of
$(\alpha\beta\gamma)$ one can write,
\begin{eqnarray}
\sum_m
D_{m'm}^\ell(\alpha\beta\gamma)D_{m''m}^{\ell*}(\alpha\beta\gamma)&=&\sum_m
D_{m'm}^\ell(\alpha\beta\gamma)D_{mm''}^\ell(-\gamma-\beta-\alpha)\\
&=&D_{m'm''}^\ell(000)=\delta_{m'm''}
\end{eqnarray}
So one gets,
\begin{equation}
\tilde C_\ell^\mathrm{rot}=\tilde C_\ell.
\end{equation}

\section{The Correlation Matrix}
\label{app:cor}

To do fast likelihood analysis with $\tilde C_\ell$ one needs to be
able to calculate $\VEV{\tilde C_\ell}$ and the correlations $\VEV{\tilde
C_\ell\tilde C_{\ell'}}$ fast. Calculating the average $\VEV{\tilde
C_\ell}$ by formula (\ref{eq:kernrel}) using the analytic expression (\ref{eq:kernexp}) for the
kernel is not very fast. It turns out that a faster way of evaluating
the kernel is by using
direct integration (summation on the pixelised sphere) and then, as shown in Appendix (\ref{app:rec}), recurrence.
By
means of an integral, one can then write the $\tilde a_{\ell m}$ as
(now assuming that ${\mathbf{\hat n}}_0$ is on the north pole),
\begin{eqnarray}
\tilde a_{\ell m}&=&\sum_{\ell'm'}a_{\ell'm'}\int G({\mathbf{\hat n}})Y^*_{\ell
m}({\mathbf{\hat n}})Y_{\ell'm'}({\mathbf{\hat n}})d{\mathbf{\hat n}}\nonumber\\
&=&\sum_{\ell'm'}a_{\ell'm'}\int G(\theta)\lambda_{\ell
m}(\theta)\lambda_{\ell'm'}(\theta)d\cos{\theta}\underbrace{\int
\mathrm{e}^{-\mathrm{i}\phi(m-m')}d\phi}_{2\pi\delta_{mm'}}\nonumber\\
&=&\sum_{\ell'}a_{\ell'm}2\pi\int G(\theta)\lambda_{\ell
m}(\theta)\lambda_{\ell'm}(\theta)d\cos{\theta}\nonumber\\
\label{eq:palm}
&\equiv&\sum_{\ell'}a_{\ell' m}h(\ell,\ell',m),
\end{eqnarray}
where the last line defines $h(\ell,\ell',m)$ and $\lambda_{\ell
m}(\theta)$ is given by,
\begin{equation}
Y_{\ell m}(\theta,\phi)=\lambda_{\ell m}(\theta)\mathrm{e}^{-\mathrm{i}\phi m}.
\end{equation}
Using this form, one gets,
\begin{equation}
\VEV{\tilde C_\ell}=\frac{1}{2\ell+1}\sum_{\ell'}C_{\ell'}\sum_m
h^2(\ell,\ell',m).
\end{equation}
To obtain this expression, ${\mathbf{\hat n}}_0$ was on the north
pole, but as was shown, the $\VEV{\tilde C_\ell}$s are rotationally
invariant, that is $\VEV{\tilde C_\ell}$ remains the same if one rotates
the Gabor window so that it is centred on the north pole.\\

When using real CMB data, the observed temperature map is always
pixelised. So an integral over the sphere has to be replaced by a sum
over pixels. In this case, the formula for $h(\ell,\ell',m)$ has to be
replaced by
\begin{equation}
h(\ell,\ell',m)=\sum_jG_j\lambda_{\ell m}^j\lambda_{\ell'm}^j\Delta_j,
\end{equation}
where the index $j$ is the pixel number replacing the angle $\theta$ and $\Delta_j$ is the area of pixel $j$. Using
a pixelisation scheme
like HEALPix \cite{healpix} which has a structure of azimuthal rings
going from the north to the south pole with $N_r$ pixels in ring $r$ and
equal area for each pixel $\Delta_j=\Delta$ this can be written as 
\begin{eqnarray}
\nonumber
h(\ell,\ell',m)&=&\Delta\sum_r\sum_{p=0}^{N_r-1}G_r\lambda_{\ell
m}^r\lambda_{\ell'm}^r,\\
\label{eq:hsum}
&=&\Delta\sum_rN_rG_r\lambda_{\ell
m}^r\lambda_{\ell'm}^r.
\end{eqnarray}
Here the sum over pixels is split into a sum over rings $r$ and a sum
over the pixels in each ring $p$. The first sum goes over all rings
which have $\theta<\theta_\mathrm{C}$.

Using this expression for the $\tilde a_{\ell m}$ one can now find the
correlation matrix
\begin{equation}
\VEV{\tilde C_\ell\tilde C_{\ell'}}=\sum_{mm'}\frac{\VEV{\tilde a_{\ell
m}^*\tilde a_{\ell m}\tilde a_{\ell'm'}^*\tilde
a_{\ell'm'}}}{(2\ell+1)(2\ell'+1)}.
\end{equation}
In this expression one can use relation (\ref{eq:palm}) to get,
\begin{eqnarray}
\VEV{\tilde C_\ell\tilde
C_{\ell'}}&=&\frac{1}{(2\ell+1)(2\ell'+1)}\sum_{mm'}\sum_{LL'KK'}\VEV{a_{Lm}^*a_{L'm}a_{Km'}^*a_{K'm'}}\\
&&\times h(\ell,L,m)h(\ell,L',m)h(\ell',K,m')h(\ell',K',m')\\
&=&\frac{1}{(2\ell+1)(2\ell'+1)}\\
&\times&\sum_{mm'}\sum_{LL'KK'}[\VEV{a_{Lm}^*a_{L'm}}\VEV{a_{Km'}^*a_{K'm'}}\\
&&+\VEV{a_{Lm}^*a_{Km'}^*}\VEV{a_{L'm}a_{K'm'}}\\
&&+\VEV{a_{Lm}^*a_{K'm'}}\VEV{a_{L'm}a_{Km'}^*}]\\
&\times& h(\ell,L,m)h(\ell,L',m)h(\ell',K,m')h(\ell',K',m').
\end{eqnarray}
Clearly the first term is just the product $\VEV{\tilde C_\ell}\VEV{\tilde
C_{\ell'}}$, and the two last terms are equal (using
$a_{Km'}^*=a_{Km'}(-1)^{m'}$ and $a_{K'm'}=(-1)^{m'}a^*_{K'm'}$) so one
gets,
\begin{equation}
\label{eq:cormat}
M_{ij}=\frac{2}{(2\ell+1)(2\ell'+1)}\sum_m\left(\sum_LC_Lh(\ell_i,L,m)h(\ell_j,L,m)\right)^2.
\end{equation}
This is one of the main results of this paper since the formula allows
one to analytically calculate the correlation matrix needed for
likelihood analysis. Another main result is the recurrence deduced in
appendix (\ref{app:rec}) which allows fast evaluation of the
$h(\ell,\ell',m)$ functions and thereby this correlation matrix.\\

By using the binning of the power spectrum described in equation (\ref{eq:binning}), the correlation matrix can be
calculated faster if it is written as
\begin{equation}
\label{eq:scorsimp}
M^\mathrm{S}_{ij}=\sum_b\sum_{b'}D_bD_{b'}\chi(b,b',i,j),
\end{equation}
where $\chi(b,b',i,j)$ is given as,
\begin{eqnarray}
\nonumber \chi(b,b',i,j)&\equiv&\frac{2}{(2\ell_i+1)(2\ell_j+1)}\\
\nonumber&\times&\sum_m\left(\sum_{l\epsilon
b}B_\ell^2\ell(\ell+1)h(\ell,\ell_i,m)h(\ell,\ell_j,m)\right)\\
\label{eq:chis}
&&\times\left(\sum_{l\epsilon
b'}B_\ell^2\ell(\ell+1)h(\ell,\ell_i,m)h(\ell,\ell_j,m)\right),
\end{eqnarray}
which is precomputed. The sums over $\ell$ here go over the $\ell$
values in each specific bin $b$. One sees that computing the
likelihood takes of the order $(N^\mathrm{bin})^2(N^\mathrm{in})^2$ operations whereas the
precomputation of the factor $\chi(b,b',i,j)$ goes as $\ell_\mathrm{max} (N^\mathrm{in})^2 N_m$
where $N_m$ is the number of $m$ values used. Note that the multipole
coefficients of the beam $B_\ell$ are also included. The reason is
that the input data is always affected by the beam and this is
corrected for by using the beam convolved full sky power spectrum
$C_\ell B_\ell^2$.\\

The sum over $m$ in the expressions for the covariance
matrix and $\VEV{\tilde C_\ell}$ can be limited. The $h$-functions are
rapidly
decreasing for increasing $m$ for Gaussian and top-hat windows. For Gaussian Gabor windows it seems that one
can cut the sums over $m$ at $N_m=200$ to high accuracy. For top-hat
windows, the sum should be extended to $N_m=400$.\\

\section{Including White Noise}
\label{app:noise}
In this appendix the total $\VEV{\tilde C_\ell}$ and the correlation matrix including contributions from white noise is
found analytically.
We assume that each pixel $j$ has a noise temperature denoted by
$n_j$, with the following properties,
\begin{equation}
\VEV{n_j}=0, \ \ \ \VEV{n_jn_{j'}}=\delta_{jj'}\sigma^2_j,
\end{equation}
where $\sigma_j$ is the noise variance in pixel $j$. Then one has the
following expressions for the $a_{\ell m}$ and $C_l$ (we use
superscript $N$ for noise quantities),
\begin{eqnarray}
a_{\ell m}^\mathrm{N}&=&\sum_j n_j Y^{j*}_{\ell m} \Delta_j\\
\VEV{a_{\ell m}^\mathrm{N}{a_{\ell' m'}^\mathrm{N}}^*}&=&\sum_{jj'}\VEV{n_jn_{j'}}Y_{\ell m}^j{Y_{\ell
m}^{j'}}^*\Delta_j\Delta_{j'}=\sum_j\sigma_j^2Y_{\ell m}^j{Y_{\ell' m'}^j}^*\Delta_j^2\\
\VEV{C_\ell^\mathrm{N}}&=&\frac{1}{2\ell+1}\sum_m\VEV{a_{\ell m}^\mathrm{N}{a_{\ell
m}^\mathrm{N}}^*}=\frac{1}{4\pi}\sum_j\sigma_j^2\Delta_j^2.
\end{eqnarray}
Here $Y_{\ell m}^j$ is the Spherical Harmonic of the pixel centre of
pixel $j$.
For the windowed coefficients, one gets similarly,
\begin{eqnarray}
\tilde a_{\ell m}^\mathrm{N}&=&\sum_j G_j n_j \Delta_j Y_{\ell m}^{j*}\\
\tilde C_\ell^\mathrm{N}&=&\frac{1}{4\pi}\sum_j\sigma_j^2G_j^2\Delta_j^2
\end{eqnarray}
The next step is to find the noise correlation matrix,
\begin{eqnarray}
\VEV{\tilde C_\ell^\mathrm{N} \tilde
C_{\ell'}^\mathrm{N}}&=&\frac{1}{(2\ell+1)(2\ell'+1)}\sum_{mm'}\sum_{jj'kk'}\Delta_j\Delta_{j'}\Delta_k\Delta_{k'}
\\
&\times&\VEV{n_jn_{j'}n_kn_{k'}}G_jG_{j'}G_kG_{k'}Y_{\ell m}^jY_{\ell
m}^{j'}Y_{\ell'm'}^kY_{\ell'm'}^{k'}\\
&=&\VEV{C_{\ell}^\mathrm{N}}\VEV{C_{\ell'}^\mathrm{N}}+M_{\ell\ell'}^\mathrm{N},
\end{eqnarray}
where $M_{\ell\ell'}^\mathrm{N}$ can be written as,
\begin{equation}
M_{\ell\ell'}^\mathrm{N}=\frac{2}{(2\ell+1)(2\ell'+1)}\sum_{mm'}\left(\sum_j\Delta_j^2G_j^2\sigma_j^2{Y_{\ell
m}^jY_{\ell' m'}^j}^*\right)^2\\
\end{equation}
For pixelisation schemes like HEALPix, the expression can be evaluated fast using FFT. This is apparent
when one writes the sum over pixels as a double sum over rings and
pixels per ring.
\begin{equation}
\sum_j\Delta_j^2G_j^2\sigma_j^2{Y_{\ell
m}^jY_{\ell' m'}^j}^*=\sum_r\Lambda_{\ell
m}^r\Lambda_{\ell'm'}^r\sum_{p=0}^{N_r-1} \mathrm{e}^{-\mathrm{i}\phi_p(m-m')}\Delta^2G_r^2\sigma_{r,p}^2.
\end{equation}
In the case of an axisymmetric noise model, this expression becomes even
easier which is apparent writing this as
\begin{equation}
\sum_r\Lambda_{\ell
m}^r\Lambda_{\ell'm'}^r\Delta^2G_r^2\sigma_r^2\underbrace{\sum_{p=0}^{N_r-1}
\mathrm{e}^{-\mathrm{i}\phi_p(m-m')}}_{N_r\delta_{mm'}}=\Delta\sum_r\Lambda_{\ell
m}^r\Lambda_{\ell'm}^r\underbrace{\Delta G_r^2\sigma_r^2}_{G'_r}\equiv h'(\ell,\ell',m).
\end{equation}
The sum is equivalent to the previous expression for $h(\ell,\ell',m)$
(equation \ref{eq:hsum}) with a new window $G'_r$. This motivates
the definition of $h(\ell,\ell',m,m')$ such that
\begin{equation}
\label{eq:noisecormat}
M_{\ell\ell'}^\mathrm{N}=\frac{2}{(2\ell+1)(2\ell'+1)}\sum_{mm'}h'^2(\ell,\ell',m,m'),
\end{equation}
where
\begin{equation}
h'(\ell,\ell',m,m')\equiv \Delta\sum_jG'_jY_{\ell m}^jY_{\ell'm'}^j,
\end{equation}
where $G'_j=\Delta G_r^2\sigma^2_{r,p}$. These
function can also be calculated using the recursion which we deduce in appendix (\ref{app:rec}). Note that the
noise correlation matrix usually is diagonally dominant and calculating only the elements close to the diagonal
suffices and speeds up the calculations.

One can then find the total correlation matrix, splitting it up into
one part due to signal, one part due to noise and a cross term,
\begin{eqnarray}
\tilde a_{\ell m}&=&\tilde a_{\ell m}^\mathrm{S}+\tilde a_{\ell m}^\mathrm{N}\\ 
\tilde C_\ell&=&\frac{1}{2\ell+1}\sum_m \VEV{\tilde a_{\ell m} \tilde
a_{\ell'm}}=\tilde C_\ell^\mathrm{S}+\tilde C_\ell^\mathrm{N}+\tilde C_\ell^\mathrm{X}\\
\VEV{\tilde C_\ell}&=&\VEV{\tilde C_\ell^\mathrm{S}}+\VEV{\tilde C_\ell^\mathrm{N}}\\
\tilde C_\ell^\mathrm{X}&=&\frac{1}{2\ell+1}\sum_m\left(a_{\ell m}^\mathrm{S} a_{\ell
m}^\mathrm{N*}+a_{\ell m}^\mathrm{N} a_{\ell m}^\mathrm{S*}\right),
\end{eqnarray}
where the assumption that there is no correlation between
signal and noise was used. One can then see that the correlation matrix
can be written in a similar manner,
\begin{equation}
\VEV{\tilde C_{\ell_i}\tilde C_{\ell_j}}-\VEV{\tilde C_{\ell_i}}\VEV{\tilde
C_{\ell_j}}=M_{ij}^\mathrm{S}+M_{ij}^\mathrm{N}+\VEV{\tilde C_{\ell_i}^\mathrm{X} \tilde C_{\ell_j}^\mathrm{X}}.
\end{equation}
This is another major result of this paper showing the full
correlation matrix of $\tilde C_\ell$ including noise.
One can write the cross term as,
\begin{eqnarray}
\VEV{\tilde C_\ell^\mathrm{X} \tilde C_{\ell'}^\mathrm{X}}&=&4\sum_{mm'}\frac{\VEV{\tilde a_{\ell
m}^\mathrm{S} \tilde a_{\ell'm'}^\mathrm{S*}}\VEV{\tilde a_{\ell m}^\mathrm{N}\tilde
 a_{\ell'm'}^\mathrm{N*}}}{(2\ell+1)(2\ell'+1)}\\
\label{eq:crosscormat}
&=&4\sum_m\frac{\VEV{\tilde a_{\ell m}^\mathrm{S}\tilde a_{\ell'm}^\mathrm{S*}}\VEV{\tilde a_{\ell
m}^\mathrm{N}\tilde a_{\ell'm}^\mathrm{N*}}}{(2\ell+1)(2\ell'+1)},
\end{eqnarray}
where the relation $\VEV{\tilde a_{\ell m}^\mathrm{S}\tilde
 a_{\ell'm'}^\mathrm{S*}}=\delta_{mm'}\VEV{\tilde a_{\ell m}^\mathrm{S}\tilde
 a_{\ell'm}^\mathrm{S*}}$ was used.
From the above, one can see that these two factors can be written as,
\begin{eqnarray}
\VEV{\tilde a_{\ell m}^\mathrm{S}\tilde a_{\ell' m}^\mathrm{S*}}&=&\sum_{\ell''}C_{\ell''}h(\ell,\ell'',m)h(\ell',\ell'',m),\\
\VEV{\tilde a_{\ell m}^\mathrm{N}\tilde a_{\ell'm}^\mathrm{N*}}&=&\sum_iG_i^2Y_{\ell
m}^iY_{\ell' m}^i\Delta_i^2\sigma_i^2,\\
&=&h'(\ell,\ell',m).
\end{eqnarray}

Again using the binning in equation (\ref{eq:binning}), one can write the signal-noise cross correlation matrix
similar to equation (\ref{eq:scorsimp}) as
\begin{equation}
M^\mathrm{X}_{ij}\equiv\VEV{\tilde C_{\ell_i}^\mathrm{X}\tilde C_{\ell_j}^\mathrm{X}}=\sum_kD_b\chi'(b,i,j),
\end{equation}
where
\begin{equation}
\label{eq:chix}
\chi'(b,i,j)\equiv\frac{2}{(2\ell_i+1)(2\ell_j+1)}\sum_m\left(\sum_{l\epsilon
b}B_\ell^2\ell(\ell+1)h(\ell,\ell_i,m)h(\ell,\ell_j,m)\right)h'(\ell_i,\ell_j,m).
\end{equation}

\section{Derivatives of the Likelihood}
\label{app:devlik}

In the minimisation of the likelihood, one also needs
the first and second derivative of
the log-likelihood with respect to the bin values $D_b$ described in equation (\ref{eq:binning}). These can be
found to be,
\begin{equation}
\frac{\partial L}{\partial D_b}=Tr({\mt{A}}_b)+{\mathbf{ f}}^\mathrm{T}{\mathbf{ h}}_b+2\frac{\partial
{\mathbf{ d}}^\mathrm{T}}{\partial D_b}{\mathbf{ f}}
\end{equation}
\begin{eqnarray}
\frac{\partial^2 L}{\partial D_b\partial
D_{b'}}&=&-Tr({\mt{A}}_b{\mt{A}}_{b'})+Tr({\mt{C}}^{-1}\frac{\partial^2 {\mt{C}}}{\partial
D_b \partial D_{b'}})+2{\mathbf{ h}}_b^\mathrm{T}{\mt{C}}^{-1}{\mathbf{ h}}_{b'}\\
&&-2{\mathbf{
h}}^\mathrm{T}_b{\mathbf{ g}}_{b'}-2{\mathbf{ h}}_{b'}^\mathrm{T}{\mathbf{ g}}_b-{\mathbf{ f}}^\mathrm{T}\frac{\partial^2 {\mt{C}}}{\partial
D_b \partial D_{b'}}{\mathbf{ f}}+2{\mathbf{ f}}^\mathrm{T}{\mathbf{ k}}_{bb'}+2\frac{\partial
{\mathbf{ d}}^\mathrm{T}}{\partial D_b}{\mathbf{ g}}_{b'}
\end{eqnarray}
We have used the following definitions,
\begin{eqnarray}
{\mt{A}}_b&=&{\mt{C}}^{-1}\frac{\partial {\mt{C}}}{\partial D_b}\\
{\mathbf{ h}}_b&=&\frac{\partial {\mt{C}}}{\partial D_b}{\mt{C}}^{-1}{\mathbf{ d}}\\
{\mathbf{ g}}_b&=&{\mt{C}}^{-1}\frac{\partial {\mathbf{ d}}}{\partial D_b}\\
{\mathbf{ f}}&=&{\mt{C}}^{-1}{\mathbf{ d}}\\
{\mathbf{ k}}_{bb'}&=&\frac{\partial^2 {\mathbf{ d}}}{\partial D_b\partial D_{b'}}
\end{eqnarray}

Here the derivatives of ${\mathbf{ d}}$ are,
\begin{equation}
\frac{\partial d_i}{\partial D_b}=-\frac{\partial \VEV{\tilde{C_{\ell_i}}}}{\partial
D_b}=\frac{-1}{2\ell_i+1}\sum_m\biggl(\sum_{L'}h(\ell_i,L',m)^2\frac{\partial
C_{L'}}{\partial D_b}B_\ell^2\biggr),
\end{equation}
\begin{equation}
\frac{\partial^2 d_i}{\partial D_b\partial D_{b'}}=-\frac{\partial^2
\VEV{\tilde{C_{\ell_i}}}}{\partial D_b\partial
D_{b'}}=\frac{-1}{2\ell_i+1}\sum_m\biggl(\sum_{L'}h(\ell_i,L',m)^2\frac{\partial^2
C_{L'}}{\partial D_b \partial D_{b'}}B_\ell^2\biggr).
\end{equation}
Obviously for our binning, the double derivative of ${\mathbf{ d}}$ disappears.\\

\section{Correlation between Different Patches}
\label{app:multiple}

Suppose one has two axisymmetric Gabor windows, $G^\mathrm{A}(\mathbf{\hat n})$ and
$G^\mathrm{B}(\mathbf{\hat n})$, centred at two different positions $A$ and $B$ on the
sky. Suppose also that the rotation operators $\hat D^\mathrm{A}$ and $\hat
D^\mathrm{B}$ will rotate these patches so that the centres are on the north
pole. Considering patch $A$, one can define,
\begin{equation}
\tilde a^\mathrm{A}_{\ell m}=\int G^\mathrm{A}_0(\mathbf{\hat n})\left[\hat D^\mathrm{A}
T(\mathbf{\hat n})\right]Y_{\ell m}(\mathbf{\hat n}),
\end{equation}
where $G_0^\mathrm{A}$ is the window $G^\mathrm{A}$ rotated to the north pole.
Since $T(\mathbf{\hat n})=\sum_{\ell m}a_{\ell m}Y_{\ell m}(\mathbf{\hat n})$, one gets
that
\begin{equation}
\hat D^\mathrm{A}T(\mathbf{\hat n})=\sum_{\ell m}a_{\ell m}\sum_{m'}D_{m'm}^{lA}Y_{\ell
m'}(\mathbf{\hat n}).
\end{equation}
Here the $D_{m'm}^\ell$ coefficients are described in appendix
(\ref{app:rotmat}). One now gets,
\begin{eqnarray}
\tilde a^\mathrm{A}_{\ell
m}&=&\sum_{\ell'm'}a_{\ell'm'}\sum_{m''}D_{m''m'}^{\ell'A}h^\mathrm{A}(\ell,\ell',m)\delta_{mm''}\\
&=&\sum_{\ell'm'}a_{\ell'm'}D_{mm'}^{\ell'A}h^\mathrm{A}(\ell,\ell',m),
\end{eqnarray}
where $h^\mathrm{A}(\ell,\ell',m)$ is just the $h(\ell,\ell',m)$ function for
the Gabor window $G^\mathrm{A}(\mathbf{\hat n})$.

The next step is to find the correlations between $\tilde
C^\mathrm{A}_\ell$ and $\tilde C^\mathrm{B}_\ell$, defined for patch $A$ as,
\begin{equation}
\tilde C^\mathrm{A}_\ell=\sum_m\frac{\tilde a^\mathrm{A}_{\ell m}\tilde a^\mathrm{A*}_{\ell m}}{2\ell+1}.
\end{equation}
Following the procedure we used for a single patch one gets,
\begin{eqnarray}
\nonumber
\VEV{\tilde C_\ell^\mathrm{A} \tilde
C_{\ell'}^\mathrm{B}}&=&\frac{1}{(2\ell+1)(2\ell'+1)}\sum_{mm'}\VEV{\tilde a_{\ell
m}^\mathrm{A*} \tilde a_{\ell m}^\mathrm{A} \tilde a_{\ell'm'}^\mathrm{B*} \tilde
a_{\ell'm'}^\mathrm{B}}\\
\label{eq:spot2spot}
&=&\VEV{\tilde C_\ell^\mathrm{A}}\VEV{\tilde
C_{\ell'}^\mathrm{B}}+\frac{2\sum_{mm'}|\VEV{\tilde a_{\ell
m}^\mathrm{A*} \tilde a_{\ell'm'}^\mathrm{B}}|^2}{(2\ell+1)(2\ell'+1)}.
\end{eqnarray}
One can use the expression for $\tilde a^\mathrm{A}_{\ell m}$ to find,
\begin{eqnarray}
\VEV{\tilde a^\mathrm{A}_{\ell m}\tilde a^\mathrm{B*}_{\ell
m}}&=&\sum_{\ell''m''}\sum_{L''M''}\VEV{a_{\ell''m''}a^*_{L''M''}}D_{mm''}^{\ell''A}D_{m'M''}^{L''B*}\\
&\times&h^\mathrm{A}(\ell,\ell'',m)h^\mathrm{B}(\ell',L'',m')\\
&=&\sum_{\ell''}C_{\ell''}\underbrace{\left(\sum_{m''}D_{mm''}^{\ell''A}D_{m'm''}^{\ell''B*}\right)}_{D_{mm'}^{\ell''}(\Delta)}h^\mathrm{A}(\ell,\ell'',m)h^\mathrm{B}(\ell',\ell'',m')\\
\label{eq:spot2spotalm}
&=&\sum_{\ell''}C_{\ell''}d_{mm'}^{\ell''}(\Delta)h^\mathrm{A}(\ell,\ell',m)h^\mathrm{B}(\ell',\ell'',m'),
\end{eqnarray}
where $\Delta$ is the angel between the centres of the
patches. Relations from appendix (\ref{app:rotmat}) were used here.

The next step is to see what happens when noise is introduced. We assume that the
noise is uncorrelated. The noise in pixel $j$ is $n_j$ and
$\VEV{n_jn_{j'}}=\delta_{jj'}\sigma^2_j$. From above one has,
\begin{equation}
\tilde C_\ell^\mathrm{A}=\tilde C_\ell^\mathrm{AS}+\tilde C_\ell^\mathrm{AN}+\tilde
C_\ell^\mathrm{AX},
\end{equation}
where
\begin{eqnarray}
\tilde C_\ell^\mathrm{AN}&=&\sum_m\frac{\tilde a^\mathrm{NA}_{\ell m}\tilde
a^\mathrm{NA*}_{\ell m}}{2\ell+1}\\
\tilde C_\ell^\mathrm{AX}&=&\sum_m\frac{\tilde a^\mathrm{NA}_{\ell m}\tilde
a^\mathrm{SA*}_{\ell m}}{2\ell+1}\\
\tilde a^\mathrm{AN}_{\ell m}&=&\sum_jG^\mathrm{A}_jn^\mathrm{A}_jY_{\ell m}^j,
\end{eqnarray}
where the last sum is over pixels, $G^\mathrm{A}_j$ and $n^\mathrm{A}_j$ being the window
and noise for pixel $j$ respectively.

The correlation between the two patches then becomes,
\begin{equation}
\VEV{\tilde C_\ell^\mathrm{A}\tilde C_{\ell'}^\mathrm{B}}-\VEV{\tilde C_\ell^\mathrm{A}}\VEV{\tilde
C_{\ell'}^\mathrm{B}}=M_{\ell\ell'}^\mathrm{S}+M_{\ell\ell'}^\mathrm{N}+\VEV{\tilde C_\ell^\mathrm{AX}\tilde
C_{\ell'}^\mathrm{BX}},
\end{equation}
where,
\begin{eqnarray}
M_{\ell\ell'}^\mathrm{S}&=&\frac{2}{(2\ell+1)(2\ell'+1)}\sum_{mm'}|\VEV{\tilde a_{\ell
m}^\mathrm{AS*} \tilde a_{\ell'm'}^\mathrm{BS}}|^2\\
M_{\ell\ell'}^\mathrm{N}&=&\frac{2}{(2\ell+1)(2\ell'+1)}\sum_{mm'}|\VEV{\tilde a_{\ell
m}^\mathrm{AN*} \tilde a_{\ell'm'}^\mathrm{BN}}|^2.
\end{eqnarray}

Finally,
\begin{equation}
\VEV{\tilde C_\ell^\mathrm{AX}\tilde
C_{\ell'}^\mathrm{BX}}=\frac{4}{(2\ell+1)(2\ell'+1)}\sum_{mm'}\VEV{\tilde a_{\ell
m}^\mathrm{AS*}\tilde a_{\ell'm'}^\mathrm{BS}}\VEV{\tilde a_{\ell m}^\mathrm{AN*}\tilde a_{\ell'm'}^\mathrm{BN}}.
\end{equation}

Now one needs an expression for $\VEV{\tilde a_{\ell m}^\mathrm{AN}\tilde
a_{\ell'm'}^\mathrm{BN*}}$. One gets,
\begin{equation}
\VEV{\tilde a_{\ell m}^\mathrm{AN}\tilde
a_{\ell'm'}^\mathrm{BN*}}=\sum_{jj'}G^\mathrm{A}_jG^\mathrm{B}_{j'}\VEV{n^\mathrm{A}_jn^\mathrm{B}_{j'}}Y^j_{\ell m}Y^{j'}_{\ell'm'}.
\end{equation}
Here there are only correlations between overlapping pixels. If there
are no overlapping pixels between the patches, this term is
zero. Otherwise this can be written as a sum over the overlapping
pixels
\begin{equation}
\VEV{\tilde a_{\ell m}^\mathrm{AN}\tilde
a_{\ell'm'}^\mathrm{BN*}}=\sum_{j}G^\mathrm{A}_jG^\mathrm{B}_j\sigma_j^2Y^j_{\ell m}Y^{j'}_{\ell'm'}.
\end{equation}

\end{appendix}

\end{document}